\documentclass[12pt]{article}
\usepackage{amscd, amsmath}
\usepackage{amsthm, amssymb}
\usepackage{mathtools}
\usepackage{scalerel}
\usepackage{graphicx}
\usepackage[bottom,flushmargin]{footmisc}
\usepackage{mathrsfs}
\usepackage[format=plain, labelfont=it]{caption}
\usepackage{subcaption}
\usepackage{float}
\usepackage{array}
\usepackage{tabularx}
\usepackage{booktabs}
\usepackage{lscape}
\usepackage{accents}

\newcommand{\linebr}{\\[0.2em]}

\DeclareMathAlphabet{\mathpzc}{OT1}{pzc}{m}{it}

\newtheorem*{assumption(A)}{Condition (A)}

\theoremstyle{remark}

\theoremstyle{remark}

\numberwithin{equation}{section}

\newcommand{\CC}{\mathbb{C}}
\newcommand{\Lie}{\mathcal L}

\newcommand{\VV}{\mathscr V}
\newcommand{\HH}{\mathcal H}
\newcommand{\mph}{|\phi|}
\newcommand{\tphi}{\tilde{\phi}}
\newcommand{\FF}{\mathcal F}
\newcommand{\tg}{\tilde{g}}
\newcommand{\tbeta}{\tilde{\beta}}

\newcommand{\SU}{\mathrm{SU}(3)}
\newcommand{\su}{\mathfrak{su}(3)}
\newcommand{\Utwo}{\mathrm{U}(2)}
\newcommand{\utwo}{\mathfrak{u}(2)}
\newcommand{\sutwo}{\mathfrak{su}(2)}
\newcommand{\kk}{\mathfrak{k}}

\newcommand{\dd}{{\mathrm d}}
\newcommand{\vol}{\mathrm{vol}}
\newcommand{\Vol}{\mathrm{Vol}}
\newcommand{\diag}{\mathrm{diag}}

\DeclareMathOperator{\Tr}{Tr}
\DeclareMathOperator{\Ad}{Ad}
\DeclareMathOperator{\ad}{ad}
\DeclareMathOperator{\Real}{Re}

\DeclareMathOperator{\grad}{grad}
\DeclareMathOperator{\divergence}{div}

\newcommand{\beq}{\begin{equation}}
\newcommand{\eeq}{\end{equation}}
\newcommand{\bal}{\begin{align}}
\newcommand{\eal}{\end{align}}

\newcommand{\bmatr}{\begin{bmatrix}}
\newcommand{\ematr}{\end{bmatrix}}

\newcommand*\conj[1]{\overbar{#1}}
\newcommand{\overbar}[1]{\mkern 1.5mu\overline{\mkern-1.5mu#1\mkern-1.5mu}\mkern 1.5mu}
\newcommand{\hb}{\overbar{h}}

\newcommand{\LL}{{\scaleto{L}{4.8pt}}}
\newcommand{\RR}{{\scaleto{R}{4.8pt}}}
\newcommand{\LLL}{{\scaleto{L}{3.8pt}}}
\newcommand{\RRR}{{\scaleto{R}{3.8pt}}}
\newcommand{\PPP}{{\scaleto{P}{3.1pt}}}
\newcommand{\MMM}{{\scaleto{M}{3.1pt}}}
\newcommand{\KKK}{{\scaleto{K}{3.1pt}}}

\makeatletter

\renewenvironment{bmatrix}
  {\left[\mkern3.5mu\env@matrix}
  {\endmatrix\mkern3.5mu\right]}

\def\blfootnote{\xdef\@thefnmark{}\@footnotetext}
\makeatother

\begin{document}

\begin{titlepage}
\title{\bf Higher-dimensional routes \\ to the Standard Model bosons}
\vskip -70pt

\vspace{3cm}

\author{{Jo\~ao Baptista}}  
\date{May 2021}

\maketitle

\thispagestyle{empty}
\vspace{0.5cm}
\vskip 20pt
{\centerline{{\large \bf{Abstract}}}}
\noindent
In the old spirit of Kaluza-Klein, we consider a spacetime of the form $P = M_4 \times K$, where $K$ is the Lie group $\SU$ equipped with a left-invariant metric that is not fully right-invariant. This metric has a ${\rm U}(1) \times \SU$ isometry group, corresponding to the massless gauge bosons, and depends on a parameter $\phi$ with values in a subspace of $\su$ isomorphic to $\CC^2$. It is shown that the classical Einstein-Hilbert Lagrangian density $R_P - 2 \Lambda$ on the higher-dimensional manifold $P$, after integration over $K$, encodes not only the Yang-Mills terms of the Standard Model over $M_4$, as in the usual Kaluza-Klein calculation, but also a kinetic term $|\dd^A \phi|^2$ identical to the covariant derivative of the Higgs field. For $\Lambda$ in an appropriate range, it also encodes a potential $V(| \phi|^2)$ having absolute minima with $|\phi_0|^2 \neq 0$, thereby inducing mass terms for the remaining gauge bosons. The classical masses of the resulting Higgs-like and gauge bosons are explicitly calculated as functions of the vacuum value $|\phi_0|^2$ in the simplest version of the model. In more general versions, the classical values of the strong and electroweak gauge coupling constants are given as functions of the parameters of the left-invariant metric on $K$.

\end{titlepage}

\tableofcontents

\newpage 

\section{Introduction}
  
Traditional Kaluza-Klein theories propose to replace four-dimensional Minkowski spacetime $M_4$ with a higher-dimensional product manifold $P \!=\! M_4 \times K$, where the internal space $K$ is a Lie group or a homogeneous space with very small volume. The proposed Lorentzian metric on P is not the simple product of the metrics on $M_4$ and $K$, but has non-diagonal terms that can be interpreted as the observed gauge fields on $M_4$. Geometrically, the projection $P \to M_4$ should be a Riemannian submersion with fibre $K$. 

The original Kaluza-Klein choice $K = \mathrm{U}(1)$ has the remarkable feature that geodesics on $P$ project down to paths on $M_4$ satisfying the Lorentz law for moving electric charges. For general choices of $K$, it can be shown that a natural quantity on $P$, namely its scalar curvature $R_P$, can be written as a sum of components that include the individual scalar curvatures of $M_4$ and $K$ and, more remarkably, the norm $|F_A|^2$ of the gauge field strength. Since the scalar curvature is also the Lagrangian density for general relativity, it follows that the Einstein-Hilbert action on the higher-dimensional $P$ produces, after projection down to $M_4$, two of the essential ingredients of physical field theories in four dimensions: the Einstein-Hilbert and the Yang-Mills Lagrangians on $M_4$.

Kaluza-Klein theories, however, do present challenging difficulties when interpreted simply as higher-dimensional versions of general relativity, i.e. as dynamical field theories for a metric tensor on $P$ that satisfies the full Einstein equations on the higher-dimensional space. Although unifying and appealing, the direct extension of general relativity to higher dimensions seems to imply the existence of many unobserved scalar fields satisfying complicated equations of motion with few physically reasonable solutions. The new fields generally do not bear much resemblance to the well-known field content of the Standard Model. Moreover, following the interpretation of fermions in Kaluza-Klein theory as zero modes of the Dirac operator on the internal space K, there does not seem to be a good choice of Riemannian manifold $K$ able to deliver the necessary zero modes in the chiral representations appearing in the Standard Model. For reviews and discussions of Kaluza-Klein theory from different viewpoints, see for instance \cite{Bailin, Duff, Witten81, Witten83, WessonOverduin, Coq, Hogan, Bleecker}. Some of the early original references are \cite{Original}, with much more complete lists given in the mentioned reviews.

The plan of the present investigation is to dig deeper into some of the geometrical aspects of the Kaluza-Klein framework and suggest that, besides the curvature $|F_A|^2$ of the gauge fields, there are other natural objets in a Riemannian submersion that resemble the field content of the Standard Model. For example, when the fibre $K$ is a Lie group equipped with a left-invariant metric, the second fundamental form of the fibres, denoted by $S$, generates terms in the four-dimensional Lagrangian sharing notable similarities with the covariant derivative of a Higgs field. See for instance the general formula \eqref{NormT2} for the norm $|S|^2$, whose quadratic terms in the gauge fields are what is needed to generate the gauge bosons' mass. When $K$ is chosen to be the group $\SU$ equipped with a specific family of left-invariant metrics, denoted by $g_\phi$, then the terms generated by $S$ contain the precise covariant derivative $\dd^A \phi$ that appears in the Standard Model, namely a $\CC^2$-valued Higgs field coupled to the electroweak gauge fields through the correct representation.

In the companion study \cite{Baptista}, we suggest possible ways to integrate fermions in this picture. For an internal space $K = \SU$, we regard fermions as spinorial functions on the 12-dimensional spacetime $M_4 \times K$ with a prescribed behaviour along $K$. A complete generation of fermionic fields can then be encoded in the 64 complex components of a single higher-dimensional spinor. Moreover, the vertical behaviour of this spinor can be chosen so that, after fibre-integration over K, the resulting Dirac kinetic terms in four dimensions couple to the $\mathfrak{u}(1) \oplus \sutwo \oplus \su$ gauge fields in the exact chiral representations present in the Standard Model. Perhaps one could think of the prescribed vertical behaviour as a sort of elementary, spinorial oscillation along the compact direction $K$.

\subsubsection*{Decomposing the higher-dimensional scalar curvature} 
%\addcontentsline{toc}{subsection}{Decomposing the higher-dimensional scalar curvature}

This second part of the Introduction gives a brief description of the calculations that motivate the present study. Let $\beta$ be an $\Ad_{\SU}$-invariant inner-product on the Lie algebra $\su$. Using the left-translations on the group, this product can be extended to a left-invariant metric on the whole manifold $K = \SU$. The Ad-invariance of $\beta$ guarantees that the resulting metric is bi-invariant on $K$, i.e. it has isometry group $\SU \times \SU$. In this study we will consider a deformation $g_\phi$ of the product $\beta$ that extends to $K$ as a left-invariant metric that is no longer bi-invariant, but has the smaller isometry group $\rm{U}(1) \times \SU$. To do that, observe that any matrix in the Lie algebra $\su$ can be uniquely written as
\beq \label{AlgebraDecomposition1}
v \ = \ \bmatr - \Tr(v') &  - (v'')^\dag \linebr v'' & v' \ematr \ ,
\eeq
where $v'$ is an anti-hermitian matrix in $\utwo$ and $v''$ is a vector in $\CC^2$. This determines a vector space decomposition $\su \simeq \utwo \oplus \CC^2$ that is orthogonal with respect to $\beta$. Identifying $v'$ and $v''$ with their images in $\su$, the deformed inner-product on this space is defined by the three equations
\bal \label{DefinitionMetric2}
g_\phi (u',v') \ &= \ \beta (u',v')  \\
g_\phi (u',v'') \ &= \ \beta ( \,[u', v''] ,\, \phi)  \nonumber \\
g_\phi (u'',v'') \ &= \ \beta (u'',v'')  \nonumber \ .
\end{align}
The deformation parameter is a vector $\phi \in \CC^2$ after identification with the matrix
\beq 
 \bmatr  & - \phi^\dag  \\  \phi &  \ematr \quad \in \ \su \ .
\eeq
As in the usual Kaluza-Klein framework, the left-invariant metric $g_K = g_\phi$ on the internal space $K$ can be combined with a metric $g_M$ and one-forms $A$ on Minkowski space to define a submersive metric $g_P$ on the higher-dimensional space $P =  M_4 \times K$. In our case, there are two one-forms $A_\LL$ and $A_\RR$ on $M_4$ with values in the Lie algebra $\su$. Using a basis $\{ e_j \}$ of $\su$, they can be decomposed as $A_\LL^j \, e_j$ and $A_\RR^j \, e_j$. Now denote by $e_j^\LL$ and $e_j^\RR$ the extensions of $e_j$ as left and right-invariant vector fields on $K$, respectively. We can construct a one-form $A$ on $M_4$ with values in the space of invariant vector fields on $K$ by the formula
\[
A(X) \ := \ \textstyle{\sum_{\, j}} \ A_\LL^j (X) \, e_j^\LL \; - \; A^j_\RR (X) \, e_j^\RR 
\]
for all tangent vectors $X \in TM$. Then the higher-dimensional metric on $P$ is defined by
\bal
g_P (V, V) \ &:= \ g_K (V, V)    \nonumber \\
g_P (X, V) \ &:= \  -\, g_K (A(X),\,  V)   \nonumber \\
g_P (X, X) \ &:= \ g_M (X, X)  \; + \;  g_K (A(X) \, , \, A(X))         \ , 
\end{align}
for all $X \in TM$ and all vertical vectors $V \in TK$. This fully determines the higher-dimensional metric. In this study we will investigate the scalar curvature of the metric $g_P$. A standard result in Riemannian submersions \cite{Besse} says that it can be decomposed as 
\beq   
R_P \ = \ R_M \ + \ R_K \ -\  |\FF|^2 \ - \ |S|^2 \ - \ |N|^2 \ - \ 2\, \check{\delta} N  \nonumber  \ .
\eeq
Here $R_M$ and $R_K$ are the scalar curvatures of the metrics $g_M$ and $g_K$, respectively; $|\FF|^2$ is the component that originates the Yang-Mills terms $|F_A|^2$ in the usual Kaluza-Klein calculation; the tensor $S$ is the second fundamental form of the fibres $K$, also called shape operator; the vector $N$ is the trace of $S$, which is a horizontal vector in $TP$ usually called the mean curvature vector of the fibres. On a Riemannian submersion, the tensor $S$ vanishes precisely if the all the fibres $K$ are geodesic submanifolds of $P$. In this case all the fibres will be isometric to each other. The vector $N$ can be thought of as the gradient in $P$ of the volume of the fibres, which may vary as one moves along the base $M_4$. Thus, vanishing $N$ means that all internal spaces have the same volume.

Since the metric $g_P$ can be written as a function of $g_K$, $g_M$ and the one-forms $A_\LL$ and $A_\RR$, the same must be true for all the terms of the scalar curvature $R_P$.  
Now fix the choice of internal metric $g_K = g_\phi$. If we assume that the one-form $A_\RR$ has values in the full $\su$ but $A_\LL$ has values in the smaller electroweak subalgebra $\utwo \subset \su$, then the integral has the following schematic result:
\begin{multline}    
\int_K \big(\,R_P \, - \, 2\, \Lambda_P \,) \ \vol_{g_\phi}  \ = \  -\, \bigg[ \,  \frac{1}{4}\,  B_\phi \, \left(\, |F_{A_\LLL}|^2 \, + \, |F_{A_\RRR}|^2 \, \right) \,+ \,C_\phi \,  \big|\dd^{A_\LLL} \phi \big|^2 \,   \linebr 
+ \, D_\phi \,  \big|\dd\, |\phi|^2 \, \big|^2   
\ + \, U( \mph^2) \, + \, 2 \,\Delta_M f_\phi  \  \bigg]  \, \Vol(K, \beta)  \nonumber \ .
\end{multline}
The coefficients $B_\phi$, $C_\phi$, $D_\phi$ and $f_\phi$ are functions of the norm $|\phi|^2$ in $\CC^2$ that will be explicitly computed later. Thus, the integral's result is a Lagrangian density on $M_4$ that contains: 1) strong and electroweak Yang-Mills terms; 2) the norm $\big|\dd^{A_\LLL} \phi \big|^2$ of a covariant derivative coupling the field $\phi \in \CC^2$ to the electroweak gauge fields $A_\LL$, but not to the strong force gauge fields $A_\RR$; 3) a total derivative term $\Delta_M f_\phi$ that does not affect the four-dimensional equations of motion; 4) a potential term
\[
 U( \mph^2) \ := \ ( 2\, \Lambda_P \, -\, R_{g_\phi} \,- \, R_M )\ f_\phi \ ,
\]
involving the scalar curvature $R_K = R_{g_\phi}$ and the volume density $f_\phi$ of the internal space; 5) finally, a term proportional to the norm of the derivative $\dd\, |\phi|^2$ that only affects the equations of motion of $\phi$ and the mass of the Higgs-like boson. In the simplest versions of the model, it can be shown that when the constant $2 \Lambda_P - R_M$ is larger than a certain critical value, the potential $U( \mph^2)$ has absolute minima for $|\phi|^2 \neq 0$ and explodes to positive infinity when $|\phi|^2$ approaches the boundary value $1/4$. Overall, the result of the fibre-integral over $K$ is a density in $M_4$ remarkably similar to the bosonic part of the Standard Model Lagrangian.

Sections 2 and 3 of this study are dedicated to the calculations necessary to arrive at the four-dimensional Lagrangian density described above, after fibre-integration of the higher-dimensional scalar $R_P  -  2 \Lambda_P$. Section 4 starts from this Lagrangian on $M_4$ and calculates the classical masses of the associated Higgs-like and gauge-bosons as a functions of the ``vacuum value'' of $|\phi_0|^2$. Section 5 describes a more precise version of the model, where the deformation $g_\phi$ of the internal metric depends on additional parameters that essentially correspond to the three gauge coupling constants of the Standard Model. In addition, it discusses some of the important questions that are not sufficiently clarified or even addressed here, such as the mass in this model of the four additional gauge bosons present in the full $\SU \times \SU$ gauge theory, and the stability of vacuum configurations of the form $g_P = g_M \times g_\phi$ under the full higher-dimensional Einstein equations of motion. The discussion in this study also does not encompass the fundamental quantum aspects of the Standard Model.

\newpage

\section{A left-invariant metric on $\SU$}

\subsection*{Decomposition of $\su$} 
\addcontentsline{toc}{subsection}{Decomposition of $\su$}

Consider the eight-dimensional Lie group $K\! =\! \SU$ and the group homomorphism $\iota: \Utwo \to K$ defined by
\beq
\iota(a) \ = \ \bmatr (\det a)^{-1} &   \\   & a \ematr \ .
\eeq 
This map induces an inclusion of Lie algebras $\iota: \utwo \to \su$ that is denoted by the same symbol:
\beq \label{AlgebraInclusionHomomorphism}
\iota(v') \ = \ \bmatr - \Tr(v') &   \\   & v' \ematr \ .
\eeq
Any matrix $v$ in $\su$ can be uniquely written as in \eqref{AlgebraDecomposition1}, where $v'$ is a matrix in $\utwo$ and $v''$ is a vertical vector with two complex components. This defines a decomposition of $\su$ and an isomorphism of real vector spaces
\beq \label{AlgebraDecomposition2}
\iota: \utwo \oplus \CC^2 \ \longrightarrow \  \su \ ,
\eeq
which extends \eqref{AlgebraInclusionHomomorphism} and is still denoted by the same symbol. This decomposition of $\su$ is clearly orthogonal with respect to the usual Ad-invariant inner product on the space:
\beq \label{DecompositionInnerProduct}
\beta_0 (u,v)  \ :=  \ \Tr(u^\dag\, v) \ = \   \conj{\Tr(u')} \, \Tr(v') \, + \, \Tr[\, (u')^\dag \,v'\, ] \, + \, 2 \, \Real \left[ (u'')^\dag v''\, \right]  \ . 
\eeq
When acting on vectors in the summand subspaces, the Lie bracket of $\su$ satisfies the simple relations
\bal
 [\,\utwo, \, \utwo\,] \ &= \ \sutwo \ \subset \ \utwo  \\
 [\,\CC^2, \, \CC^2\, ]          \   &=\  \utwo    \nonumber \\
 [\,\utwo, \, \CC^2\,]    \  &= \ \CC^2  \nonumber \ ,
 \end{align}
where we have denoted $\iota(\utwo)$ and $\iota(\CC^2)$ simply by $\utwo$ and $\CC^2$, as will be often done ahead. The adjoint action of any group element $a \in \Utwo$  on the algebra $\su$ can then be written as
\beq \label{AdjointAction}
\Ad_{\iota(a)} (v) \ = \ \bmatr -\Tr(v') &  - [\,(\det a)\, a\, v'' \,]^\dag \\[0.6em]   (\det a)\, a\, v''  &  \Ad_a (v') \ematr  \ .
\eeq
Observe that the action of $\Utwo$ on the vector $v''$ in $\CC^2$ coincides with the action of the same group on the Higgs field $\phi$ in the Standard Model, having the hypercharge necessary to absorb the fermionic hypercharges in the Yukawa coupling terms (see \cite{Hamilton}, for instance).

The decomposition $\utwo \oplus \CC^2$ of the matrix space $\su$ can also be thought of as an eigenspace decomposition with respect to the involution
\beq \label{DefinitionTheta}
v \ \longmapsto \  \Ad_\theta v \  =\  \theta \,v \, \theta 
\eeq
defined by the diagonal matrix $\theta \, := \, \diag (1, -1, -1)$ in $\SU$. The involution $\Ad_\theta$ has eigenvalue $+1$ on the subspace $\iota(\utwo)$ and eigenvalue $-1$ on the subspace $\iota(\CC^2)$ of $\su$.

\subsection*{General properties of left-invariant metrics} 
\addcontentsline{toc}{subsection}{General properties of left-invariant metrics}

In the next few paragraphs we introduce notation and mostly describe standard properties of left-invariant metrics on a Lie group. See for instance \cite{Milnor, BD}.
As a vector space, the Lie algebra of a group is the tangent space to the group at the identity element. A vector $v$ in the Lie algebra $\kk$ can be extended to a vector field on the group $K$ in two canonical ways, as a left-invariant vector field $v^\LL$ or as a right-invariant field $v^\RR$. They satisfy
\beq
(L_h)_\ast (v^\LL) \ = \ v^\LL \qquad \qquad \qquad (R_h)_\ast (v^\RR) \ = \ v^\RR
\eeq
for all group elements $h \in K$, where $L_h (h') = h\,h'$ and $R_h (h') = h'\,h$ denote the left and right-multiplication automorphisms on the group. The one-parameter flows on $K$ associated to these vector fields can be written in terms of the exponential map $\exp: \mathfrak{k} \to K$ as
\beq
\Phi_t^{v^\LLL} (h) \ = \ h \, \exp (t\,v) \qquad \qquad \qquad  \Phi_t^{v^\RRR} (h) \ = \ \exp (t\,v) \, h \ .
\eeq
The explicit expressions for the flows can be used to show that the Lie brackets of invariant vector fields are also invariant on $K$ and satisfy
\beq \label{BracketsInvariantFields}
[u^\LL, \, v^\LL ] \ = \ [u, \, v ]^\LL_\kk  \qquad \qquad  [u^\RR, \, v^\RR ] \ = \ - [u, \, v ]^\RR_\mathfrak{k}   \qquad \qquad   [u^\LL, \, v^\RR ] \ = \ 0 \ ,
\eeq
where the bracket $[\, .\, , \, . \,]_\kk$ in the Lie algebra is just the commutator of matrices in the case of matrix Lie groups. Just as with vectors, any tensor in the Lie algebra $\kk$ can be extended to a left or right-invariant tensor field on $\kk$. For example, given an inner product $g$ on $\kk$, it can be extended to a left-invariant metric on $K$ by decreeing that the product of left-invariant vector fields should have the same value everywhere on $K$ and coincide with $g$ at the identity element of the group, thus $g(u^\LL, v^\LL) = g(u,v)$. In the opposite direction, every left-invariant metric on $K$ is fully determined by its restriction to the Lie algebra. When a left-invariant metric is applied to right-invariant vector fields the result is a function on $K$ that is not constant in general, but still simple enough:
\bal \label{ProductInvariantFields}
 g(u^\LL, v^\RR)\, |_h \ =\  g(u, \, \Ad_{h^{-1}} v)   \qquad \qquad    g(u^\RR, v^\RR)\, |_h \ =\  g(\Ad_{h^{-1}} u, \, \Ad_{h^{-1}} v)
\end{align}
for all elements $h \in K$ and all vectors $u, v$ in the Lie algebra. The preceding observations are enough to recognize that right-invariant fields are always Killing vector fields for left-invariant metrics on K, since 
\beq
( \Lie_{w^\RRR} g) (u^\LL, v^\LL) \ = \  \Lie_{w^\RRR} \left(  g(u^\LL,v^\LL)  \right) - g([w^\RR, u^\LL],v^\LL) - g(u^\LL, [w^\RR, v^\LL]) \ = \ 0 \ .
\eeq
The same is not true for general left-invariant vector fields, since
\bal \label{LieDerivativeMetric1}
( \Lie_{w^\LLL} g) (u^\LL, v^\LL) \ &= \  \Lie_{w^\LLL} \left(  g(u^\LL,v^\LL)  \right) - g([w^\LL, u^\LL],v^\LL) - g(u^\LL, [w^\LL, v^\LL])    \nonumber  \linebr
 &= \ - g([w, u],v) - g(u, [w, v]) \ 
\end{align}
entails that the Lie derivative $\Lie_{w^\LLL} g$ may be a non-zero left-invariant tensor on $K$. In the special case when $g$ is an Ad-invariant inner-product on $\kk$, then $g(u^\RR, v^\RR)$ is also a constant function on $K$ and the metric $g$ is both left and right-invariant. In this case left-invariant vector fields are Killing as well. These are called bi-invariant metrics on the group and, when $K = \SU$, they coincide with minus the Killing form, up to a constant factor.

If a left-invariant vector field $v^\LL$ is indeed Killing, then the usual Killing condition in terms of the Levi-Civita connection implies that, for any other invariant field $u^\LL$,
\beq
g(\nabla_{v^\LLL} v^\LL, \, u^\LL) \ = \ -\, g(\nabla_{u^\LLL} v^\LL, \, v^\LL)  \ = \ -\, \frac{1}{2} \ \Lie_{u^\LLL}  \big[ \, g(v^\LL, \, v^\LL) \, \big] \ = \ 0 \ .
\eeq
Thus, we must have that $\nabla_{v^\LLL} v^\LL$ vanishes as a vector field on $K$. In particular, the flow lines $t \mapsto  h \, \exp(t v)$ generated by the field $v^\LL$ are affinely parameterized geodesics on $K$.

The Riemannian volume form $\vol_g$ of a left-invariant metric $g$ is always a left-invariant differential form on the group. In the case of connected, unimodular Lie groups, such as our $K = \SU$, it is also a right-invariant form, even though the metric itself may not be right-invariant. Thus, we always have here
\beq
(L_h)^\ast \,\vol_g \ = \ (R_h)^\ast \,  \vol_g \ = \ \vol_g \ ,
\eeq
and the bi-invariant volume form of $g$ coincides, up to normalization, with the Haar measure on $K$. Standard results on invariant integration on Lie groups \cite{BD} then say that, for any smooth function $f(h)$ on $K$ and any fixed element $h'$ in the group,
\beq
\int_{h\in K} f(h) \, \vol_g \ = \ \int_{h\in K} f(h'\, h) \, \vol_g \ = \ \int_{h\in K} f(h\, h') \, \vol_g \ = \ \int_{h\in K} f(h^{-1}) \, \vol_g  \ .
\eeq
So the variable of integration can be changed by left-multiplication, right-multiplication or inversion without changing the result. This invariance extends to other automorphism of the Lie group, such as matrix transposition or matrix conjugation in the case of our $K=\SU$:
\beq
\int_{h\in K} f(h) \, \vol_g \ = \ \int_{h\in K} f(h^T) \, \vol_g \ = \ \int_{h\in K} f(\conj{h}) \, \vol_g  \ = \ \int_{h\in K} f(h^\dag) \, \vol_g  \ .
\eeq
These invariance properties of integrals can be used to show, for instance, that for any vector $v$ in the Lie algebra of a simple Lie group we have
\beq \label{VanishingAdIntegral}
\int_{h\in K} \Ad_h (v) \, \vol_g \  =  \ 0 \ .
\eeq
This is true because the result of the integral is an Ad-invariant vector in the Lie algebra,
\beq
\int_{h\in K} \Ad_h (v) \, \vol_g \  =  \   \int_{h\in K} \Ad_{h'h} (v) \, \vol_g \  =  \ \Ad_{h'} \left( \int_{h\in K} \Ad_h (v) \, \vol_g \right)   , \nonumber
\eeq
and hence  belongs to the centre of the algebra, which only contains the zero element in the case of simple groups. In particular, it follows that left and right-invariant vector fields look orthogonal to each other after integration over $K$, since \eqref{ProductInvariantFields} and \eqref{VanishingAdIntegral} imply that
\beq \label{ProductInvariantVectors1}
\int_{h\in K}  g(u^\LL, v^\RR)\ \vol_g \  =  \ 0 \ 
\eeq
for all vectors $u$ and $v$ in $\kk$ and for all left-invariant metrics. The integral over $K$ of inner-products of the form $g(u^\LL, v^\LL)$ is also easy to compute, since these are constant functions on $K$, by definition of left-invariant metric. So
\beq \label{ProductInvariantVectors2}
\int_{h\in K}  g(u^\LL, v^\LL)\ \vol_g \  =  \  g(u, v)\ \Vol (K, g)  \ .
\eeq
The integral over $K$ of the product $g(u^\RR, v^\RR)$ is not immediate in general, although it does follow from the second equality in \eqref{ProductInvariantFields} that it is Ad-invariant and hence proportional to the Cartan-Killing product on the simple algebra $\kk$. This happens because the second integral in
\beq \label{ProductInvariantVectors5}
\int_{h\in K}  g(u^\RR, v^\RR)\ \vol_g \  =  \  \int_{h\in K}  g(\Ad_{h^{-1}} u, \, \Ad_{h^{-1}} v)\ \vol_g \ \  \propto \   \Tr ({\rm{ad}}_u \, {\rm{ad}}_v) \, \Vol (K, g)  \ 
\eeq
is explicitly averaging the pull-back metric $\Ad^\ast_{h^{-1}} g$ over $K$, and hence is invariant under a change of variable $h \to h' h$ for any fixed group element $h' \in K$.

Finally, the Ricci curvature of a left-invariant metric is also a left-invariant tensor on $K$. This implies that the scalar curvature is constant on the group. Its value can be expressed in terms of a $g$-orthonormal basis $\{u_j\}$ of the Lie algebra $\kk$ through the formula
\beq \label{GeneralScalarCurvature}
R_g \ = \ - \sum_{i,j} \  \frac{1}{4} \, g \left( [u_i, u_j],\, [u_i, u_j] \right)  \,  + \,  \frac{1}{2} \, g([\,u_i, [u_i, u_j]\, ] ,\, u_j) \ .
\eeq
This expression is valid for unimodular Lie groups and is a special case of a well-known formula for the scalar curvature of homogeneous spaces (e.g. see chapter 7 of \cite{Besse}).

\subsection*{A family of metrics on $\SU$}
\addcontentsline{toc}{subsection}{A family of metrics on $\SU$}

Start by considering the general bi-invariant metric on $K = \SU$, determined by its restriction to $\su$ and unique up to a positive real constant $\lambda$:
\beq \label{definitionbeta}
\beta (u,v) \ := \ \lambda\, \beta_0 (u,v) \ = \ \lambda \, \Tr(u^\dag \, v) \ .
\eeq
We want to deform this metric and break its bi-invariance using a parameter $\phi \in \CC^2$.
It was noted in \eqref{AlgebraDecomposition2} that there exists an isomorphism of vector spaces $\iota: \utwo \oplus \CC^2 \to \su$ that takes any element $\phi \in \CC^2$ to the matrix 
\beq \label{phi}
\iota (\phi) \ = \ \bmatr  & - \phi^\dag  \\  \phi &  \ematr \quad \in \ \su \ .
\eeq
We use the parameter $\iota(\phi)$, together with decomposition \eqref{AlgebraDecomposition1}, to define a new inner-product $g_\phi$ on $\su$ through the general formula
\bal \label{DefinitionMetric1}
g_\phi (u,v) \ &:= \ \beta (u,v) \, +\, \beta \left( \,[u', v''] + [v', u''] ,\, \phi\, \right)  \linebr
                    & \ = \ \beta (u,v) \, + \, \beta \left(\, [\Ad_\theta u ,\, v] , \, \phi \, \right)  \ . \nonumber
\end{align}
Here we have simplified the notation by omitting the isomorphism $\iota$ to write $\phi$, $v'$ and $v''$ instead of the respective $\su$ matrices $\iota(\phi)$, $\iota(v')$ and $\iota(v'')$. We will do this often below, writing formulae such as $v = v' + v''$ and regarding the components as elements of $\su$.

A first observation is that the deformed product $g_\phi$ coincides with $\beta$ when restricted to the subspace $\utwo$ of $\su$, since both $u''$ and $v''$ vanish in this case. For similar reasons, the two products coincide when restricted to the subspace $\CC^2$. It is only in products mixing both subspaces that $g_\phi$ differs from $\beta$. It is clear that the inner-product $g_\phi$ can be equally characterized by the three identities \eqref{DefinitionMetric2}, which show, in passing, that the two subspaces of $\su$ are no longer orthogonal. Using the Ad-invariance of $\beta$, it can be readily verified that the orthogonal complement to $\utwo$ in $(\su, g_\phi)$ is the subspace
\beq \label{orthogonal1}
\utwo^\perp \ = \  \left\{ v'' \,+ \,[\phi,\, v''] \, : \ v'' \in \CC^2   \right\}  \ ,
\eeq
while the orthogonal complement to $\CC^2$ is the subspace
\beq \label{orthogonal2}
(\CC^2)^\perp \ = \  \left\{ v' \,+ \, [v' , \, \phi] \, : \ v' \in \utwo   \right\} \ .
\eeq
The Ad-invariant product $\beta$ is positive-definite, so the new product $g_\phi$ will maintain that property if the parameter $\phi \in \CC^2$ is sufficiently small. For larger $\phi$ it may become an indefinite product. It can be shown that $g_\phi$ is positive-definite if and only if the vector $\phi = [\phi_1\ \, \phi_2]^T$ in $\CC^2$ satisfies
\[
|\, \phi \, |^2_{\CC^2}\  =\  |\phi_1|^2 \, + \, |\phi_2|^2  \ < \ \frac{1}{4} \ .
\]
We will always assume that the parameter is in this range. 

By construction, the new product $g_\phi$ is not Ad-invariant in $\su$. However, its transformation is simple enough when the adjoint action is restricted to elements in the subgroup $\iota(\Utwo)$ of $\SU$, which always preserve the decomposition $\utwo \oplus \CC_2$ of $\su$. If we take any element $a \in \Utwo$, it follows from \eqref{DefinitionMetric2} and the Ad-invariance of $\beta$ that
\[
\left[ (\Ad_{\iota(a)^{-1}})^\ast \,g_\phi \right]  \, (u', v'') \ = \ \beta ( \,[\Ad_{\iota(a)^{-1}} u', \ \Ad_{\iota(a)^{-1}} v''] ,\, \phi) \ =  \ \beta (  \,[u', v''] ,\, \Ad_{\iota(a)} \phi ) \ .
\]
Combining with expression \eqref{AdjointAction} for $\Ad_{\iota(a)}$, we conclude that $g_\phi$ transforms as
\beq
(\Ad_{\iota(a)^{-1}})^\ast \,g_\phi \ = \ g_{(\det a) a\, \phi} \  
\eeq
for any $a \in \Utwo$. In other words, when the subgroup $\Utwo$ of $\SU$ acts on the product $g_\phi$ through the co-adjoint action, the parameter $\phi$ simply rotates in $\CC^2$ in a representation analogous to the Higgs field one.

In the section on general left-invariant metrics, we saw that inner-products of the form $g(u^\RR, v^\RR)$ have an integral over the group $K$ that is proportional to the Ad-invariant product of the vectors $u$ and $v$ in $\su$. We will now calculate the constant of proportionality for the case $g = g_\phi$. It follows from \eqref{ProductInvariantFields}, the definition of $g_\phi$ and the Ad-invariance of $\beta$ that
\beq \label{ProductInvariantVectors4}
g_\phi (u^\RR, v^\RR) \ |_{h} \ = \  \beta (u, v)  \, + \, \beta\big(\,[\Ad_{\theta  h^{-1}}u, \Ad_{h^{-1}} v ] , \, \phi \, \big) \ ,
\eeq
for any $h \in \SU$. Since the volume form $\vol_{g_\phi}$ is bi-invariant, the integral of the second term must be invariant under the change of variable $h \to h\theta$, where $\theta = \diag(1, -1, -1)$. Thus,
\bal 
\int_{h \in K} \, \beta\big([\Ad_{\theta  h^{-1}}u, \Ad_{h^{-1}} v ] , \phi \big) \ \vol_{g_\phi} \  &= \  \int_{h \in K} \, \beta\big([\Ad_{\theta  (h \theta)^{-1}}u, \Ad_{(h\theta)^{-1}} v ] , \, \phi \big)  \ \vol_{g_\phi} \nonumber \linebr
 &= \  \int_{h \in K} \, \beta\big([\Ad_{\theta  h^{-1}}u, \Ad_{h^{-1}} v ] , \, \Ad_{\theta} \phi \big)  \ \vol_{g_\phi} \nonumber \linebr 
 &= \  - \, \int_{h \in K} \, \beta\big([\Ad_{\theta  h^{-1}}u, \Ad_{h^{-1}} v ] , \, \phi \big)  \ \vol_{g_\phi} \nonumber \ . 
\end{align}
This shows that the integral of the second term is zero, and so 
\beq \label{ProductInvariantVectors3}
\int_{K} \,  g_\phi (u^\RR, v^\RR)  \ \vol_{g_\phi} \  = \  \int_{K} \,  \beta (u, v)  \ \vol_{g_\phi} \  = \ \beta (u, v) \, \Vol (K, g_\phi)  \  . 
\eeq
This means that the inner-product of right-invariant vector fields, after integration over $K$, is completely blind to the deformation of the metric caused by the parameter $\phi$. The integrals over $K$ of inner-products of the form $g_\phi (u^\LL, v^\RR)$ and $g_\phi (u^\LL, v^\LL)$ have already been calculated in \eqref{ProductInvariantVectors1} and \eqref{ProductInvariantVectors2}, respectively.

\subsection*{Killing vector fields of $g_\phi$}
\addcontentsline{toc}{subsection}{Killing vector fields of $g_\phi$}

The inner-product $g_\phi$ on the Lie algebra $\su$ can be extended to a left-invariant metric on the group $\SU$, as described before. The right-invariant vector fields $u^\RR$ will then be Killing fields of $g_\phi$ for every vector $u \in \su$. The same is not true for the left-invariant fields $u^\LL$ when $\phi \neq 0$, since expression \eqref{LieDerivativeMetric1} says that the Lie derivative of the metric is given by
\bal \label{LieDerivativeMetric2}
(\Lie_{u^\LLL} \,  g_\phi) (v^\LL, v^\LL) \ &= \   2 \, g_\phi (\, [v, u], \, v\,) \linebr
                                                                &= \ 2\, \beta \big( \,[v', v''],\,  [u', \phi]\, \big) \ + \ 2\, \beta \big( \,[[v', u''], v'] \, + \, [[u'', v''], v''],\,  \phi \, \big)  \nonumber \ .
\end{align}
The second equality is obtained after inserting the definition of $g_\phi$ and using both the Ad-invariance of $\beta$ and the Jacobi identity for the Lie bracket on $\su$. The left-invariant vector field $u^\LL$ will be Killing precisely if the right-hand side of \eqref{LieDerivativeMetric2} vanishes for all vectors $v$ in $\su$. A closer investigation of this condition (see appendix A.2) shows that it can be fulfilled only if $u'' = 0$. This means that only vectors in the subspace $\iota(\utwo)$ of $\su$ can originate left-invariant Killing fields. But for such a vector $u$ the Killing condition reduces to
\[
\beta \big( \,[v', v''],\,  [u, \phi]\, \big) \ = 0  \qquad {\rm{for\ all \ vectors}}\  v\in \su \ ,
\]
and this can be satisfied only if the bracket $[u, \phi]$ vanishes in $\su$. Finally, the results of appendix A.1 show that any such $u$ must in fact be proportional to the $3 \times 3$ block-diagonal matrix
\beq \label{GammaVector}
\gamma_\phi \ := \  \frac{i}{\sqrt{3}}  \bmatr -1 &    \linebr     & 2 I_2 -  3 \, |\phi|^{-2}\, \phi\, \phi^\dag  \ematr \ ,
\eeq
where $I_2$ denotes the $2 \times2$ identity matrix and $|\phi|^2$ denotes the canonical norm in $\CC^2$. 

The conclusion is that there is precisely one left-invariant Killing vector field for the Riemannian metric $g_\phi$, up to normalization, whenever $\phi \neq 0$. This field is the left-invariant extension of the vector $\gamma_\phi$ that sits inside the subalgebra $\iota(\utwo)$ of $\su$. Adding to it the space of all right-invariant fields $u^\RR$ on $\SU$, which are always Killing and satisfy $[\gamma_\phi^\LL, \, u^\RR] = 0$, we conclude that the algebra of Killing vector fields of $g_\phi$ can be identified with a subalgebra $\mathfrak{u}(1) \oplus \su$ of the full space $\su \oplus \su$ of translation-invariant vector fields on $\SU$.

\subsection*{Orthonormal basis and volume form of $g_\phi$}
\addcontentsline{toc}{subsection}{Orthonormal basis and volume form of $g_\phi$}

The aim of this section is to write down an explicit $g_\phi$-orthonormal basis of $\su$ in terms of a $\beta$-orthonormal basis of the same space. This will allow us to express the volume form $\vol_{g_\phi}$ in terms of the volume form $\vol_\beta$ and, at the end, calculate the Riemannian volume of the internal space $(\SU , g_\phi)$.

Let $\{  u_0, \ldots, u_3, w_1, \ldots, w_4 \}$ be a $\beta$-orthonormal basis of $\su = \iota(\mathfrak{u}(1) \oplus \sutwo \oplus \CC^2)$ such that the vectors $\{w_j\}$ span the subspace $\iota(\CC^2)$ of $\su$; the vectors $\{u_1, u_2, u_3\}$ span the subspace $\iota(\sutwo)$; and $u_0$ is the vector
\beq
u_0 \ = \  \frac{1}{\sqrt{6\, \lambda}} \,  \diag(-2i, i, i) \ = \ \frac{1}{\sqrt{6\, \lambda}} \,  \iota(i I_2)
\eeq
spanning $\iota(\mathfrak{u}(1))$. The positive factor $\lambda$ comes from definition \eqref{definitionbeta} of $\beta$ and ensures that $u_0$ has $\beta$-norm equal to 1. Since the restriction of $g_\phi$ to the subspace $\iota(\CC^2)$ coincides with the restriction of $\beta$, the four vectors $\{w_j\}$ are $g_\phi$-orthonormal and can be included in the desired basis. The remaining vectors $\{u_j\}$, however, are not $g_\phi$-orthogonal to the $\{w_j\}$, so have to be modified in order to complete the $g_\phi$-orthonormal basis. With this purpose, start by recalling from \eqref{orthogonal2} that the vectors in $\su$ that are $g_\phi$-orthogonal to the subspace $\iota(\CC^2)$ are of the form $u' + [u', \phi]$, with $u'$ in $\iota (\utwo)$. Moreover, one can check that the metric $g_\phi$ satisfies a nice identity when acting on vectors of this form, provided $u'$ is in the smaller subspace $\iota (\sutwo)$, namely
\beq \label{NiceIdentityMetric}
g_\phi \big(\, u' + [u', \phi], \ v' + [v', \phi] \, \big) \ = \ (1- |\phi|^2) \, \beta (u', v')    \ 
\eeq
for all vectors $u', v' \in \iota(\sutwo)$, where $|\phi|^2$ denotes the canonical $\CC^2$-norm of $\phi$. Thus, if $u'$ is $\beta$-orthogonal to $v'$, the shifted vectors $u' + [u', \phi]$ and $v' + [v', \phi]$ will automatically be $g_\phi$-orthogonal to each other, besides being $g_\phi$-orthogonal to the $\{w_j\}$. It follows that the vectors
\beq \label{orthonormalbasis1}
v_j \ := \  \frac{1}{\sqrt{1 - |\phi|^2}} \, \left( u_j + [u_j , \, \phi] \right) \qquad {\rm{for\ }}\  j = 1, 2, 3,  
\eeq
can be added to the $w_j$ to form a $g_\phi$-orthonormal set of vectors $\{  v_1, v_2, v_3, w_1, \ldots, w_4 \}$ in $\su$. At this point we only need one more vector to complete the desired basis, and it should have a non-zero component along the subspace $\iota(\mathfrak{u}(1))$ of $\su$. Defining the vector in $\iota(\utwo)$
\[
u_\phi  \ := \ \frac{1}{3}\, \big[ \, \iota(i I_2) \, - \, 2 \, \sqrt{3}\, |\phi|^2 \,\gamma_\phi  \, \big] \ ,
\]
we will choose for $v_0$ the normalized version of the combination 
\[ 
u_\phi + [u_\phi, \phi] = u_\phi + \iota(i \phi) \ ,
\]
as this automatically ensures orthogonality to the subspace $\iota(\CC^2)$, and hence to the $\{w_j\}$. An explicit calculation shows that
\bal  \label{orthonormalbasis2}
v_0 \  &:= \ \frac{3}{\sqrt{ 6 \lambda (1-|\phi|^2)(1 - 4|\phi|^2)}}  \, \big( \, u_\phi + [u_\phi, \phi] \, \big) \linebr
         &\, = \ \sqrt{ \frac{1 - |\phi|^2}{1 - 4 |\phi|^2}} \, u_0 \ + \ \frac{\sqrt{3}}{\sqrt{ 2 \lambda (1-|\phi|^2)(1 - 4|\phi|^2)}} \, \iota\big( 2i \, \phi \phi^\dag \,  - \, i |\phi|^2 I_2  \, +\,  i \phi \big) \nonumber
\end{align}
does the job of completing the $g_\phi$-orthonormal basis $\{  v_0, \ldots, v_3,  w_1, \ldots, w_4 \}$ of $\su$. The explicit form of this basis will be used in many calculations ahead.

To compute the volume form $\vol_{g_\phi}$, consider the exterior product of vectors in $\su$ and start by observing that
\beq
v_1 \wedge v_2 \wedge v_3 \wedge w_1 \wedge \dots \wedge w_4 \ = \ (1 - |\phi|^2)^{- 3/2} \, u_1 \wedge u_2 \wedge u_3 \wedge w_1 \wedge \dots \wedge w_4  \ . \nonumber
\eeq
This follows from definition \eqref{orthonormalbasis1} of $v_j$ after noticing that the vectors $[u_j , \, \phi]$ are in the four-dimensional subspace $\iota(\CC^2)$ of $\su$, and therefore have zero exterior product with the top product $w_1 \wedge \dots \wedge w_4$ of that subspace. For the same reason, the second term in the bottom line of \eqref{orthonormalbasis2} is in the subspace $ \iota(\sutwo \oplus \CC^2)$ and therefore has vanishing exterior product with $u_1 \wedge u_2 \wedge u_3 \wedge w_1 \wedge \dots \wedge w_4$. Taking only the first term of $v_0$ into account then yields
\beq
v_0 \wedge \dots \wedge v_3 \wedge w_1 \wedge \dots \wedge w_4 \ = \ (1 - |\phi|^2)^{-1} (1 - 4 |\phi|^2)^{-1/2}  \, u_0 \wedge \dots \wedge u_3 \wedge w_1 \wedge \dots \wedge w_4  \ . \nonumber
\eeq
Since $\{  v_0, \ldots, v_3,  w_1, \ldots, w_4 \}$ is a $g_\phi$-orthonormal basis of $\su$, the top exterior product of its vectors is dual to the volume form $\vol_{g_\phi}$. For the same reason, 
the product $u_0 \wedge \dots \wedge u_3 \wedge w_1 \wedge \dots \wedge w_4$ is dual to the volume form $\vol_\beta$. This implies that the two volume forms on $\su$ are related simply by
\beq \label{VolumeForm}
\vol_{g_\phi} \ = \ (1 - |\phi|^2) \,\sqrt{1 - 4 |\phi|^2} \, \, \vol_\beta \ = \  \lambda^4 \, (1 - |\phi|^2) \, \sqrt{1 - 4 |\phi|^2} \, \, \vol_{\beta_0} \ ,
\eeq 
where in the last equality we opted to flesh out the scale factor $\lambda$ appearing in definition \eqref{definitionbeta} of the Ad-invariant product $\beta$.

The relations between the volume forms written above allow us to express the Riemannian volume of the left-invariant metric $g_\phi$ on $K = \SU$ in terms of the volume of the bi-invariant metrics $\beta$, 
\beq
\Vol \big(K, \, g_\phi \big) \ := \ \int_K \vol_{g_\phi} \  = \ (1 - |\phi|^2) \sqrt{1 - 4 |\phi|^2} \ \Vol \big(K,\, \beta \big) \ .
\eeq
But the volume of $\SU$ equipped with the Cartan-Killing metric
\[
- \Tr ({\rm{ad}}_u \, {\rm{ad}}_v) \ = \ 6\, \Tr(u^\dag v) \ = \ 6 \, \lambda^{-1} \, \beta (u, v)
\]
is known to be equal to $\sqrt{3} \, (12)^4 \pi^5$ (see \cite{Abe}, for instance). Therefore, after performing the necessary rescaling to $\beta$, we finally get that
\beq \label{VolumeK}
\Vol \big(K, \, g_\phi \big) \ = \   \sqrt{3}\,  (2\,\lambda)^4 \,\pi^5 \, (1 - |\phi|^2) \, \sqrt{1 - 4 |\phi|^2} \ .
\eeq
Thus, the volume of the internal manifold $K$ is controlled both by the overall scaling factor $\lambda$ and by the norm $|\phi|^2$ of the $\CC^2$-parameter in the metric $g_\phi$. The volume is maximal for $\phi = 0$, i.e. for the bi-invariant metric on $K$, and then tends to zero as the parameter $|\phi|^2$ approaches the critical value $1/4$ at which the metric $g_\phi$ stops being positive-definite. In a model with dynamical $\phi$, one would certainly wish to have a potential $V(|\phi|^2)$ that explodes when $|\phi|^2$ approaches $1/4$, and therefore prevents the internal metric from ever becoming non-definite. The presence of such a potential is a nice feature of the Lagrangian densities studied ahead.

\subsection*{Scalar curvature of $g_\phi$}
\addcontentsline{toc}{subsection}{Scalar curvature of $g_\phi$}

The aim of this section is to present a formula for the scalar curvature $R_{g_\phi}$ of the metric $g_\phi$ on the group $K = \SU$. The scalar curvature of $K$ is one of the components of the scalar curvature of the higher-dimensional spacetime $P = M_4 \times K$, so will appear in the higher-dimensional Lagrangian density. Our calculation of $R_{g_\phi}$ uses the standard formula \eqref{GeneralScalarCurvature}, from \cite{Besse}, applied to the particular $g_\phi$-orthonormal basis $\{  v_0, \ldots, v_3,  w_1, \ldots, w_4 \}$ of $\su$ that was constructed in the previous section. It is a rather long calculation, so in this section we will write down only the final result and its main intermediate components, which would deserve to be checked independently.

\begin{figure}[H]
\centering
\includegraphics[scale=0.4]{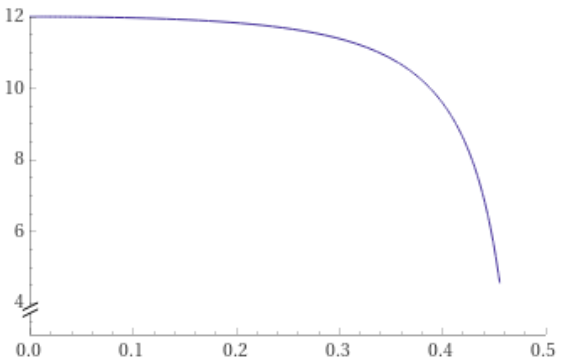}
\vspace*{-.3cm}
\caption{Scalar curvature $R_{g_\phi}$ as a function of $\mph$ at $\lambda=1$. \protect\footnotemark}
\label{fig:ScalarCurvature}
\end{figure}

We start by stating the final result of the calculation. It says that the scalar curvature of the left-invariant metric $g_\phi$ on $\SU$ is given by

\beq \label{ScalarCurvature}
R_{g_\phi} \ = \ \frac{3\, (\,4 - 25\, \mph^2 + 33\, \mph^4 - 8 \,\mph^6 \,)}{\lambda\, (1-\mph^2)^2 \, (1 - 4\mph^2)} \ ,    % Formula corrected in arXiv-v2.
\eeq
where $\mph^2$ is the canonical norm in $\CC^2$ of the parameter $\phi$ and $\lambda$ is the positive scaling factor appearing in definitions \eqref{definitionbeta} and \eqref{DefinitionMetric1} of the metrics $\beta$ and $g_\phi$. Observe that $R_{g_\phi}$ only depends on the norm of the vector $\phi$, not on its orientation, and when $\phi= 0$ we recover the positive scalar curvature $R_\beta = 12 / \lambda$ of the bi-invariant metric $\beta$ on $\SU$. In the limit where $\mph^2$ approaches the critical value $1/4$, at which $g_\phi$ stops being positive-definite and the volume of $\SU$ collapses to zero, the scalar curvature $R_{g_\phi}$ explodes to infinity. The numerator in \eqref{ScalarCurvature} takes a negative value when $\mph^2 = 1/4$, so $R_{g_\phi}$ tends to minus infinity in this limit. The change of sign of $R_{g_\phi}$ occurs at $\mph^2 = 0.221$, approximately. A visual profile of the scalar curvature as $\mph$ ranges from 0 to 1/2, with the choice $\lambda=1$, is given in figure \ref{fig:ScalarCurvature}.
\footnotetext{Figure generated with the free online version of Wolfram Alpha.}
We will now detail some of intermediate results that lead to \eqref{ScalarCurvature}. The general formula \eqref{GeneralScalarCurvature} for the scalar curvature of left-invariant metrics uses an orthonormal basis of $\su$, which we denote here by $\{ e_i\}$, and presents it as a sum of two terms. In the case of the metric $g_\phi$, the separate value of these two components is calculated to be
\bal
 - \frac{1}{2} \,  \sum_{i, j=1}^8 \, g([\,e_i, [e_i, e_j]\, ] ,\, e_j) \ &= \ - \frac{1}{2} \,  \sum_{i}^8 \Tr ({\rm{ad}}_{e_i} \, {\rm{ad}}_{e_i}) \ = \ 
  3 \sum_{j=1} ^8 \Tr(e_i^\dag \, e_i)     \nonumber \linebr
&= \ \frac{12\, (2 - 9\mph^2 +9 \mph^4 - 2\mph^6)}{\lambda\, (1-\mph^2)^2 \, (1 - 4\mph^2)}   \ ,
\end{align}
for the term proportional to the contraction of the Cartan-Killling form on $\su$, while the term that sums the norms of all the commutators is given by     
\beq
 - \frac{1}{4} \, \sum_{i,j=1}^8 \, g_\phi \big([e_i, \, e_j], \, [e_i, e_j] \big) \ = \  - \frac{3\, ( \, 4 - 11 \mph^2 + 3\mph^4 \, ) }{\lambda\, (1-\mph^2)^2 \, (1 - 4\mph^2)} \ . % Formula corrected in arXiv-v2.
\eeq
The calculation of this second sum is more laborious than that of the first term, so we will also write down five partial results that originate it. Choosing as $g_\phi$-orthonormal basis the set of vectors $\{  v_0, \ldots, v_3,  w_1, \ldots, w_4 \}$ described in the previous section, the sum of the norms of all commutators is obtained from the following partial sums:
\bal
\sum_{i=1}^4 \sum_{j > i}  \, g_\phi \big([w_i, \, w_j], \, [w_i, w_j] \big) \ &= \  \frac{6}{\lambda}  \quad;   \nonumber \linebr
\sum_{k=1}^3 \sum_{l > k}  \, g_\phi \big([v_k, \, v_l], \, [v_k, v_l] \big) \ &= \  \frac{ 3 \, (2 + \mph^4) }{\lambda\, (1-\mph^2)^2}    \quad;   \nonumber \linebr    % Formula corrected in arXiv-v2.
\sum_{k=1}^3  \, g_\phi \big([v_0, \, v_k], \, [v_0, v_k] \big) \ &= \  \frac{ 3\,  \mph^4 \, (9 - 8\mph^2) }{\lambda\, (1-\mph^2)^2 \, (1 - 4\mph^2)}  \quad;    \nonumber \linebr
\sum_{j=1}^4  \, g_\phi \big([v_0, \, w_j], \, [v_0, w_j] \big) \ &= \  \frac{ 3 \, (2 + 5 \mph^2) }{\lambda\, (1-\mph^2) (1 - 4\mph^2)}   \quad;   \nonumber \linebr
\sum_{k=1}^3 \sum_{j=1}^4  \, g_\phi \big( [v_k, \, w_j], \, [v_k, w_j] \big) \ &= \  \frac{ 3\, (2 + 5 \mph^2) }{\lambda \, ( 1-\mph^2)}  \quad.  \nonumber 
\end{align}
These are the intermediate components that give rise to the formula for the scalar curvature $R_{g_\phi}$.

\newpage

\section{Lagrangians and fibre-integrals on $M_4 \times \SU$}

\subsection*{Submersive metrics on $M_4 \times \SU$ and their scalar curvature}
\addcontentsline{toc}{subsection}{Submersive metrics on $M_4 \times \SU$ and their scalar curvature}

The first objective of this section is to define the metric $g_P$ on the higher-dimensional $P = M_4 \times K$ that will be used to write Lagrangian densities on that spacetime. As usual in Kaluza-Klein theories, in order to account for the gauge fields on Minkowski space, one should go beyond the product ``vacuum'' metric $(g_M , \, g_K)$ and consider metrics on $P$ with non-diagonal terms. We will also spend a few paragraphs recalling the formula for the scalar curvature of a Riemannian submersions and establishing the associated notation.

Let $\pi$ denote the natural projection $\pi: P \to M$. The tangent space to $P$ at any given point $p=(x,h)$ has a distinguished subspace $\VV_p$ defined by the kernel of the derivative map $\pi_\ast: T_p P \to T_x M$. This is called the vertical subspace of the projection $\pi$. When $P$ is a simple product of manifolds, it can be identified with the tangent space to the internal manifold $T_h K$. If we are also given a metric $g_P$ on $P$, the $g_P$-orthogonal complement to $\VV_p$ is called the horizontal subspace $\HH_p$ of the tangent space $T_pP$. Then we have a decomposition
\beq \label{HorizontalDistribution}
T_p P = \HH_p \oplus \VV_p   \, \qquad \quad  X \ = \  X^\HH  \ +  \ X^\VV \ ,
\eeq
and every tangent vector $E\in T_p P$ can be written as a sum of the respective components. By definition of submersion, the derivative $\pi_\ast$ must induce an isomorphism of vector spaces $\HH_p \simeq T_x M$. If this isomorphism is an isometry at every point $p\in P$, that is, if the product $(g_P)_p (X^\HH , X^\HH )$ is equal to $(g_M)_x (\pi_\ast X, \pi_\ast X)$ for every $p \in P$ and every vector $X\in T_p P$, then the projection $\pi$ is called a Riemannian submersion. For this kind of submersions, the metric $g_P$ on the higher-dimensional space is completely determined by the metrics $g_K$ and $g_M$, together with the rule \eqref{HorizontalDistribution} to decompose tangents vectors into their horizontal and vertical components. In fact, we can write
\beq \label{DefinitionBundleMetric}
g_P (X, X)  \, |_{(x,h)}\ =   \ g_M (\pi_\ast \, X, \pi_\ast \, X) \, |_x \ + \ g_K( X^\VV, \, X^\VV ) \, |_h  \ .
\eeq
The rule \eqref{HorizontalDistribution} to decompose tangent vectors at every point, i.e. the definition of the horizontal distribution $\HH \subset TP$, is called a Ehresmann connection on the submersion. It is equivalent to the more pervasive notion of $K$-connections on a $K$-principal bundle in the special case where the distribution $\HH$ is invariant under right-multiplication on $K$.

In this study we will consider metrics on $P$ determined by Ehresmann connections that can be written down using two one-forms $A_\LL$ and $A_\RR$ on $M_4$ with values in the Lie algebra $\su$. These one-forms can be coupled to the invariant vector fields $e^\LL$ and $e^\RR$ on the group in order define the horizontal and vertical components of any vector $X \in T_pP$:  
\bal \label{DefinitionHorizontalDistribution}
X^\VV  \ &:= \  \sum_j \, A_\RR^j ( \pi_\ast \, X) \, e_j^\RR   \ - \  \sum_j \, A^j_\LL (\pi_\ast \, X) \, e_j^\LL    \linebr
X^\HH  \ &:= \  X \ + \  \sum_j \, A^j_\LL (\pi_\ast \, X) \, e_j^\LL   \ - \ \sum_j \, A_\RR^j ( \pi_\ast \, X) \, e_j^\RR \ ,   \nonumber
\end{align}
where $\{ e_j\}$ denotes any basis of $\su$. Since $\pi_\ast X$ is a vector tangent to $M_4$, it can be contracted with the one-forms $A^j_\LL$ and $A^j_\RR$ to define the coefficients of the vertical vector fields $e_j^\LL$ and $e_j^\RR$ evaluated at $p$. The minus signs are inserted to later obtain the usual formulae for the curvature. The one-forms $A_\LL$ and $A_\RR$ do not define a traditional $\SU$-connection on $P$, although they could be used to define a principal connection on a $\SU \! \times \! \SU$-bundle over $M_4$. In practice, we will think of them as determining the horizontal distribution and metric $g_P$ on $P$, and never work with that second bundle.

A second point of order seems appropriate now. The main aim of this investigation is trying to reproduce the bosonic terms of the Standard Model Lagrangian using a higher-dimensional, Kaluza-Klein-like route. Since the Lie algebra associated to the classic Standard Model gauge fields is $\utwo \oplus \su$, not the more symmetrical $\su \oplus \su$, in most of this study we will assume that the one-form $A_\LL$ in the definition of $g_P$ has values in the subalgebra $\utwo$ of $\su$, and then calculate to see if this produces densities in $M_4$ similar to the terms present in the classical electroweak Lagrangian. Nonetheless, this constraint on $A_\LL$, and hence on the metric $g_P$, is not natural from a geometrical point of view and calls for further justification . In section 5.3 we discuss the natural possibility of having a form $A_\LL$ with values in the full $\su \simeq \utwo \oplus \CC^2$, but with very massive and still unobserved bosons associated to the components in the subspace $\CC^2$.

Returning to the description of submersive metrics, it follows from groundwork in \cite{ONeill} that the scalar curvature of the higher-dimensional metric $g_P$, defined by \eqref{DefinitionBundleMetric}, can be written as a sum of components
\beq \label{DecompositionScalarCurvature}
R_P \ = \ R_M \, + \, R_K \, - \, |\mathcal{F}|^2 \, - \, |S|^2  \ - \  |N|^2 \, -\, 2\,\check{\delta} N \, 
\eeq
where $R_M$ and $R_K$ denote the scalar curvatures of $g_M$ and $g_K$, respectively, and $\mathcal{F}$, $S$ and $N$ are tensors on $P$ that we now describe (see chapter 9 of \cite{Besse}\footnote{The notation here differs from that in \cite{ONeill, Besse} in the following points: the tensor called $A$ in \cite{ONeill, Besse} is called here $\FF$, to avoid confusion with the gauge fields; the tensor called $T$ in \cite{ONeill, Besse} is called here $S$, to avoid confusion with the energy-momentum tensor.}). 

Let $\nabla$ denote the Levi-Civita connection of the metric $g_P$; let $U$ and $V$ denote vertical vector fields on $P$; let $X$ and $Y$ denote horizontal vector fields on $P$. Then $S$ is the linear map $\VV \times \VV \to \HH$ that extracts the horizontal component of the covariant derivative of vertical fields,
\beq \label{tensorT}
S_U V  \ := \   (\nabla_U V)^\HH  \ .
\eeq
Since $U$ and $V$ are tangent vectors to the fibre $K$, the map $S$ can be identified with the second fundamental form of the fibres immersed in $P$. When $S$ vanishes, all the fibres are geodesic submanifolds of $P$ and are isometric to each other \cite{Hermann, Besse}. On its turn, $\FF$ is the linear map $\HH \times \HH \to \VV$ that extracts the vertical component of the covariant derivative of horizontal fields,
\beq \label{tensorF}
\mathcal{F}_X\, Y \ := \ (\nabla_{\!X} Y)^\VV \ =  \ \frac{1}{2} \,  [X, Y]^\VV  \ .
\eeq
where the second equality is a standard result for torsionless connections \cite{ONeill, Besse}. When $\FF$ vanishes, all the Lie brackets of horizontal fields vanish, and hence $\HH$ is an integrable distribution. It is clear from the respective definitions that both $S$ and $\FF$ are $C^\infty$-linear when their arguments are multiplied by smooth functions on $P$. The vector field $N$ is perpendicular to the fibres and is defined simply by
\beq \label{vectorN}
N \ := \ \sum_j \, S_{V_j} V_j  \ ,
\eeq
where $\{V_j\}$ is a $g_K$-orthonormal basis for the vertical space. So $N$ can be identified with the mean curvature vector of the fibres of $P$. The norms of all these objects are defined by
\bal \label{NormsTensors}
|\FF|^2 \ &:= \ \sum_{\mu,\nu}\  g_K \big(\, \FF_{X_\mu} X_\nu  \, , \, \FF_{X_\mu} X_\nu \, \big)  \linebr
|S|^2 \ &:= \ \sum_{i,j }\  g_M \big(\, \pi_\ast \, S_{V_i} V_j  \, , \, \pi_\ast \, S_{V_i} V_j \, \big)   \nonumber  \linebr
|N|^2 \ &:= \ g_P (N, N) \ = \ g_M (\pi_\ast N , \pi_\ast N)  \nonumber \ .
\end{align}
where $\{X_\mu\}$ stands for a $g_M$-orthonormal basis of the horizontal space, isomorphic to the tangent space $TM$. Finally, the scalar $\check{\delta} N$ is just the negative trace
\beq \label{DivergenceN}
\check{\delta} N \ = \ - \sum_\mu \, g_P\big(\nabla_{X_\mu} N, X_\mu \big) \ . %= \ - \sum_\mu \, g_M( \pi_\ast\, \nabla_{X_\mu} N,  \pi_\ast \, X_\mu)
\eeq
The purpose of the next few sections will be to calculate explicitly all the terms of $R_P$, integrate them over the fibre $K$ and analyze the resulting terms in the four-dimensional Lagrangian.

\subsection*{Yang-Mills terms on $M_4$}
\addcontentsline{toc}{subsection}{Yang-Mills terms on $M_4$}

The content of the standard Kaluza-Klein calculation is that the Yang-Mills terms for the gauge field strength on Minkowski space can be obtained from the term $| \FF |^2$ contained in  the scalar curvature of the higher-dimensional metric. In this section we will verify how this works in the case of the metric $g_P$ on $P = M\times K$, as determined by the metrics $g_M$ and $g_K = g_\phi$ on the factors and by the horizontal distribution defined in \eqref{DefinitionHorizontalDistribution}. Everything develops as expected, with a bonus at the end saying that, after fibre-integration over $K$, the Yang-Mills terms for the subalgebra $\utwo \oplus \su$ of gauge fields are independent of the orientation of the parameter $\phi$ in the metric $g_\phi$, and thus are broadly similar to the Yang-Mills terms that would be obtained from the bi-invariant metric $\beta$ on $\SU$. This is a nice feature to have, since the Yang-Mills terms of the Standard Model Lagrangian do not involve the orientation of the Higgs field.

Let $X$ and $Y$ be tangent vectors to $M_4$, which can also be regarded as tangent vectors to $P$ satisfying $\pi_\ast X = X$. We will simplify the notation of \eqref{DefinitionHorizontalDistribution} and write the horizontal component of $X$ as
\bal
X^\HH  \ &:= \  X \ + \  A^j_\LL (X) \, e_j^\LL   \ - \ A_\RR^j (X) \, e_j^\RR \  = \  X \ + \ A(X)  \ ,
\end{align}
where $A$ can be regarded as a one-form on $M_4$ with values in the invariant vertical fields of $P$. Then the tensor $\FF$ of \eqref{tensorF} satisfies 
\bal \label{ExplicitTensorF}
2\, \, \FF_{X^\HH} Y^\HH \ &= \ [X^\HH, \, Y^\HH]^\VV       \nonumber   \linebr
&= \ \big\{ \, [X, \, Y]^\HH \ - \ A([X, Y]) \ + \ [A(X), Y] \ + \ [X, A(Y)] \ + \ [A(X), A(Y)] \, \big\}^\VV   \nonumber   \linebr
&= \ \big\{  \, (\dd_M  A) (X, Y) \ + \ [A(X), A(Y)] \, \big\}^\VV  \nonumber   \linebr
&= \  F^j_{A_\LLL} (X, Y) \, e_j^\LL \ - \ F^k_{A_\RRR} (X, Y)\, e_k^\RR  \  \ , 
\end{align}
where in the last equality we have used the Einstein summation convention and defined the coefficients
\bal \label{DefinitionCurvature}
F_{A_\RRR}^j (X, Y) \ &:= \ \dd A^j_\RR (X, Y) \ + \ A^i_\RR (X)\, A^k_\RR (Y) \, [e_i, e_k]^j  \ \linebr
F_{A_\LLL}^k (X, Y) \ &:= \ \dd A^k_\LL (X, Y) \ + \ A^i_\LL (X)\, A^j_\LL (Y) \, [e_i, e_j]^k  \nonumber \ .
\end{align}
The derivation of the third equality in \eqref{ExplicitTensorF} uses the standard formula for the exterior derivative of a one-form $\omega$:
\beq
\dd \omega (u, v)  \ = \ \Lie_u [\, \omega(v) \, ] \ - \ \Lie_v [\, \omega (u) \, ] \ - \ \omega ([u, v]) \ , 
\eeq
while the derivation of the fourth equality uses the properties \eqref{BracketsInvariantFields} of the brackets of invariant vector fields on $K$. For example, the decomposition of $\FF$ into separate components $F_{A_\LLL}^j$ and $F_{A_\RRR}^k$ is due to the commutation $[e_j^\LL, e_k^\RR] = 0$ of left and right-invariant vector fields.

To explicitly write down the norm $|\FF|^2$, let $\{X_\mu\}$ denote a basis for the tangent space $TM$. It follows from \eqref{ExplicitTensorF} combined with \eqref{NormsTensors} that
\beq \label{NormF}
|\FF|^2 \ = \ \frac{1}{4} \, g_M^{\mu \nu} \, g_M^{\sigma \rho}  \ \, g_\phi \Big( \ (F^j_{A_\LLL})_{\mu \sigma} \, e_j^\LL \ - \ (F^k_{A_\RRR})_{\mu \sigma} \, e_k^\RR \ , \   (F^j_{A_\LLL})_{\nu \rho} \, e_j^\LL \ - \ (F^k_{A_\RRR})_{\nu \rho} \, e_k^\RR \  \Big) \, .
\eeq
Even though the metric $g_M$ and the curvature coefficients $F^j_{A}$ only depend on the coordinate $x \in M$, the norm of $\FF$ is not a constant function along $K$, since the inner-products $g_\phi (e_j^\LL, e_k^\RR)$ and $g_\phi (e_j^\RR, e_k^\RR)$ do depend on the coordinate $h \in K$. 

The expression for $|\FF|^2$ is significantly simplified if we integrate over $(K, \vol_{g_\phi})$, as we have already seen that the integrals of $g_\phi (e_j^\LL, e_k^\RR)$ are equal to zero, while the integrals of $g_\phi (e_j^\LL, e_k^\LL)$ and $g_\phi (e_j^\RR, e_k^\RR)$ are proportional to the volume of $K$. In fact, combining expression \eqref{NormF} with the integrals of products of invariant vector fields calculated in \eqref{ProductInvariantVectors1},  \eqref{ProductInvariantVectors2} and \eqref{ProductInvariantVectors3}, one obtains
\begin{multline} \label{IntegralNormF}
\int_K \, |\FF|^2 \ \vol_{g_\phi} \ = \ \frac{1}{4} \, g_M^{\mu \nu} \, g_M^{\sigma \rho}  \ \Big\{ \   g_\phi(e_j, e_k)\,  (F^j_{A_\LLL})_{\mu \sigma} \, (F^k_{A_\LLL})_{\nu \rho}  \\ + \  \beta(e_j, e_k)\, (F^j_{A_\RRR})_{\mu \sigma} \, (F^k_{A_\RRR})_{\nu \rho} \  \Big\} \ \Vol (K, g_\phi) \ .
\end{multline}
Observe how the coefficients in front of the curvature components $F_{A_\RRR}$ depend solely on the bi-invariant metric $\beta$, and not on the whole metric $g_\phi$ as one could presume from \eqref{NormF}. In the case where the one-forms $A_\LL$ have values in the electroweak subalgebra $\utwo$ of $\su$, the coefficient $g_\phi(e_j, e_k)$ in front of the curvature components $F_{A_\LLL}$ will also be equal to $\beta(e_j, e_k)$, since the metric $g_\phi$ coincides with $\beta$ when restricted to $\utwo$. So for the restricted gauge field algebra $\utwo \oplus\su$, the expression for the norm of $\FF$ is
\begin{multline} \label{IntegralNormF2}
\int_K \, |\FF|^2 \ \vol_{g_\phi} \ = \ \frac{1}{4} \, g_M^{\mu \nu} \, g_M^{\sigma \rho}  \ \bigg\{ \   \sum_{j, k = 1}^4 \beta(e_j, e_k)\,  (F^j_{A_\LLL})_{\mu \sigma} \, (F^k_{A_\LLL})_{\nu \rho}  \\ + \  \sum_{j, k = 1}^8 \beta(e_j, e_k)\, (F^j_{A_\RRR})_{\mu \sigma} \, (F^k_{A_\RRR})_{\nu \rho} \  \bigg\} \ \Vol (K, g_\phi) \ .
\end{multline}
This scalar density on $M_4$ broadly coincides with the Yang-Mills terms of the Standard Model Lagrangian. The fact that the coefficients in front of the curvature terms appear with the Ad-invariant product $\beta$, and not with its deformation $g_\phi$, seems to be a relevant and positive point, since the coupling constants of the strong and weak gauge fields in the Standard Model do not depend on the orientation of the Higgs field inside $\CC^2$. However,  integral \eqref{IntegralNormF2} does depend on the norm $|\phi|^2$, for instance through the overall factor $\Vol (K, g_\phi)$, which will also appear in the integrals of the remaining terms of the higher-dimensional scalar curvature $R_P$.

\subsection*{Fibres' second fundamental form and Higgs covariant derivatives}
\addcontentsline{toc}{subsection}{Fibres' second fundamental form and Higgs covariant derivatives}

Let $U$ and $V$ be vertical vector fields on the submersion $\pi : P \to M$ and let $\nabla$ be a metric connection on the tangent bundle $TP$. In a submersion, the Lie bracket of vertical fields is always vertical \cite{Besse}, so for torsionless connections $\nabla$ it is clear that $S_U V$ is symmetric,
\beq
S_U V \ = \    (\nabla_U V)^\HH   \ = \  \big( \ \nabla_V U + [U, V] + {\mathrm{Tor}}^{\nabla} (U, V) \ \big)^\HH \  = \  S_V U \ .
\eeq
Observe that it is not strictly necessary to start with a torsionless connection in order to obtain a symmetric $S$. It is enough to demand that ${\mathrm{Tor}}^{\nabla} (U, V)$ be a vertical vector field whenever $U$ and $V$ are vertical. This will be the case whenever ${\mathrm{Tor}}^{\nabla} (U, V)$ is proportional to the bracket $[U, V]$, for instance. Be that as it may, we will still assume in the calculations ahead that $\nabla$ is the Levi-Civita connection. Thus, using the definition of $S_U V$ and the fact that $\nabla$ is a torsionless metric connection, one can write for every vector $X \in TM \subset TP$:
\bal
g_P \big(\,S_U V, \,  X\, \big) \ &= \ g_P \big(\,\nabla_U V, \, X^\HH \, \big)  \ 
= \ \Lie_U \big[\, g_P ( V, \, X^\HH) \,\big] \, - \, g_P \big(\,V, \,  \nabla_U  X^\HH \,) \nonumber \linebr
&= \ - \ g_P \big( \, V, \, \nabla_{X^\HH} U \, + \, [U, X^\HH] \, \big) \nonumber \ .
\end{align}
But $S_U V$ is symmetric in $U$ and $V$, so using again that $\nabla$ is a metric connection,
\bal \label{ExplicitTensorT}
2\, g_P \big(\, S_U V, \,  X \, \big) \ &=   \ g_P \big(\, S_U V, \,  X \, \big) \ + \ g_P \big(\, S_V U, \,  X \, \big)                  \nonumber \linebr
&= \ - \ \Lie_{X^\HH} \big[\, g_P ( U, \, V) \,\big] \, - \,  g_P \big( \, V, \, [U, X^\HH] \, \big) \, - \, g_P \big( \, U, \, [V, X^\HH] \, \big) \qquad \nonumber  \linebr
&= \ - \ (\Lie_{X^\HH} \,g_P ) \big(U, \, V  \big) \ , 
\end{align}
where the last equality is a general identity of Lie derivatives. This expression provides a concise relation between the tensor $S_UV$ and the horizontal Lie derivatives of the submersion metric $g_P$. Now suppose that the vertical vector fields $U$ and $V$ are left-invariant on $K$, and hence can be written as $u^\LL$ and $v^\LL$, respectively. Since the metric on the fibres is also left-invariant, the product $g_P ( u^\LL, \, v^\LL)$ is constant along $K$, so
\[
\Lie_{X^\HH} \big[\, g_P ( u^\LL, \, v^\LL) \,\big] \ = \ \Lie_{X} \big[\, g_P ( u^\LL, \, v^\LL) \,\big] \ .
\]
Combining the definition \eqref{DefinitionHorizontalDistribution} of $X^\HH$ with the usual results \eqref{BracketsInvariantFields} for the brackets of invariant vector fields, we also obtain that
\beq  \label{Aux3.2}
g_P \big( \, u^\LL, \, [X^\HH, v^\LL] \, \big) \ = \ \sum_j \, A^j_\LL (X) \ g_P \big(  u^\LL, \, [e_j, \, v]^\LL  \big) \ = \ g_P \big( u^\LL, \, [A_\LL (X),  v]^\LL  \big) \ .
\eeq 
The one-form $A_\RR$ does not appear in this expression because the brackets $[e_j^\RR, v^\LL]$ always vanish on $K$. Thus, in the case of left-invariant vertical fields, expression \eqref{ExplicitTensorT} can be rewritten as
\beq \label{ExplicitTensorT2}
2\, g_P ( \, S_{u^\LLL} v^\LL, \,  X) \ =  \ - \, \Lie_{X} \big[\, g_P ( u^\LL,  v^\LL) \,\big]  -  g_P \big( \, v^\LL,  [\, A_\LL (X),  \, u \, ]^\LL \, \big)  -  g_P \big( \, u^\LL,  [\, A_\LL (X),  \, v \, ]^\LL \, \big)      \nonumber  \, .
\eeq
To progress any further we have to be more specific about the restriction to the fibres of the submersion metric $g_P$, which so far we have called $g_K$ and only assumed to be left-invariant, and say how this restriction can vary when we move across different fibres over $M_4$. 

We will choose, of course, $g_K$ to be the metric $g_\phi$ defined in \eqref{DefinitionMetric1} and studied in section 2. We will assume that the parameter $\phi \in \CC^2$ of the metric can change when one moves across the fibres of $P$, and so $\phi$ becomes a dynamical variable in four-dimensions that we will try to identify with the Higgs field. Furthermore, at this point we will also admit the possibility that the parameter $\phi$ affects the metric $g_\phi$ not only through the second term of \eqref{DefinitionMetric1}, but also through the positive scale factor $\lambda$ of \eqref{definitionbeta}. More precisely, we admit that $\lambda = \lambda (|\phi|^2)$ may be a non-trivial function of the norm $|\phi|^2$. Such a dependence does not change any of the calculations done so far in this study. Using the definition \eqref{DefinitionMetric1} of the metric $g_\phi$ it is then clear that, for any vector $X$ tangent to $M$,
\beq \label{Aux3.1}
\Lie_{X} \big[\, g_\phi ( u^\LL,  v^\LL) \,\big] \ = \ \beta \big(\, [u', v''] + [v', u''], \, \dd \phi (X) \,\big) \ + \ ( \Lie_X \log \lambda) \  g_\phi ( u,  v) \ .
\eeq
Moreover, an algebraic calculation using the Ad-invariance of $\beta$ and the Jacobi identity for the Lie brackets says that, for any vector $z'$ in the subspace $\iota(\utwo)$ of $\su$, we have
\bal
g_\phi \big( \, v,  [\, z',   u \, ] \, \big) \, + \, g_\phi \big( \, u,  [\, z',  v \, ] \, \big)   \ &= \ \beta \big( \, [v',  [ z', u'' ] ]  \, + \, [[ z', u' ], v'' ]    \nonumber \\  & \qquad \qquad \qquad \quad \ \ 
+ \, [u', [ z', v'' ]] \, +\, [[ z', v' ], u'' ], \ \phi \, \big) \,    \nonumber \linebr
&= \  \beta \big( \, [z',  [ v', u'' ] ]  +  [[ v'', u' ], z' ] , \, \phi \, \big)             \nonumber \linebr
&= \  \beta \big( \, [ v', u'' ]  +  [ u', v'' ], \, [z', \phi] \, \big)       \ .      \nonumber 
\end{align}
In particular, when the one-form $A_\LL$ has values in the same subspace $\iota(\utwo)$, we can substitute $z' = A_\LL(X)$ to obtain
\beq
g_\phi \big( \, v,  [\, A_\LL (X),   u \, ] \, \big) \, + \, g_\phi \big( \, u,  [\, A_\LL (X),  v \, ] \, \big)   \ = \ \beta \big( \, [ v', u'' ]  +  [ u', v'' ], \, [A_\LL (X), \phi] \, \big)  \nonumber \ .
\eeq
Combining this expression with \eqref{Aux3.1}, we finally conclude that the choice of metric $g_K = g_\phi$ in the internal space leads to the result
\beq \label{ResultT}
2\, g_P ( \, S_{u^\LLL} v^\LL, \,  X)  \ =\   -\,  \beta \big(\, [u', v''] + [v', u''], \, \dd^A \phi (X) \,\big) \, - \, ( \Lie_X \log \lambda) \  g_\phi ( u,  v) \ ,
\eeq
with the implicit definition of covariant derivative
\beq \label{CovariantDerivativePhi1}
\dd^A \phi (X) \ := \ \dd \phi (X) \, + \,  [\, A_\LL (X),  \phi \, ]  \qquad \in \su \ . 
\eeq
Here we should be more careful, perhaps, and explicitly insert back the vector space isomorphism $\iota: \utwo \oplus \CC^2 \to \su$ that identifies the parameter $\phi \in \CC^2$ with a matrix in $\su$. If we do this, the covariant derivative $\dd^A \phi (X)$ can be written more completely in two different ways:
\bal
\dd^A \phi (X) \ &:= \ [\dd\, \iota(\phi)]  (X)  \, + \,  [\, A_\LL (X),  \iota(\phi) \, ]     \label{CovariantDerivativePhi2}  \linebr
&= \ \iota \bigg( \, \dd \phi (X) \ +  \ \sum_{j=1}^4 A_\LL^j (X) \, \rho_{e_j} (\phi) \, \bigg) \ ,  \label{CovariantDerivativePhi3}
\end{align}
where $\rho: \utwo \times \CC^2 \to \CC^2$ is the Lie algebra representation associated to the $\Utwo$-representation $\phi \to  (\det a) \, a\,  \phi$ on the space $\CC^2$ coming from \eqref{AdjointAction}. The first line represents the covariant derivative of the vector $\iota(\phi)$ in $\su$, hence the bracket is the commutator of matrices in $\su$. The second line represents the $\iota$-image of the covariant derivative of the vector $\phi \in \CC^2$ associated to the indicated $\Utwo$-representation. 

It should be mentioned again that the last two expressions for the covariant derivative $\dd^A \phi$ are valid only for gauge fields $(A_\LL, A_\RR)$ with values in the subalgebra $\utwo \oplus \su$ of the more symmetric $\su \oplus \su$. Moreover, as remarked after \eqref{AdjointAction}, the $\Utwo$-representation and the covariant derivative of $\phi$ are consistent with those attributed to the Higgs field in the Standard Model Lagrangian, having the hypercharge necessary to absorb the fermionic hypercharges in the Yukawa coupling terms. The previous calculation, more specifically the comment after \eqref{Aux3.2}, also provides a geometrical model to understand why the Higgs field couples to the electroweak gauge fields $A_\LL$ but not to the strong force fields $A_\RR$.

\subsection*{Norm of the second fundamental form}
\addcontentsline{toc}{subsection}{Norm of the second fundamental form}

In this section we will calculate the norm $|S|^2$ of the second fundamental form of the fibres in the higher-dimensional spacetime $P = M \times K$. The metric $g_P$ on $P$ is the one described in the last section. We will use \eqref{NormsTensors} as definition of norm and result \eqref{ResultT} as a working formula for the tensor $S$. In particular, since the latter formula is valid only for gauge fields with values in the Standard Model subalgebra $\utwo \oplus \su$ of the larger $\su \oplus \su$, the same applies to the results obtained in this section. Our calculation of $|S|^2$ uses the $g_\phi$-orthonormal basis $\{  v_0, \ldots, v_3,  w_1, \ldots, w_4 \}$ of $\su$ that was constructed in section 2. Since it is a rather long calculation, here we will write down only the final result and its main intermediate components, which would deserve to be checked independently.

We start by stating the final result of the calculation. It says that the squared-norm $|S|^2$ is a constant function along the fibres of $P$ that descends to the following function on the base $M_4$:
\begin{multline} \label{NormT}
|S|^2 \ = \  \frac{3\, (1 - 2\mph^2)}{(1 - \mph^2) \, (1 - 4\mph^2)} \ | \dd^A \phi|^2 \ + \  \frac{ 3\,  (\, 3 - 8\mph^2 + 8 \mph^4 \, )  }{2\, (1 - \mph^2)^2 \, (1 - 4\mph^2)^2}\ |\dd \, \mph^2\, |^2 \\[0.5em]
+ \ \sum_{\mu, \nu} \ g_M^{\mu \nu} \, (\partial_\nu \log \lambda) \ \partial_\mu \Big\{   \log \Big[\,  \lambda^2 \, (1 - \mph^2)   \sqrt{1 - 4\mph^2}  \ \Big]  \Big\} \ ,
\end{multline}
where $\mph^2$ is the canonical $\CC^2$-norm of the parameter of the metric $g_\phi$; the covariant derivative $\dd^A \phi$ is that of \eqref{CovariantDerivativePhi3} and also has values in $\CC^2$; the function $\lambda (\mph^2)$ is the scale factor of $g_\phi$ that, in the last section, we admitted as possibly non-constant and dependent on $\mph^2$ only. Since it is constant on the fibres, the fibre-integral of $|S|^2$ is just
\beq
\int_K \, |S|^2 \ \vol_{g_\phi} \ = \   |S|^2 \ \Vol (K, g_\phi) \ ,
\eeq
and hence the terms induced in the four-dimensional Lagrangian can be read directly from \eqref{NormT} and the volume formula \eqref{VolumeK}. 

The first salient point coming from \eqref{NormT} is that this part of the Lagrangian density has a rather elaborate dependence on $\mph^2$, even if we take $\lambda (\mph^2)$ to be a constant function. Thus, the four-dimensional equation of motion of the parameter $\phi \in \CC^2$ will be more involved than that of the traditional Higgs field in the Standard Model. 

The second salient point is the emergence in the Lagrangian of a term proportional to $| \dd^A \phi|^2$, with a coefficient function that is always positive in the usual range of $\mph^2 < 1/4$. In particular, if the ``vacuum'' value of the parameter $\phi$ is non-zero, i.e. if the ``vacuum'' metric of the internal space $\SU$ is not bi-invariant, then we will get non-zero mass terms for the gauge fields, just as in the usual Brout-Englert-Higgs mechanism of the Standard Model (\cite{Hamilton, Weinberg} or original references \cite{EBH}). Since the parameter $\phi$ couples to the one-form $A_\LL$ but not to the one-form $A_\RR$, as mentioned in the last section, only the former fields have mass terms. Here we are taking the one-form $A_\LL$ with values in the subalgebra $\iota(\utwo)$ of $\su$. However, we have already seen in \eqref{GammaVector} that there exists a matrix $\gamma_\phi$ in this subalgebra that commutes with $\iota (\phi)$, so the corresponding component of $A_\LL$ will not couple to $\phi$ in the covariant derivative \eqref{CovariantDerivativePhi1} and will not acquire a mass term. It is the candidate for the photon field. In short, if the ``vacuum'' metric of the internal space is not bi-invariant, the component $|S|^2$ of the higher-dimensional scalar curvature $R_P$ will naturally produce all the terms in the four-dimensional Lagrangian necessary to the emergence of the Brout-Englert-Higgs mechanism, at least from a qualitative perspective. In sections 3.7 and 5 we will address the question of finding the ``vacuum'' metric of this model.

In the remainder of this section we will give more details about the calculation leading to formula \eqref{NormT}. Let $\{ x^\mu \}$ be a coordinate system on $M_4$. Since the projection $\pi_\ast: \HH \to TM$ is an isometry, it follows from \eqref{ResultT} that
\bal \label{Aux3.3}
2\, g_P \Big(S_{u^\LLL} v^\LL, \,  \frac{\partial}{\partial x^\mu} \Big)  \ &=\  2\, g_M \Big( \pi_\ast \ S_{u^\LLL} v^\LL, \,  \frac{\partial}{\partial x^\mu} \Big)  \nonumber \linebr
 &=\   -\, H_{z_\mu} (u, v) \, - \, ( \partial_\mu \log \lambda) \  g_\phi ( u,  v) \ ,
\end{align}
where we have simplified the notation and defined the auxiliary quantities 
\bal \label{Aux3.4}
H_z (u, v) \ &:= \ \beta \big(\, [u', v''] + [v', u''], \, z \,\big) \nonumber \linebr
z_\mu  \ &:= \  \dd^A \phi \, \Big(\frac{\partial}{\partial x^\mu} \Big)
\end{align}
Combined with the definition of norm in \eqref{NormsTensors}, expression \eqref{Aux3.3} leads to
\begin{multline} \label{Aux3.5}
|S|^2 \ = \ \frac{1}{4} \sum_{\mu = 0}^3  \, g_M^{\mu \nu} \ \bigg\{ \sum_{j, k = 1}^8 \  H_{z_\mu} (e_j, e_k) \  H_{z_\nu} (e_j, e_k)     \\
 + \ 2\, (\partial_\nu \log \lambda)\, \sum_{k=1}^8\, H_{z_\mu} (e_k, e_k) \ + \ (\dim K) (\partial_\mu \log \lambda)\, (\partial_\nu \log \lambda)\ \bigg\} \, ,
\end{multline}
where $\{ e_k \}$ denotes a $g_\phi$-orthonormal basis of the Lie algebra $\su$ and the dimension of $K$ is equal to 8, of course. Now let $z \in \CC^2$ represent any vector in the subspace $\iota(\CC^2)$ of $\su$. A rather long algebraic calculation using definition \eqref{Aux3.4} yields the following general properties of the tensor $H_z (u, v)$:
\bal \label{Aux3.6}
\sum_{j,\, k = 1}^8  \big[ \,H_z (e_j, e_k)\, \big]^2 \ &= \ 12 \, \frac{ |z|^2\,  (1 - 2\mph^2)}{(1 - \mph^2) \, (1 - 4\mph^2)} \ + \ 24 \, \frac{\langle z, \phi \rangle^2 \, (3 - 8\mph^2 + 8 \mph^4)  }{(1 - \mph^2)^2 \, (1 - 4\mph^2)^2}  \nonumber \linebr
\sum_{k = 1}^8  \big[ \,H_z (e_k , e_k)\, \big]^2 \ &= \ 48 \, \frac{\langle z, \phi \rangle^2 \, (1 - 2\mph^2 + 4 \mph^4)  }{(1 - \mph^2)^2 \, (1 - 4\mph^2)^2}  \nonumber \linebr
\sum_{k = 1}^8  \,H_z (e_k , e_k) \ &= \ 12 \, \frac{\langle z, \phi \rangle \, (2\mph^2 - 1)  }{(1 - \mph^2) \, (1 - 4\mph^2)} \ .
\end{align}
Here $\langle\, \cdot \, ,\, \cdot \, \rangle$ denotes the canonical real product on $\CC^2$ and $| \, \cdot \, |$ the corresponding norm. Observe that when $z \in \CC^2$ is the derivative vector $z_\mu$ defined in \eqref{Aux3.4}, then the standard properties of the covariant derivative \eqref{CovariantDerivativePhi3}, which comes from a unitary representation in $\CC^2$, imply that
\beq
\langle z_\mu,\, \phi \rangle \ = \ \Big\langle \,  \dd_\mu \phi \ +  \ \sum_{j=1}^4 (A_\LL^j)_\mu  \, \rho_{e_j} (\phi) ,\ \phi  \, \Big\rangle \ = \ \frac{1}{2} \, \partial_\mu |\phi|^2 \ .
\eeq
It can then be easily checked that the last identity in \eqref{Aux3.6} becomes simply
\bal \label{Aux3.8}
\sum_{k = 1}^8  \,H_{z_\mu} (e_k , e_k) \ = \ 6 \, \frac{(\, \partial_\mu |\phi|^2 \,) \, (2\mph^2 - 1)  }{(1 - \mph^2) \, (1 - 4\mph^2)} \ = \  \partial_\mu\, \log \Big[\, (1 - \mph^2)^2 \, (1 - 4\mph^2) \, \Big] \ .
\end{align}
Substituting \eqref{Aux3.8} and the sums \eqref{Aux3.6} into \eqref{Aux3.5}, we get the final formula \eqref{NormT}. 
Due to length of the calculations involved in obtaining identities \eqref{Aux3.6}, we will also write down the partial sums that originated them. Choosing as $g_\phi$-orthonormal basis the set of vectors $\{  v_0, \ldots, v_3,  w_1, \ldots, w_4 \}$ described previously, the partial sums are
\bal
\sum_{k,\, l = 1}^3  \big[ \,H_z (v_k, v_l)\, \big]^2 \ &= \ \frac{12 \, \langle z, \phi \rangle^2}{ (1 - \mph^2)^2}  \nonumber \linebr
2\, \sum_{k = 1}^3  \big[ \,H_z (v_0, v_k)\, \big]^2 \ &= \ 24 \, \frac{(1-\mph^2)^2 \, \big(\, |z|^2 \mph^2 - \langle z, i \phi \rangle^2  \big) \, +\, \mph^2 \langle z, \phi \rangle^2 \, (2-\mph^2) }{(1-\mph^2)^2 \, (1 - 4\mph^2)}    \nonumber \linebr
\big[ \,H_z (v_0, v_0)\, \big]^2 \ &= \ \frac{36 \, \langle z, \phi \rangle^2}{ (1 - \mph^2)^2 \, (1 - 4\mph^2)^2}  \nonumber \linebr
2\, \sum_{k = 1}^3 \sum_{j=1}^4  \big[ \,H_z (v_k, w_j)\, \big]^2 \ &= \ \frac{ 6 \, |z|^2 }{1-\mph^2}    \nonumber \linebr
2\, \sum_{j=1}^4  \big[ \,H_z (v_0, w_j)\, \big]^2 \ &= \ 6 \, \frac{ (1- 2\mph^2)^2 \, |z|^2 \, + \, 4\, (1 - \mph^2)\, \big(\, \langle z, \phi \rangle^2 + \langle z, i \phi \rangle^2 \, \big)}{(1-\mph^2) \, (1 - 4\mph^2)}      \nonumber \linebr
\sum_{i,\, j = 1}^4  \big[ \,H_z (w_i, w_j)\, \big]^2 \ &= \ 0 \ .
\end{align}

\subsection*{Mean curvature of the fibres}
\addcontentsline{toc}{subsection}{Mean curvature of the fibres}

Among the six components of the higher-dimensional scalar curvature $R_P$, as decomposed in formula \eqref{DecompositionScalarCurvature}, only the two terms involving the mean curvature vector of the fibres --- the vector field denoted by $N$ in that formula --- have not yet been calculated here. That is the purpose of the present section.

Having in mind definition \eqref{vectorN} of the horizontal field $N$, we start by taking the trace of \eqref{Aux3.2} and the formula below it. Let $\{ e_k\}$ denote a $g_K$-orthonormal basis of the Lie algebra $\su$, where $g_K$ is the left-invariant metric on the fibre that includes the point $p \in P$. Then the trace of \eqref{Aux3.2} evaluated at $p$ is identically zero,
\bal
\sum_k  g_P \big( \, e_k^\LL, \, [X^\HH, e_k^\LL] \, \big) \ &= \ \sum_{k, j} \, A^j_\LL (X) \ g_P \big(  e_k^\LL, \, [e_j, \, e_k]^\LL  \big) \ = \  \sum_{k, j} \, A^j_\LL (X) \ g_K \big(  e_k, \, [e_j, \, e_k]  \big) \nonumber \linebr
&= \ \sum_{j} \, A^j_\LL (X) \ \Tr( \ad_{e_j} ) \ = \ 0 \ , \nonumber
\end{align}
where the second equality used the left-invariance of $g_K$, while the third equality used that $\SU$ is an unimodular group, which implies that the $\ad_v$ transformations in the Lie algebra are all traceless. Therefore, combining the definition of $N$ with the trace of the formula below  \eqref{Aux3.2}, we get that, for any vector $X \in TM \subset TP$,
\bal \label{ExplicitVectorN1}
2\, g_P ( \, N, \,  X) \ &=  \ \sum_k \, 2\, g_P ( \, S_{e_k^\LLL}\, e_k^\LL, \,  X) \ =  \ - \, \sum_k \, \Lie_{X} \big[\, g_P ( e_k^\LL,  \, e_k^\LL) \,\big]    \nonumber \linebr 
&=  \ - \, \Lie_{X} \Big[ \sum_k \, g_K ( e_k, \, e_k) \,\Big]    \    \ .
\end{align}
When reading this expression, it is important to keep in mind that the vertical metric $g_K$ may vary across different fibres, while the basis $\{ e_k\}$ was defined to be $g_K$-orthonormal only at the fibre that includes the point $p \in P$. In particular, the functions $g_K ( e_k, \, e_k)$ have value 1 at $p$ but need not be constant when moving across the fibres. The mean curvature vector $N$ is essentially the gradient on $P$ of the sum of these functions.

In the particular case of the vertical metric $g_K = g_\phi$, one can write an explicit expression for $N$ in terms of the derivatives of $|\phi|^2$. Taking $\{ x^\mu \}$ to be a coordinate system on $M_4$, it follows from \eqref{Aux3.3} and \eqref{Aux3.8} that
\bal \label{ExplicitVectorN2}
 g_M (\pi_\ast\, N, \,  \frac{\partial}{\partial x^\mu}) \ &= \  \sum_k \,  g_P \Big(S_{e_k^\LLL} \, e_k^\LL \, , \,  \frac{\partial}{\partial x^\mu} \Big)  \nonumber \linebr
 &=\   - \, \frac{1}{2} \, (\dim K) \, ( \partial_\mu \log \lambda) \ -\ \frac{1}{2} \, \sum_k\,  H_{z_\mu} (e_k, e_k)    \nonumber \linebr
 &=\   - \, \partial_\mu\, \log \Big[\, \lambda^4 \,  (1 - \mph^2) \, \sqrt{1 - 4\mph^2} \, \Big]   \ .
\end{align}
This means that $\pi_\ast \, N$ is minus the gradient vector in $M_4$ of the logarithmic function appearing in \eqref{ExplicitVectorN2}. Observe that the argument of the logarithm is precisely the function that appears in formula \eqref{VolumeForm} for the volume form $\vol_{g_\phi}$ in the group $K$:
\beq \label{DefinitionF}
f_\phi \ := \  \lambda^4 \, (1 - \mph^2) \, \sqrt{1 - 4\mph^2} \ = \ \frac{\vol_{g_\phi}}{\vol_{\beta_0}} \ .
\eeq
It can be regarded either as a function on the base $M_4$ or as a function on $P$ that is constant along the fibres. Since the projection $\pi_\ast: \HH \to TM$ is an isometry, for any vector $E$ tangent to P we have
\beq
g_P (N, \, E) \, = \, g_P (N, \, E^\HH) \, = \, g_M (\pi_\ast\, N, \,  \pi_\ast\, E) \, = \, - \, \Lie_{\pi_\ast  E}\, (\, \log f_\phi \,) \, = \, - \, \Lie_{E}\, (\, \pi^\ast\, \log f_\phi \,) \ ,  \nonumber
\eeq 
so we can write, equivalently,
\bal \label{NasGradient}
N \ &= \ -\, \grad_{P} \, (\pi^\ast\, \log f_\phi ) \  \nonumber \linebr
\pi_\ast \, N \ &= \ -\, \grad_{M} \, (\, \log f_\phi \,) \ ,
\end{align}
in agreement with well-known properties of the mean curvature vector in Riemannian fibrations. The norm of $N$ is then equal to the norm on the base $M_4$ of the exterior derivative of the same logarithmic function,
\beq \label{NormN2}
|N|^2 \ = \ \big| \,\dd (\log f_\phi)\, \big|^2_{g_M} \ .
\eeq
Now let $\{x^\mu\}$ stand for a coordinate system in $M_4$ and let $\{X_\mu\}$ stand for the unique $g_P$-orthonormal basis of the horizontal subspace of $TP$ such that $\pi_\ast X_\mu = \frac{\partial}{\partial x^\mu}$. Starting from definition \eqref{DivergenceN} of $\check{\delta} N$, we have
\bal    \label{DivergenceN2}
\check{\delta} N \ &= \ - \sum_\mu \, g_P\big(\nabla_{X_\mu} N, X_\mu \big) \ = \ \ - \sum_\mu \, g_M \Big(\pi_\ast \, (\nabla_{X_\mu} N) , \, \frac{\partial}{\partial x^\mu} \Big)  \nonumber \linebr
&= \ - \sum_\mu \, g_M \Big(\, \Big(\nabla^M_{\frac{\partial}{\partial x^\mu}} \, \pi_\ast N \Big) , \,  \frac{\partial}{\partial x^\mu} \Big)
 \ = \ - \divergence_M \,(\pi_\ast N)  \nonumber \linebr
 &= \ \Delta_M \, \big(\, \log f_\phi \, \big) \ , 
\end{align}
where $\divergence_M$ and $\Delta_M$ stand for the divergence of a vector field and for the Laplacian of a function on $M_4$, respectively. The third equality uses a standard relation between the Levi-Civita connection $\nabla$ on $P$ and the Levi-Civita connection $\nabla^M$ on $M$, valid for all Riemannian submersions (see page 240 of \cite{Besse}, for instance).

It is clear from expressions \eqref{NormN2} and  \eqref{DivergenceN2} that the mean curvature components $|N|^2$ and $\check{\delta} N$ of the scalar curvature $R_P$, unlike its other components $|S|^2$ and $|\FF|^2$, are completely independent of the one-forms $A_\LL$ and $A_\RR$ that participate in the definition of the higher-dimensional metric $g_P$. They are only sensitive to the variation of the volume \eqref{VolumeK} of the internal space $K$ as one moves around the four-dimensional base $M_4$.

\subsection*{Lagrangian densities on $M_4 \times \SU$}
\addcontentsline{toc}{subsection}{Lagrangian densities on $M_4 \times \SU$}

The purpose of this section is to bring together the work of the last few sections. We want to write down the Lagrangian density in four dimensions that emerges from the fibre-integral of the scalar curvature of the higher-dimensional metric $g_P$. This scalar curvature $R_P$ was decomposed in \eqref{DecompositionScalarCurvature} into a sum of natural terms, including the scalar curvatures of $K$ and $M$; the Yang-Mills term $|\FF|^2$; the norm $|S|^2$ of the fibres' second fundamental form; and the norm and divergence of the fibres' mean curvature vector field $N$. Defining the higher-dimensional Lagrangian density
\beq    \label{DensityP}
{\mathscr L}_P \ :=  \  \frac{1}{2\, \kappa_P}\,  (R_P \, - \, 2\, \Lambda_P) \ ,
\eeq
where $\kappa_P$ and $\Lambda_P$  are real constants, we can integrate it over $K$ using the explicit formulae \eqref{ScalarCurvature}, \eqref{IntegralNormF}, \eqref{NormT}, \eqref{NormN2} and \eqref{DivergenceN2} to obtain the four-dimensional density ${\mathscr L}_M$. The result is that, after fibre-integration, the conceptually simple density on $P$ cascades down to a more complicated but familiar group of terms in four dimensions, once the components of the metric $g_P$ are separated from each other and are identified with four-dimensional bosonic fields: 
\bal     \label{DensityM}
{\mathscr L}_M \ &= \  \frac{1}{2\, \kappa_P}\, \int_K \big(\,R_P \, - \, 2\, \Lambda_P \,) \ \vol_{g_\phi}  \linebr
&= \  \frac{1}{2\, \kappa_P}\, \int_K (\,R_M + R_K - |\FF|^2 - |S|^2 - |N|^2 - 2\, \check{\delta} N - 2\, \Lambda_P \, \big) \ \vol_{g_\phi}  \nonumber  \linebr
&= \ \frac{1}{2\, \kappa_P}\, \Big[ \, R_M \, f_\phi \, - \,  \frac{1}{4}\,  B_\phi \, \left(\, |F_{A_\LLL}|^2_{\beta_0} \, + \, |F_{A_\RRR}|^2_{\beta_0} \, \right) \,- \,C_\phi \,  \big|\dd^{A_\LLL} \phi \big|^2 \, - \, D_\phi \,  \big|\dd\, |\phi|^2 \, \big|^2    \nonumber  \linebr 
& \qquad \qquad \qquad \qquad  \qquad \qquad  \qquad \qquad \qquad \qquad \ \, - \, V( \mph^2) \, - \, 2 \,\Delta_M f_\phi  \  \Big]  \, \Vol(K, \beta_0)  \nonumber \ .
\end{align}
Here $\beta_0$ is the Ad-invariant product on the Lie algebra $\su$ defined in \eqref{definitionbeta}. It does not depend on $\mph^2$. The term proportional to the Laplacian $\Delta_M f_\phi$ is a total derivative on $M_4$, so does not contribute to the classical equations of motion in four dimensions. The coefficient functions $f$, $B$, $C$ and $D$ do depend on $\mph^2$ and are collected below: 
\bal \label{CoefficientFunctions}
f_\phi \ &:=  \ \lambda^4\, (1 - \mph^2) \, \sqrt{1 - 4 \mph^2}      \linebr
B_\phi \ &:=  \ \lambda  \, f_\phi       \nonumber   \linebr
C_\phi \ &:= \ \frac{ 3\, \lambda^4 \, (1- 2 \mph^2)}{ \sqrt{1 - 4 \mph^2}}   \nonumber   \linebr
D_\phi \ &:= \ \lambda^4 \, \frac{12\, + \, 15 \,(1- 2 \mph^2)^2}{8\, (1 - \mph^2) \, (1 - 4 \mph^2)^{3/2}} \ - \ \frac{7}{8} \ f_\phi^{-1}\,  \Big( \frac{\dd f_\phi}{ \dd \mph^2} \Big)^2  \nonumber \ .
\end{align}
Recall that we admit the possibility of $\lambda = \lambda( \mph^2)$ being a constant or being an arbitrary positive function of $\mph^2$. Finally, the potential term that does not depend on the gauge fields or on the derivatives of $\phi$ is given by
\beq \label{DefinitionPotential}
V( \mph^2) \ := \ (2\, \Lambda_P \,  - \,  R_{g_\phi} \, ) \, f_\phi   \ , 
\eeq
where the scalar curvature $R_{g_\phi}$ of $K$ is explicitly given in \eqref{ScalarCurvature} and is depicted in figure \ref{fig:ScalarCurvature} of section 2. Inspecting this figure and the dependence of $R_{g_\phi}$ on $\mph^2$, it is clear that the potential $V$ will explode to positive infinity when $\mph^2$ approaches the value $1/4$ from below. This is good news, since at $\mph^2 =  1/4$ the deformed metric $g_\phi$ stops being positive-definite, and we now see that it takes infinite energy to deform the bi-invariant metric on $K$ to such an extent. The detailed behaviour of $V( \mph^2)$ for smaller values of $\mph^2$, however, will depend on the value of the constant $\Lambda_P$ and on the specific dependence $\lambda( \mph^2)$ that is chosen. For instance, in the next section we will see that if $\lambda$ is constant, then the potential $V( \mph^2)$ will have absolute minima with $\mph^2 \neq 0$ whenever the real constant $\lambda\, \Lambda_P$ is larger than $13/2$. This suggests that the bi-invariant metric on $K$ need not be the lowest-energy configuration of the system whenever $\Lambda_P$ is positive, and that deformed metrics such as $g_\phi$ may be a better model for the classical ``vacuum'' geometry of the internal space $K$.

The explicit form of the function $B_\phi$ given above, in \eqref{CoefficientFunctions}, is a direct consequence of the Yang-Mills term \eqref{IntegralNormF}, definition \eqref{definitionbeta} and the relation between volume forms on $K$ that says that $\vol_{g_\phi}$ is equal to $f_\phi \, \vol_{\beta_0}$. Likewise, the coefficient function $C_\phi$ can be directly read from formula \eqref{NormT} for the norm $|S|^2$ and the relation between the two volume forms. The calculation of $D_\phi$ is slightly less immediate, as it combines contributions from $|S|^2$, $|N|^2$ and $\check{\delta} N$. The details will not be reproduced here, but the main intermediate steps can be summarized as follows. The general identity for the scalar Laplacian
\beq 
\Delta (\log f) \ = \ f^{-1} \, \Delta f \ - \  |\grad\, (\log f)|^2 \ , \nonumber
\eeq
combined with \eqref{NormN2} and \eqref{DivergenceN2}, implies that
\beq  \label{Aux3.9}
\big(\, |N|^2 \, + \, 2\, \check{\delta} N \, \big)\, \vol_{g_\phi} \  = \ \big(\, 2 \, \Delta_M f_{\phi} \, -  \,  f^{-1}_{\phi} \, \big| \dd f_{\phi} \big|^2 \, \big)\, \vol_{\beta_0} \ .
\eeq
At the same time, the third term in expression \eqref{NormT} for $|S|^2$ can be rewritten as
\begin{multline} \label{Aux3.10}
\sum_{\mu, \nu} \ g_M^{\mu \nu} \, (\partial_\nu \log \lambda) \ \partial_\mu \Big\{   \log \Big[\,  \lambda^2 \, (1 - \mph^2)   \sqrt{1 - 4\mph^2}  \ \Big]  \Big\} \  \vol_{g_\phi} \ = \\
\frac{1}{8} \, \Big\{ \,  f^{-1}_{\phi} \, \big| \dd f_{\phi} \big|^2   \ - \    \big| \,\dd \log ( \lambda^{-4}\, f_\phi ) \,  \big|^2 \, f_\phi \, \Big\} \  \vol_{\beta_0} \ .
\end{multline}
Then it is clear that the last term of $D_\phi$ and the last term of ${\mathscr L}_M$ result from the simple sum of  \eqref{Aux3.9} with \eqref{Aux3.10}. On the other hand, the last term on the right-hand side of \eqref{Aux3.10} can be combined with the second term in formula \eqref{NormT} for $|S|^2$ to obtain the first term in the expression for $D_\phi$.

Before ending this section, we will briefly discuss other possible choices to define the density ${\mathscr L}_P$ on the higher-dimensional manifold $P$. The choice \eqref{DensityP} comes about as the higher-dimensional analogue of the Einstein-Hilbert Lagrangian for general relativity, of course. As in the four-dimensional case, the cosmological constant term $\Lambda_P$ is not particularly natural here, although it helps to obtain potentials $V( \mph^2)$ having minima with $\phi \neq 0$. Unlike the four-dimensional case, however, the structure of the higher-dimensional submersion $\pi: P \to M_4$ provides additional natural functions on $P$, besides the scalar curvature of the metric $g_P$, which {\it a priori} could be combined with $R_P$ to define other variants of the density ${\mathscr L}_P$. We are talking about the fibres' second fundamental form and mean curvature, of course. For instance, if we add to ${\mathscr L}_P$ any linear combination of the scalar functions $|N|^2$ and $\check{\delta} N$, it is clear from the previous discussion that the Einstein-Hilbert and Yang-Mills terms in four dimensions will not be affected, and neither will the potential $V( \mph^2)$ and the coefficient $C_\phi$ of the Higgs covariant derivative. Only the function $D_\phi$ will change, and this will in general be reflected in a different value for the classical mass of the Higgs particle, as will be discussed in section 4. 

For example, a particularly nice combination of the scalar curvature $R_P$ with the two functions $|N|^2$ and $\check{\delta} N$ is
\bal   \label{DensityP2}
W_{P} \ &:= \ R_{P} \, + \, \frac{33}{32}\, |N|^2 \, + \, \frac{11}{4}\, \check{\delta} N     \linebr
&= \ R_M \,+\, R_K \, - \,  |\FF|^2  \, - \, |S|^2 \, + \, \frac{1}{32} \, |N|^2 \,  + \,  \frac{3}{4} \, \check{\delta} N      \nonumber \ .
\end{align}
Indeed, if $\Omega: \, P \to \mathbb{R}^+$ is any positive function with constant values on the fibres and $\tilde{g}_P :=  \Omega^2\, g_P$ is the corresponding Weyl transformation, it is shown in appendix A.3 that the function $\tilde{W}_{P}$ calculated for the rescaled metric satisfies the simple relation
\[
\tilde{W}_{P} \ = \ \Omega^{-2}\ W_{P} \ .
\]
This contrasts with the complicated behaviour of $R_P$ under the same Weyl transformations. Here we focus on rescalings that are constant on the fibres, i.e. on scaling functions $\Omega$ that are  pull-backs to $P$ of arbitrary functions on the base $M_4$. A more general rescaling on $P$ would spoil its structure as a Riemannian submersion. If we use $W_P$ instead of the scalar curvature $R_P$ to define the density ${\mathscr L}_P$, then fibre-integration over $K$ yields the following Lagrangian in four dimensions:
\bal     \label{DensityM2}
\hat{{\mathscr L}}_M \ &= \ \frac{1}{2\, \kappa_P}\, \int_K \big(\,W_{P} \, - \, 2\, \Lambda_P \,) \ \vol_{g_\phi}  \linebr
&= \ \frac{1}{2\, \kappa_P}\, \Big[ \, R_M \, f_\phi \, - \,  \frac{1}{4}\,  B_\phi \, \left(\, |F_{A_\LLL}|^2_{\beta_0} \, + \, |F_{A_\RRR}|^2_{\beta_0} \, \right) \,- \,C_\phi \,  \big|\dd^{A_\LLL} \phi \big|^2 \, - \, \hat{D}_\phi \,  \big|\dd\, |\phi|^2 \, \big|^2    \nonumber  \linebr 
& \qquad \qquad \qquad \qquad  \qquad \qquad  \qquad \qquad \qquad \qquad \ \, - \, V( \mph^2) \, + \, \frac{3}{4} \,\Delta_M f_\phi  \  \Big]  \, \Vol(K, \beta_0)  \nonumber \ ,
\end{align}
where the potential $V$ and the coefficient functions $f_\phi$, $B_\phi$ and $C_\phi$ remain the same as in \eqref{DefinitionPotential} and \eqref{CoefficientFunctions}, respectively, while the function $\hat{D}_\phi$ is slightly changed to
\bal \label{CoefficientFunctions2}
\hat{D}_\phi \ &:= \ \lambda^4 \, \frac{12\, + \, 15 \,(1- 2 \mph^2)^2}{8\, (1 - \mph^2) \, (1 - 4 \mph^2)^{3/2}} \ + \ \frac{27}{32} \ f_\phi^{-1}\,  \Big( \frac{\dd f_\phi}{ \dd \mph^2} \Big)^2  \ .
\end{align}
Compared to the function $D_\phi$ of \eqref{CoefficientFunctions}, the new $\hat{D}_\phi$ has the advantage of being manifestly positive for $\mph^2 < 1/4$. As will be seen in section 4, this property guarantees that the radial component of the field $\phi (x) \in \CC^2$ will have non-negative mass independently of the choice of function $\lambda (\mph^2)$. This is not always true in the case of the first density ${\mathscr L}_M$.

\subsection*{Vacuum configurations and Higgs-like potentials}
\addcontentsline{toc}{subsection}{Vacuum configurations and Higgs-like potentials}

In this section we will consider ``vacuum'' configurations where the metric $g_P$ is taken to be a product metric $(g_M, g_{\phi})$ on $M_4 \times \SU$ with vanishing gauge fields $A_\LL$ and $A_\RR$, constant $\phi$ and constant scalar curvature $R_M$. We want to analyze the profile of the potential that subsists in the Lagrangian densities ${\mathscr L}_M$ and $\hat{{\mathscr L}}_M$ in these configurations, and want to check whether it can have absolute minima for non-zero values $\phi$, as this would lead to spontaneous symmetry breaking and mass generation for the gauge fields of the model. For a broader discussion about vacuum configurations see also section 5. 

The terms that subsist in the four-dimensional Lagrangians with vanishing gauge fields and constant $\phi$ define a potential:
\begin{multline} \label{Potential}
U(\mph^2)\ := \ V( \mph^2) \, - \,  R_M \, f_\phi  \ = \ 3\, \lambda^{3}\  \frac{- 4 + 25\, \mph^2 - 33\, \mph^4 + 8 \,\mph^6}{(1-\mph^2) \, \sqrt{1 - 4\mph^2}} \linebr
 + \ 2\, \lambda^{4} \, (\,  \Lambda_P  - R_M /  2\, )\, (1-\mph^2) \, \sqrt{1 - 4\mph^2} \ ,             % Formula corrected in arXiv-v2.
\end{multline}
where we have used formula \eqref{ScalarCurvature} for the scalar curvature $R_{g_\phi}$ and the definition of the volume density $f_\phi$. For Minkowski space we have of course $R_M = 0$. We allow the scale factor $\lambda$ of the metric $g_\phi$ to be any positive function $\lambda( \mph^2)$.

Consider the simpler case where $\lambda( \mph^2) = \lambda_0$ is a positive constant. Then the profile of the potential, up to rescaling, depends on the single parameter
\beq \label{DefinitionParameterA}
a\ := \  \lambda_0 \, (\Lambda_P \, - \, \frac{1}{2} \, R_M )  \ , 
\eeq
which is assumed to be constant on the vacuum $M_4$. At the point $|\phi| = 0$, corresponding to the bi-invariant metric on $K$, the potential has the value $2 \, \lambda_0^3 \, (a - 6)$, whereas it clearly diverges in the limit $|\phi|^2 \to 1/4$. Observe that if the constant $a$ is positive and large, the second term of the potential will decrease as $\mph$ grows, and somewhere inside the interval $[0, 1/2 [$ this might just balance the increase of the first term in order to define a minimum with $|\phi| \neq 0$. Due to the presence of high-degree polynomials, it does not seem possible to give an analytic expression for these minima as a function of the parameter $a$, but we may try to illustrate the situation with numerical plots. 
Start by defining
\beq \label{SimplifiedPotential}
\hat{V}_a( x) \ := \ 3\ \frac{- 4 + 25\, x^2 - 33\, x^4 + 8 \,x^6}{(1-x^2) \, \sqrt{1 - 4x^2}} \ + \ 2a\, (1-x^2) \, \sqrt{1 - 4x^2} \ ,    % Formula corrected in arXiv-v2.
\eeq
for a real variable $x$, and taking the derivative
\bal
\hat{V}'_a( x) \ &:= \ \frac{6\,x\,(13 - 92\, x^2 + 205\, x^4 - 162 \,x^6 + 48\, x^8 )}{(1-x^2)^2 \, (1 - 4x^2)^{3/2}} \ + \ a\, \frac{12 x (2x^2 -1)}{ \sqrt{1 - 4x^2}}  \nonumber \linebr    
&=: x\, \big[ \, v_1(x) \, + \, a\, v_2(x) \, \big] \ .  \nonumber     % Formula corrected in arXiv-v2.
\end{align}      
Then any stationary point of $\hat{V}_a$, apart from the obvious $x=0$, will satisfy the equation $a = - v_1(x) / v_2(x)$. So we can plot the right-hand side to find out how many stationary points exist for each value of the parameter $a$.
\begin{figure}[H]
\centering
\includegraphics[scale=0.65]{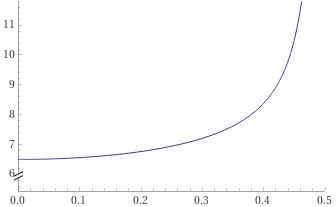}
\vspace*{-.3cm}
\caption{Auxiliary function $-v_1(x) \, / \, v_2(x)$.}
\label{fig:MinimaPotential}
\end{figure}
It follows from this graphic that when $a \leq 6.5$ the function $\hat{V}_a (x)$ has no stationary point in the interval $[0, 1/2 [$ besides $x=0$. When $a$ is larger than $6.5$, the potential is stationary at exactly one other positive point $x_0 (a)$ that increases monotonously with $a$ and approaches the boundary $x = \mph = 1/2$ as the parameter $a$ tends to infinity. The stationary points $ \pm \, x_0 (a)$ are actually absolute minima of the potential $\hat{V}_a (x)$ in the open interval $] \! - \!1/2,\, 1/2 [$, as follows from the graphics below.
\begin{figure}[H]
\centering
\caption{Potential $\hat{V}_a (x)$ for $a \leq 6.5$: single minimum at $x =0$.}
\begin{subfigure}{.5\textwidth}
  \centering
  \caption*{$a = 4$}  
  \includegraphics[width=.8\linewidth]{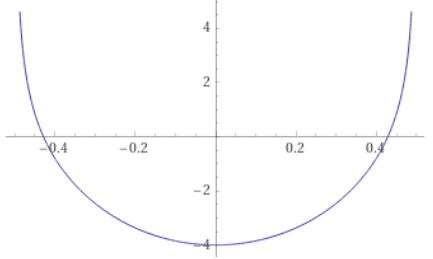}
  \label{fig:sub1}
\end{subfigure}%
\begin{subfigure}{.5\textwidth}
  \centering
  \caption*{$a = 6.5$}  
  \includegraphics[width=.8\linewidth]{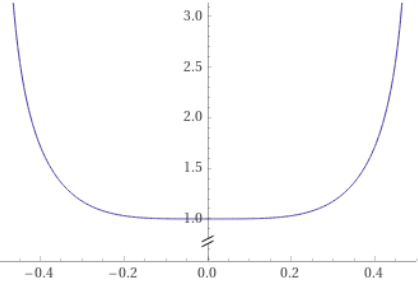}
  \label{fig:sub2}
\end{subfigure} 
\label{fig:singleminimum}
\end{figure}
\begin{figure}[H]
\centering
\caption{Potential $\hat{V}_a (x)$ for $a > 6.5$: minima with $x \neq 0$. \protect\footnotemark}
\begin{subfigure}{.5\textwidth}
  \centering
  \caption*{$a = 7$}
  \includegraphics[width=.78\linewidth]{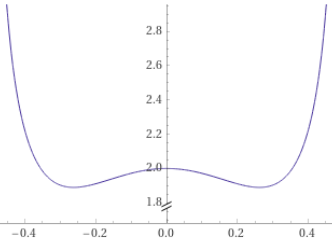}
  \label{fig:sub3}
\end{subfigure}% 
\begin{subfigure}{.5\textwidth}
  \centering
  \caption*{$a = 10$}  
  \includegraphics[width=.85\linewidth]{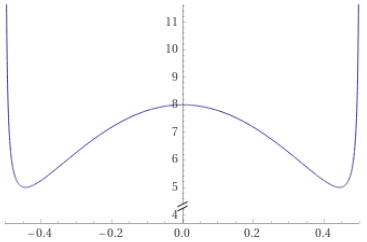}
  \label{fig:sub4}
\end{subfigure}
\label{fig:doubleminima}
\end{figure}
\footnotetext{Figures generated with the free online version of Wolfram Alpha.}
Thus, the potential $\hat{V}_a (x)$ coming from the fibre-integral of the higher-dimensional density  $R_P - 2 \Lambda_P$ can have a double-well profile, similar to the usual Higgs potential, whenever its parameter is in the half-line $a > 6.5$. In this case the potential's absolute minima occur for $x \neq 0$. Since the variable $x$ is just $\mph$, we conclude that there are relatively natural Kaluza-Klein-like models where the bi-invariant metric on the group $K$ is not the lowest-energy configuration of the system. Perhaps a deformed metric such as $g_\phi$, exhibiting manifest left-right asymmetry, could also be considered as a model of the classical ``vacuum'' geometry of the internal space $K$. See also the discussion in section 5.

The potential depicted in the previous graphics was written in \eqref{SimplifiedPotential} under the assumption that the scale factor $\lambda$ of the metric $g_\phi$ --- the factor appearing in definitions \eqref{definitionbeta} and \eqref{DefinitionMetric1} --- is just a constant $\lambda_0$. This is certainly the simplest choice. However, as mentioned before, one can also consider definitions of $g_\phi$ that include a generalized scale factor depending on $\mph^2$, and the explicit calculations of the previous sections were open to this possibility. A non-trivial dependence $\lambda (\mph^2)$ would affect the formulae for the scalar curvature $R_{g_\phi}$ and volume coefficient $f_\phi$ as functions of $\mph^2$, and hence would certainly affect the shape of the potential $V(\mph^2)$ coming from \eqref{DefinitionPotential}. One could, for example, consider scale factors of the form
\beq \label{DefinitionLambda2}
\lambda \left(\mph^2 \right) \ = \ \lambda_0  \left[ \,  (1-\mph^2) \, \sqrt{1 - 4\mph^2} \,   \right]^q
\eeq
for some power $q$. Then the potential function $\hat{V}_{a}( x)$ defined in \eqref{SimplifiedPotential} would change to the more versatile variant
\beq   \label{DefinitionVAQ}
\hat{V}_{a, q}( x) \ := \ 3\ \frac{- 4 + 25\, x^2 - 33\, x^4 + 8 \,x^6 }{\big[ \, (1-x^2) \, \sqrt{1 - 4x^2} \, \big]^{1- 3q} } \ + \ 2a\, \big[ \,(1-x^2) \, \sqrt{1 - 4x^2} \, \big]^{1 + 4q} \ .\nonumber    % Formula corrected in arXiv-v2.
\eeq
Observe that the special choice $q = - 1/4$ would yield a volume form $\vol_{g_\phi}$ and a coefficient function $f_\phi$ completely independent of $\mph^2$, as follows from  \eqref{VolumeForm} and \eqref{DefinitionF}. While this choice could simplify parts of the four-dimensional Lagrangian ${\mathscr L}_M$, the constancy of $f_\phi$ would also prevent the appearance of potentials with minima for $\mph \neq 0$, since the potential $V$ would essentially just be minus the scalar curvature $R_{g_\phi}$, up to constants. A second interesting choice is $q = - 1/5$, since in this case the coefficient function $B_\phi$ is constant and independent of $\mph$, as follows from \eqref{CoefficientFunctions}. In other words, for $q = - 1/5$ the coefficients of the Yang-Mills terms in ${\mathscr L}_M$ do not depend on the Higgs field $\phi$, as happens in the traditional Standard Model Lagrangian. The same procedure that was described in the case of constant $\lambda$ leads to the conclusion that, in the case $q = - 1/5$, the absolute minima of $\hat{V}_{a, q}( x)$ have $x \neq 0$ whenever the parameter $a$ is larger than the value $14.5$.

More generally, for an arbitrary positive function $\lambda ( |\phi|^2)$, observe that the potential $U(\mph^2)$ written in \eqref{Potential} has finite value at $|\phi| = 0$ and diverges to positive infinity as $\mph \to 1/2$. So a sufficient condition for $U$ to have absolute minima with $|\phi| \neq 0$ is that it is a decreasing function for small $|\phi|^2$. But expanding $\lambda$ around the origin:
\beq \label{ExpansionLambda}
\lambda (|\phi|^2) \  = \ \lambda_0  \,   \left[ \ 1 \, +\,  b \,  |\phi|^2  \,+  \, d \, |\phi|^4 \,  + \, O\big(|\phi|^6 \big) \   \right] \ ,
\eeq
the corresponding expansion of $U$ is
\begin{multline} \label{ExpansionPotential}
U (|\phi|^2) \  = \ \lambda_0^3  \,   \Big[ \ 2\,(a-6) \, +\,  (39 - 6\,a + 8 \,a b - 36 \,b) \  |\phi|^2   \\
+ \ ( 18 + 12 \,a b^2 - 24\, a b  + 8 \,a d   - 36 \,b^2  + 117\, b  - 36\, d  ) \ |\phi|^4 + \, O\big(|\phi|^6 \big) \   \Big] \ .     % Formula corrected in arXiv-v2.
\end{multline}
So the potential $U(|\phi|^2)$ is a decreasing function near the origin whenever
\[
a \ > \  \ \frac{39 - 36\,b}{6 - 8 \,b} \ = \ \frac{9}{2} + \frac{6}{3 - 4 \,b} \ .
\]
Thus, for any fixed positive function $\lambda ( |\phi|^2)$ with $b \neq 3/4$, there is a wide range of values of the constant $a$ for which the potential $U$ will have absolute minima with $|\phi| \neq 0$.

\subsubsection*{Kaluza-Klein normalizations}

Four-dimensional Lagrangians determined by the higher-dimensional scalar curvature $R_P$ through Kaluza-Klein-type calculations are similar to, but never exactly equal to, the traditional Lagrangians of Einstein-Maxwell or Einstein-Yang-Mills field theories. Hence the four-dimensional equations of motion of the classical fields will also not be exactly the same. If we want that at least the linearized equations of motion around the vacuum configuration coincide with the traditional ones, then a series of standard normalizations must be established \cite{Bailin, Duff}.

If we assume that in the dynamical theory the parameter $\phi$ is always close to its vacuum value $\phi_0$, then the coefficient $f_\phi$ in front of the curvature $R_M$ in Lagrangians \eqref{DensityM} and \eqref{DensityM2} will also be approximately constant and equal to its vacuum value $f_{\phi_0}$. In this case, the scalar curvature term will resemble the usual Einstein-Hilbert term in four-dimensions, provided that the constant $\kappa_P$ satisfies the normalization condition
\beq \label{Normalization1}
 \frac{1}{2\, \kappa_P}\, f_{\phi_0} \, \Vol(K, \beta_0)  \ = \   \frac{1}{2\, \kappa_M}   \quad \ \   \iff  \ \ \quad      \kappa_P \ = \  \kappa_M \, \Vol(K, \, g_{\phi_0})    \ ,
\eeq
where $\kappa_M = 8\pi \,G \, c^{-4}$ is the Einstein gravitational constant. Again, if $\phi$ is close to its vacuum value $\phi_0$, also the potential term in \eqref{DensityM} and \eqref{DensityM2} will be approximately constant at its minimum value $V( |\phi_0|^2)$. In this case, the potential term resembles a four-dimensional cosmological constant term $\Lambda_M$ determined by
\beq   \label{Normalization2}
 \frac{1}{2\, \kappa_P} \, V( |\phi_0|^2) \, \Vol(K, \beta_0) \ = \ \frac{1}{2\, \kappa_M}\,  2\, \Lambda_M   \quad  \ \  \iff \ \  \quad  2\,  \Lambda_M \ = \ 2 \, \Lambda_P \, - \, R(g_{\phi_0}) \ .
\eeq
To normalize the Maxwell term of the photon gauge field, recall from section 2 that, among the left-invariant vector fields on $\SU$, there is a special one that is a Killing field of the metric $g_\phi$. It is generated by a vector $\gamma_\phi$ in the subalgebra $\iota(\utwo)$ of $\su$ that satisfies $[\gamma_\phi, \, \iota(\phi)] = 0$ and, up to normalization, is explicitly given by \eqref{GammaVector}. Decomposing $\su$ into the sum of the span of $\gamma_\phi$ and its orthogonal complement, the photon field $A_\LL^\gamma$ is defined as the component of $A_\LL$ with values in $\gamma_\phi$. The normalization of the field $A_\LL^\gamma$ is determined by the normalization of $\gamma_\phi$. Again, if $\phi$ is close to its vacuum value $\phi_0$, the Maxwell term in Lagrangians \eqref{DensityM} and \eqref{DensityM2} will resemble the canonical Maxwell term only if we pick the normalization $\accentset{\circ}{\gamma}_{\phi}$ of $\gamma_{\phi}$ satisfying the condition
\beq \label{Normalization3}
 \frac{1}{2\, \kappa_P} \, B_{\phi_0} \ \beta_0(\accentset{\circ}{\gamma}_{\phi_0}, \accentset{\circ}{\gamma}_{\phi_0})   \ \Vol(K, \, \beta_0) \ = \ 1 \ . \nonumber
\eeq
Using the definition of $B_{\phi_0}$, the definition of $\beta$ and the previous normalization condition \eqref{Normalization1}, this equation is equivalent to
\beq \label{Normalization4}
\frac{1}{2\, \kappa_P} \ \beta(\accentset{\circ}{\gamma}_{\phi_0}, \accentset{\circ}{\gamma}_{\phi_0})   \ \Vol(K, \, g_{\phi_0}) \ = \ 1   \quad \ \  \iff \ \   \quad \beta(\accentset{\circ}{\gamma}_{\phi_0}, \accentset{\circ}{\gamma}_{\phi_0})  \ = \ 2\,  \kappa_M\ .
\eeq
At this stage, we will not try to normalize the electroweak and strong-force fields, since the metrics $\beta$ and $g_\phi$ are not flexible enough to allow for separate normalizations of these fields. This will be addressed in section 5.2 using the metrics $\tbeta$ and $\tg_\phi$, which allow for adjustable values of the classical gauge coupling constants.

\newpage

\section{Masses of the classical fields}

\subsection*{Higgs-like particle}
\addcontentsline{toc}{subsection}{Higgs-like particle}

The purpose of this section is to calculate the classical masses associated to the fields $A_\LL$, $A_\RR$ and $\phi$ that appear in the four-dimensional Lagrangian density ${\mathscr L}_M$ written in \eqref{DensityM}. As customary \cite{Weinberg67, Weinberg, Hamilton}, the calculation is made in the approximation of weak fields that are small perturbations of the vacuum configuration defined by vanishing $A_\LL$ and $A_\RR$ and by constant $\phi = \phi_0$. We also work on Minkowski space with $R_M = 0$. Since the Lagrangian ${\mathscr L}_M$ is derived from the higher-dimensional scalar curvature, its terms do not come with the normalized coefficients that are conventional in the literature, so we will resort to the associated equations of motion to read the mass values.

Let us start with the mass of the ``Higgs particle'', that is, the mass of the radial component $r(x)$ of the field $\phi(x) \in \CC^2$. For $\phi_0 \neq 0$, we can write in the unitary gauge
\beq
\phi (x) \ = \ r(x) \, \frac{\phi_0}{|\phi_0|}
\eeq
and take the derivative
\beq
\big|\dd\, |\phi|^2 \, \big|^2 \ = \ \big|\dd\, r^2 \, \big|^2 \ = \  4 \, r^2\, g_M^{\mu \nu} \,(\partial_\mu r)\, (\partial_\nu r) \ .
\eeq
Using expression \eqref{CovariantDerivativePhi3} for the covariant derivative of fields with values in $\CC^2$, the norm that appears in ${\mathscr L}_M$ can be expanded as
\bal     \label{NormCovariantDerivative}
\big|\dd^{A_\LLL} \phi \big|^2 \ &= \ g_M^{\mu \nu} \, \Real \Big[  \,  \big( \dd^{A_\LLL} \phi \big)_\mu^\dag \,   \big( \dd^{A_\LLL}\phi  \big)_\nu  \, \Big]    \linebr
&= \  g_M^{\mu \nu} \, \left\{ \, (\partial_\mu r) (\partial_\nu r) \ + \ \frac{r^2}{|\phi_0|^2} \,  (A_\LL)_\mu^j \, (A_\LL)_\nu^k  \, \Real \big[ \, (\rho_{e_j} \phi_0)^\dag \, \rho_{e_k} \phi_0 \, \big] \,  \right\}
\ , \nonumber
\end{align}
where we have also used that $\phi_0^\dag \, \rho_{e_k} \phi_0$ is purely imaginary, since $\rho_{e_k}$ comes from a unitary action on $\CC^2$ and hence is an anti-hermitian matrix. The coefficient functions $f$, $B$, $C$ and $D$ that appear in the Lagrangian ${\mathscr L}_M$ depend on $r^2$ only, so can be written as $B_\phi = B(r^2)$, for instance. Doing this rebranding, taking the first variation of ${\mathscr L}_M$  with respect to $\delta r$ and ignoring the total derivative originated by $\Delta_M f$, yields the following equation of motion for $r(x)$:
\begin{multline}
2\, E(r^2) \, g_M^{\mu \nu} \, (\nabla_\mu \nabla_\nu r) \ + \ 2r \, E'(r^2) \, g_M^{\mu \nu} \, (\partial_\mu r) (\partial_\nu r) \ - \ \frac{r}{2} \, B'(r^2)  \, \left(\, |F_{A_\LLL}|^2_{\beta_0} \, + \, |F_{A_\RRR}|^2_{\beta_0} \, \right)  \linebr
- \ C(r^2) \, \ \frac{2r}{|\phi_0|^2} \,  (A_\LL)_\mu^j \, (A_\LL)_\nu^k  \, \Real \big[ \, (\rho_{e_j} \phi_0)^\dag \, \rho_{e_k} \phi_0 \, \big]  \ - \ 2r\, V'(r^2) \ = 0 \ ,
\end{multline}
where $E(r^2)$ stands for the combined function $C(r^2) \,+ \,4r^2\, D(r^2)$. A vacuum configuration for $g_P$ is defined as a product metric on $M \times K$ that minimizes the potential $V(|\phi|^2)$, so it is a configuration with vanishing one-forms $A_\LL$ and $A_\RR$ and a constant $\phi = \phi_0$ such that $V'(|\phi_0|^2) = 0$. Around the vacuum configuration we can decompose $r(x) = r_0 + \epsilon (x)$, with  $r_0 = |\phi_0|$ added to a small field $\epsilon (x)$, and expand
\[
r\,V'(r^2) \ = \  (r_0 + \epsilon) \,V'(r_0^2) \, + \, 2\, r_0^2\, V''(r_0^2)\, \epsilon \ + \ \cdots \ = \ 2\, r_0^2\, V''(r_0^2) \, \epsilon \ + \ \cdots \ .
\]
Also $A_\LL$ and $F_A$ will be small near the vacuum configuration, so only keeping the first order terms out of the full equation of motion yields the Klein-Gordon equation
\[
g_M^{\mu \nu} \, \nabla_\mu \nabla_\nu \, \epsilon \ - \ \frac{2\, r_0^2 \ V''(r_0^2)}{C(r_0^2) \,+ \,4\, r_0^2\ D(r_0^2)} \ \epsilon \ = \ 0 \ .
\]
Since we are working in $(- + + +)$ signature, the squared-mass of the radial field $r(x)$ can be defined as the coefficient
\beq \label{MassHiggsBoson}
 M_H^2 \ := \ ({\rm{Mass}}\ r)^2 \ = \ \frac{2\, r_0^2 \ V''(r_0^2)}{C(r_0^2) \,+ \,4\, r_0^2\ D(r_0^2)} \ .
\eeq
If $\phi_0$ is an absolute minimum of the potential $V(|\phi|^2)$, then the numerator of the squared-mass is non-negative. However, {\it a priori} nothing can be guaranteed about the denominator, since the function $D(|\phi|^2))$ defined in \eqref{CoefficientFunctions} has one negative term that depends on $f_\phi$, and hence on the chosen form of the scale factor $\lambda (|\phi|^2)$. This puts a constraint on the choice of function $\lambda (|\phi|^2)$. This does not happen for the Lagrangian $\hat{{\mathscr L}}_M$, as in this case $\hat{D}(|\phi|^2))$ is always positive in the domain $|\phi|^2 < 1/4$, as already pointed out.

\subsection*{Gauge bosons}
\addcontentsline{toc}{subsection}{Gauge bosons}

The calculations leading to a mass formula for the fields $A_\LL$ and $A_\RR$ mimic, in every essential way, the calculations usually performed in the case of the electroweak gauge fields of the Standard Model \cite{Weinberg67, Weinberg, Hamilton}. One works in the approximation where the one-forms $A_\LL$ and $A_\RR$ are small, close to their vanishing ``vacuum'' value, and the parameter $\phi \in \CC^2$ is approximately constant and equal to $\phi_0$. The terms of the four-dimensional Lagrangian ${\mathscr L}_M$ that depend on $A_\LL$ and $A_\RR$ are
\beq
- \, \left[ \, \frac{1}{4}\,  B_\phi \, \left(\, |F_{A_\LLL}|^2_{\beta_0} \, + \, |F_{A_\RRR}|^2_{\beta_0} \, \right) \,+ \,C_\phi \,  \big|\dd^{A_\LLL} \phi \big|^2  \, \right] \ \frac{\Vol(K, \beta_0)}{2\, \kappa_P}  \nonumber \ ,
\eeq
where one should keep in mind that the whole formula \eqref{DensityM} for ${\mathscr L}_M$ is valid only for one-forms $(A_\LL \, A_\RR)$ with values in the subalgebra $\iota(\utwo) \oplus \su$ of the bigger $\su \oplus \su$. This expression does not contain any quadratic terms on the fields $A_\RR$, so they have zero mass in the model. Using formula \eqref{NormCovariantDerivative} for the norm of the covariant derivative $\dd^{A_\LLL} \phi$ at constant $\phi = \phi_0$, the terms involving $A_\LL$ can be rewritten as
\beq
-  \left[ \, \frac{1}{4}\,  B_{\phi_0} \ \beta_0(e_k, e_j) \ (F_{A_\LLL}^k)^{\mu \nu}  (F_{A_\LLL}^j)_{\mu \nu}  \,+ \,C_{\phi_0} \,  \,  (A_\LL^j)^\mu \, (A_\LL^k)_\mu  \, \Real \big[ \, (\rho_{e_j} \phi_0)^\dag \, \rho_{e_k} \phi_0 \, \big]    \right]  \frac{\Vol(K, \beta_0)}{2\, \kappa_P} \nonumber \ ,
\eeq
where we have chosen a basis $\{e_k \}$ for the subspace $\iota(\utwo)$ of $\su$, while $\beta_0 $ is just the usual Ad-invariant product $\beta_0 (u, v) = \Tr (u^\dag \, v)$ on $\su$. 
Working with the Levi-Civita connection $\nabla$ on $M$ and ignoring total derivatives, the first variation of the expression above with respect to $\delta (A_\LL^j)^\mu$ leads to the equations of motion
\beq \label{EqMotionGaugeBosons}
  B_{\phi_0} \ \beta_0(e_k, e_j) \ g_M^{\mu \nu} \, g_M^{\sigma \rho} \ \nabla_\nu (F_{A_\LLL}^k)_{\mu \sigma} \ - \ 2 \, C_{\phi_0} \ g_M^{\mu \rho} \, (A_\LL^k)_\mu \,   \Real \big[ \, (\rho_{e_j} \phi_0)^\dag \, \rho_{e_k} \phi_0 \, \big]  \ = \ 0 \, .
\eeq
In the particular case where the basis $\{e_k \}$ is $\beta_0$-orthogonal $\iota(\utwo)$ and, simultaneously, diagonalizes the quadratic form $(u,v) \, \mapsto \, \Real \big[ \, (\rho_{u} \phi_0)^\dag \, \rho_{v} \phi_0 \, \big]$ on the same space, the equations of motion can be simplified to
\beq
g_M^{\mu \nu} \ \nabla_\nu (F_{A_\LLL}^k)_{\mu \sigma} \ - \ \frac{2 \, C_{\phi_0} }{B_{\phi_0} \, \beta_0(e_k, e_k) }  \ (A_\LL^k)_\sigma \,   \Real \big[ \, (\rho_{e_k} \phi_0)^\dag \, \rho_{e_k} \phi_0 \, \big]  \ = \ 0 \ , 
\eeq
where no sum over the index $k$ is intended. The usual arguments using the Lorentz condition $\partial^\mu A_\mu^k = 0$ (e.g. see \cite[section 2.7]{MS}) then say that, to first order in the fields, these equations can be simplified to the Klein-Gordon equation for gauge fields of mass
\bal
 \left[{\rm{Mass}}\  (A_\LL^k)_\mu \right]^2 \ &:= \ \frac{2 \, C_{\phi_0} \  (\rho_{e_k} \phi_0)^\dag \, \rho_{e_k} \phi_0 }{B_{\phi_0} \ \beta_0(e_k, e_k) }   \\[0.5em]
&= \  \frac{6\, (1- 2 |\phi_0|^2)  \   (\rho_{e_k} \phi_0)^\dag \, \rho_{e_k} \phi_0  }{\lambda \, (1 - |\phi_0|^2) \ (1 - 4 |\phi_0|^2)\  \beta_0(e_k, e_k) }  \ ,   \nonumber  
\end{align}
where we have used the explicit expressions \eqref{CoefficientFunctions} for the coefficient functions $B_\phi$ and $C_\phi$ evaluated at the vacuum value $\phi_0$. Recall that in this formula the vacuum vector $\phi_0$ should be regarded as an element of $\CC^2$; the squared-norm $|\phi_0|^2$ stands for the canonical norm on $\CC^2$; and $\rho_{e_k}$ is the representation of $\utwo$ on $\CC^2$ induced by the $U(2)$-representation $\phi \mapsto (\det a)\, a \phi$ on the same space. Unwinding the path that originally lead us to the representation $\rho_{e_k}$, one can also express the quadratic form $(\rho_{e_k} \phi_0)^\dag \, \rho_{e_k} \phi_0$ in $\CC^2$ as an equivalent form in $\su$. In fact, it follows from the initial expressions \eqref{DecompositionInnerProduct} and \eqref{AdjointAction} that this relation is simply
\beq    \label{IdentityQuadraticForm}
2 \, \Real \big[ \, (\rho_{e_j} \phi_0)^\dag \, \rho_{e_k} \phi_0 \, \big]  \ = \ \Tr \big( \, [e_j , \phi_0 \,]^\dag \ [e_k, \phi_0 \,] \, \big) \ ,
\eeq
where all the vectors on the right-hand side should be regarded as elements of $\su$, and $\phi_0$ should have properly been written as $\iota(\phi_0) \in \iota (\CC^2) \subset \su$. Thus, an alternative formula for the mass of the fields $A_\LL^k$ is
\beq \label{MassGaugeFields}
 \left[{\rm{Mass}}\  (A_\LL^k)_\mu \right]^2 \ = \  \frac{3\, (1- 2 |\phi_0|^2)  \  \Tr \big( \, [e_k, \phi_0]^\dag \, [e_k, \phi_0] \, \big) }{\lambda \, (1 - |\phi_0|^2) \ (1 - 4 |\phi_0|^2)\  \Tr( e_k^\dag \, e_k) }   \ .
\eeq
Again, this formula assumes that the basis $\{e_k \}$ is $\beta_0$-orthogonal in $\iota(\utwo)$ and, simultaneously, that it diagonalizes the quadratic form $(u,v) \, \mapsto \, \Tr \big( \, [u, \phi_0]^\dag \, [v, \phi_0] \, \big)$ on the same subspace of $\su$. One such basis is explicitly constructed in appendix A.1, comprising four $\beta_0$-orthogonal vectors $(\gamma_\phi,\,  z_\phi,\, w^1_\phi, \, w^2_\phi)$. If the components of the one-form $A_\LL$ on that basis are denoted by
\[
A_\LL \ = A^\gamma \, \gamma_\phi \ + \  Z \, z_\phi \ + \ W^1 \, w^1_\phi \ + \ W^2 \, w^2_\phi \ ,
\]
then the classical mass associated to each of these component fields follows directly from \eqref{MassGaugeFields} and the algebraic identities \eqref{A.5} and \eqref{A.8} of the appendix. We obtain:
\bal \label{MassGaugeBosons}
M_\gamma^2 \ &:= \  \left[{\rm{Mass}}\  (A^\gamma)_\mu \right]^2 \ = \ 0      \linebr
M_W^2 \ &:= \   \left[{\rm{Mass}}\  (W^a)_\mu \right]^2 \ = \ \frac{3\, \big(1- 2 |\phi_0|^2\big)  \, |\phi_0|^2  }{\lambda\, \big(1 - |\phi_0|^2\big) \, \big(1 - 4 |\phi_0|^2\big) }       \nonumber   \linebr 
M_Z^2 \ &:= \   \left[{\rm{Mass}}\  Z_\mu \right]^2 \ = \  4\,  M_W^2  \nonumber  \ .
\end{align}
The simple relation $M_Z = 2 \, M_W$, obtained above, seems to be a feature of the classical model described so far. However, it is significantly different from the experimental ratio $M_Z \simeq 1.13\, M_W$ observed for the masses of real $Z$ and $W$-bosons. One can point out that these are calculations for the bare masses, and all the relations are at the classical, unification energy scale, not at the experimental energy scale. But unless the running coupling constants and quantum radiative corrections come to the rescue in significant amounts --- something that will not be studied here --- this discrepancy shows that the fields $W^a_\mu$ and $Z_\mu$ described above cannot be regarded as quantitatively precise models for the real electroweak gauge fields. This situation will be improved upon in section 5.2.

That being said, the numbers and expressions obtained above do not seem to be entirely off the mark either, especially for a Lagrangian derived from a remote object such as the higher-dimensional scalar curvature. Let us consider the mass ratios $M_H \, /\, M_W$ and $M_H \, /\, M_Z$, for example. The second equation in \eqref{MassGaugeBosons} gives an explicit expression for the classical mass of the $W$-like boson in terms of the  constant $\lambda$ and the value of $|\phi|^2$ at the minimum of the potential. At the same time, formula \eqref{MassHiggsBoson} gives an expression for the mass of the Higgs-like boson in terms of similar variables, so we can try to compare the two masses. The potential $\hat{V}_{a, q}$ considered in \eqref{DefinitionVAQ} depends on two parameters. We discussed at length the simplest choice $q=0$, and then also mentioned the case $q= -1/5$. For each value of $q$ the potential depends on the second parameter $a$, defined in \eqref{DefinitionParameterA}, which also affects the ``vacuum expectation value'' $|\phi_0|$. Tables 1 and 2 register the numerically approximated values of $|\phi_0|$ and $V(|\phi_0|^2)$ as the parameter $a$ takes a sequence of naive, non-optimized values, bigger than the threshold necessary to produce double-well potentials. Formula \eqref{MassGaugeBosons} was then applied to calculate the associated mass $\lambda_0 \, M^2_W$ as the parameter varies.

Turning to the mass of the Higgs-like boson, recall that in section 3.6 we discussed two different Lagrangian densities on $P$, that lead to distinct four-dimensional Lagrangians ${\mathscr L}_M$ and $\hat{{\mathscr L}}_M$ after integration over the fibre. The two Lagrangians have the same implicit potential $V$, and hence lead to the same values of $|\phi_0|$ and $M_W$, but they differ on the coefficient function $D(|\phi|^2)$ defined in \eqref{CoefficientFunctions} and \eqref{CoefficientFunctions2}. Hence, through formula \eqref{MassHiggsBoson}, they lead to distinct masses $M_H$ of the Higgs-like boson. Tables 1 and 2 also register the numerically approximated values of $\lambda_0 \, M_H$, calculated from \eqref{MassHiggsBoson}, using the same sequence of values of the parameter $a$.\footnote{Numerical computations using the free online calculator available in https://wims.unice.fr/wims/ $\, $.} 
They are presented indirectly through the ratios $M_H / M_W$. 
\begin{table}[H]      % Table corrected in arXiv-v2.
\centering 
\begin{tabular}{ |c|  c  c c c c  c c c c|} \cline{5-10} 
\multicolumn{4}{c|}{}                &  \multicolumn{2}{c|}{\rule{0pt}{.5cm}$\lambda_0\, M^2_H$} & \multicolumn{2}{c}{$M_H \, /\, M_W$}   & \multicolumn{2}{|c|}{$M_H \, / \, M_Z$}   \\ \hline
$ \rule{0pt}{.5cm} a$   & $|\phi_0|$     & $\lambda_0^{-3} \, V(|\phi_0|^2)$       & $\lambda_0\, M^2_W$    & \multicolumn{1}{|c}{${\mathscr L}_M$}   & $\hat{{\mathscr L}}_M$  & \multicolumn{1}{|c}{${\mathscr L}_M$}  & $\hat{{\mathscr L}}_M$  &  \multicolumn{1}{|c}{${\mathscr L}_M$}   & $\hat{{\mathscr L}}_M$  \\ \hline
6.5 & 0        &    1     & 0        & 0           & 0           &        -        &       -         &         -       &      -          \\
6.51 &  0.04076 & 1.02 & 0.00501 & 0.0406 & 0.0392 & 2.85 &2.80 & 1.42 & 1.40      \\
6.55 &  0.09040  & 1.10 & 0.0251 & 0.215 & 0.182 & 2.92 & 2.69 & 1.46 & 1.34       \\
6.6 &   0.1266  & 1.19 & 0.0505 & 0.462 & 0.333 & 3.02 & 2.57 & 1.51  & 1.28      \\
6.8  &  0.2110 & 1.56 & 0.155 & 1.91 & 0.764 & 3.51 & 2.22 & 1.75 & 1.11       \\
7   &  0.2621 & 1.89 & 0.263 & 4.63 & 1.05 & 4.19 & 2.00 & 2.10 & 1.00      \\
8   &  0.3794 & 3.21 & 0.847 & -82.1 & 2.09 & -  & 1.57 & - & 0.786        \\
30  &  0.4912 &13.2  & 14.1 & -1758 & 27.6 & -  & 1.40 & - & 0.698      \\
100 &  0.4978 & 26.5 & 56.2 & -25826 & 109 & -  & 1.39 &  - & 0.697         \\ 
500 & 0.4996 & 60.8 & 296 & -704802 & 574 & -  & 1.39 & - & 0.696    \\\hline
\end{tabular}
\vspace{-.2cm}
\caption{Bosonic mass ratios for different values of the parameter $a$ when $q= 0$.}
\label{tab:table1}
\end{table}
\vspace{-.25cm}
\begin{table}[H]        % Table corrected in arXiv-v2.
\centering 
\begin{tabular}{ |c|  c  c c c c  c c c c|} \cline{5-10} 
\multicolumn{4}{c|}{}                &  \multicolumn{2}{c|}{\rule{0pt}{.5cm}$\lambda_0\, M^2_H$} & \multicolumn{2}{c}{$M_H \, /\, M_W$}   & \multicolumn{2}{|c|}{$M_H \, / \, M_Z$}   \\ \hline
$ \rule{0pt}{.5cm} a$   & $|\phi_0|$     & $\lambda_0^{-3} \, V(|\phi_0|^2)$       & $\lambda_0\, M^2_W$    & \multicolumn{1}{|c}{${\mathscr L}_M$}   & $\hat{{\mathscr L}}_M$  & \multicolumn{1}{|c}{${\mathscr L}_M$}  & $\hat{{\mathscr L}}_M$  &  \multicolumn{1}{|c}{${\mathscr L}_M$}   & $\hat{{\mathscr L}}_M$  \\ \hline
14.5 & 0        & 17      & 0        & 0           & 0           &        -        &         -       &         -       &       -         \\
14.51 & 0.01977 & 17.0 & 0.00117 & 0.00811 & 0.00811 & 2.63 & 2.63 & 1.31 & 1.31      \\
14.55 & 0.04390 & 17.1 & 0.00581 & 0.0399 & 0.0399 & 2.62 & 2.62 & 1.31 & 1.31        \\
14.6 &  0.06189 & 17.2 & 0.0116 & 0.0797 & 0.0794 & 2.62 & 2.62 & 1.31 & 1.31       \\
14.8  & 0.1059 & 17.6 & 0.0346 & 0.237 & 0.235 & 2.62 & 2.61 & 1.31 & 1.30       \\
15   & 0.1352 & 18.0 & 0.0574 & 0.392 & 0.387 & 2.61 & 2.59 & 1.31 & 1.30        \\
16   & 0.2215 & 20.0 & 0.168 & 1.13 & 1.09 & 2.59 & 2.55 & 1.30 & 1.27       \\
30   & 0.4335 & 44.8 & 1.456 & 8.88 & 8.13 & 2.47 & 2.36 & 1.23 & 1.18     \\
100  & 0.4874 & 149 & 6.89 & 40.1 & 37.0 & 2.41 & 2.32 & 1.21 & 1.16      \\
500  &  0.4980  & 648 & 37.0 & 212  & 196 & 2.39 & 2.30 & 1.20 & 1.15    \\\hline
\end{tabular}
\vspace{-.25cm}
\caption{Bosonic mass ratios for different values of the parameter $a$ when $q= -1/5$.}
\label{tab:table2}
\end{table}
The experimental values of these ratios are approximately $M_H \, /\, M_W \simeq 1.56$ and $M_H \, / \, M_Z \simeq 1.37$. Thus, a first observation is that the values in tables 1 and 2 are certainly inaccurate, but reasonably within the correct order of magnitude, even though the model does not rely on an independent parameter to adjust the mass of the Higgs-like boson. Since the classical model works with the inaccurate relation $M_Z = 2 \, M_W$, one cannot expect it to simultaneously match both experimental $M_H / M_{\bullet}$ ratios, but an hypothetical correction to that initial inaccuracy could improve the ratios correspondence as well.  This is most evident in table 2, where a slightly lighter $Z$ and a slightly heavier $W$ would bring both ratios closer to the experimental values.

In section 5.2 we will describe a version of the present model where the metrics on internal space $\beta$ and $g_\phi$ have additional deformation parameters, equivalent to the three gauge coupling constants of the Standard Model. Using these new parameters one can adjust the mass ratio $M_Z / M_W$ at will, and hence improve the adherence of the model to the experimentally observed values of the bosons' masses. The downside is that more adjustable parameters diminish the predictive usefulness of the model, of course.

An interesting facet of the formula for the mass of the $W$-boson is its relation with the volume and scalar curvature of the vacuum internal space $(K,\, g_{\phi_0})$. Direct combinations of \eqref{MassGaugeBosons} with expressions \eqref{VolumeK} and  \eqref{ScalarCurvature}, from section 2, lead to the relations
\bal \label{VolumeMassW}
\Vol \,(K,\, g_{\phi_0})  \ &= \ \frac{ \sqrt{3} \ 6^4 \ \pi^{5} \ |\phi_0|^8 \, \big(1-2\, | \phi_0|^2 \big)^4}{ M_W^{8} \, \big( 1 - |\phi_0|^2 \big)^{3}  \,  \big( 1 - 4\, |\phi_0|^2\big)^{7/2}  } \\[0.5em]
R(K,\, g_{\phi_0}) \ &= \  M_W^2 \   \frac{4 - 25\, |\phi_0|^2 + 33\, |\phi_0|^4 - 8 \,|\phi_0|^6 }{|\phi_0|^2 \, \big(1-|\phi_0|^2\big)\, \big(1- 2 |\phi_0|^2\big)} \ .  \nonumber
\end{align}
These are determined by the vacuum ``expectation value'' $|\phi_0|$ at the minima of the potential and have the merit of not depending explicitly on the unknown scaling factor $\lambda$.

\subsection*{Some numerical estimates}
\addcontentsline{toc}{subsection}{Some numerical estimates}

Consider again the vector $\gamma_\phi$ in $\su$ that generates the electromagnetic $\mathrm{U}(1)$-isometries of the metric $g_\phi$. It was defined in \eqref{GammaVector} as a function of $\phi$. The normalization condition \eqref{Normalization4} was applied in the calculations of \cite{Baptista}, section 2, and lead to a relation between the positron electromagnetic charge $e$ and the inner-product $\beta(\gamma_\phi, \gamma_\phi)$ in the approximation where $\phi$ is constant and equal to its vacuum value $\phi_0$. This relation is the first equality in
\beq
\frac{e^2}{6\, \kappa_M} \ = \ \beta( \gamma_{\phi_0}, \, \gamma_{\phi_0}) \ = \ \lambda\, \Tr(\gamma_{\phi_0}^\dag \, \gamma_{\phi_0}) \ = \ 2\, \lambda \ .
\eeq
The second equality is the definition of the product $\beta$ and the third equality follows from calculation \eqref{A.5} in the appendix. In Lorentz-Heaviside-Planck units with $c = \hbar = \varepsilon_0 = \mu_0 = 8\pi G = 1$, we have that $\kappa_M = 1$ and $e = \sqrt{4 \pi \alpha}$, where $\alpha \simeq 1/ 137$ is the fine-structure constant. Thus, we get at estimate for the scale factor $\lambda$ that appears in the definitions of the metrics $\beta$ and $g_\phi$,
\beq \label{EstimateLambda}
\lambda (|\phi_0|^2) \ = \frac{e^2}{12\, \kappa_M} \ = \ \frac{\pi\, \alpha}{3} \ .
\eeq
Using this value in formula \eqref{MassGaugeBosons} for the mass of the W-bosons, one obtains
\beq
M_W^2 \ = \  \frac{9\,  |\phi_0|^2 \, \big(\ell_P^{\,2} \, - \, 2 |\phi_0|^2\big)  }{\pi \,\alpha\, \big(\ell_P^{\,2}\, -\, |\phi_0|^2\big) \, \big(\ell_P^{\,2} \, -\, 4 |\phi_0|^2\big) } \ M_P^2 \ , 
\eeq
where we have displayed the implicit Planck length $\ell_P = \sqrt{8\pi G\hbar c^{-3}}$ and Planck mass $M_P = \sqrt{\hbar c / (8\pi G)}$, so that the equation remains valid in any system of units. Recall that $|\phi|$ refers here to the standard norm in $\CC^2$ of the vector $\phi$, which is identified with an element of $\su$, i.e. a tangent vector to the internal space $K$, and thus has the dimensions of length. But the experimental value of $M_W$ is many orders of magnitude smaller than the Planck mass, so the formula above implies that the vacuum value of the deformation $\phi$ must be very small, that is $|\phi_0| <<  \ell_P$ inside its usual domain $[0;\, \ell_P /2[$. In fact, using the experimental value of $M_W$ and calculating to lowest order in the ratio $M_W/ M_P$, we get the estimate 
\beq \label{EstimatePhi}
|\phi_0| \ \simeq \ \frac{ \sqrt{\pi \,\alpha}}{3}  \ \frac{M_W  }{M_P} \ \ell_P   \ \simeq \  1.67 \times 10^{-18} \ \ell_P \ \simeq \  1.35 \times 10^{-52} \ {\rm m} \ .
\eeq
The values of $\lambda$ and $|\phi_0|$ coming from these estimates can also be applied to formulae \eqref{VolumeK} and \eqref{ScalarCurvature}, giving the volume and scalar curvature of the vacuum metric $g_{\phi_0}$ on the internal space $K$. To lowest order in $|\phi_0|$, we obtain that
\bal \label{VolumeCurvatureEstimates}
\Vol \,(K,\, g_{\phi_0})  \ &\thicksim \  \sqrt{3} \ \pi^{5} \  \Big(\frac{2\, \pi\, \alpha  }{3} \Big)^4 \ \ell_P^{\, 8}   \ \thicksim \   (\, 0.27\  \ell_P\, )^8    
 %\ \thicksim \     (  2.2 \times 10^{-35}  {\rm m} )^8   
  \nonumber  \\[0.5em]
R(K,\, g_{\phi_0}) \ &\thicksim \  \frac{36}{\pi\, \alpha}  \ \ell_P^{\,-2}   \ .  %  \ \thicksim \    24  \times 10^{70}\ {\rm m}^{-2}         \ .  \nonumber     % Formula corrected in arXiv-v2.
\end{align}
For very small $|\phi_0|$, formula \eqref{MassHiggsBoson} for the mass of the Higgs-like boson also gets simplified. Since the coefficient function $D(|\phi_0|^2)$ is finite at the origin, to lowest order in $|\phi_0|^2$ we have the asymptotic expression
\beq
M_H^2 \ \thicksim \ \frac{2\, |\phi_0|^2\ V''(|\phi_0|^2)}{C(|\phi_0|^2)}  \ \thicksim \ \frac{2\, |\phi_0|^2\ V''(|\phi_0|^2)}{3\, \lambda^4}  \ .
\eeq
Using expansion \eqref{ExpansionPotential} of the potential $V(|\phi|^2)$, to lowest order in $|\phi_0|^2$ the second derivative is constant,
\beq \label{EstimateDerivativePotential}
V''(|\phi_0|^2) \  \thicksim \ V''(0)  \  = \  2\, \lambda_0^3 \,  (18 + 12 \,a b^2 - 24\, a b  + 8 \,a d   - 36 \,b^2  + 117\, b  - 36\, d  ) \ ,
\eeq
whereas the potential has an absolute minimum for positive but very small $|\phi_0|^2$ only if the constant $a$ is just slightly bigger than the critical value $(39 - 36\,b) / (6 - 8 \,b)$. Substituting this value of $a$ in the second derivative \eqref{EstimateDerivativePotential}, we obtain that, to lowest order in $|\phi_0|^2$,
\beq \label{AsymptoticMH}
M_H^2 \ \thicksim \  \frac{ 4\, (18 + 16 \,d - 63\, b + 30\, b^2 - 24 \,b^3) }{\lambda_0\, (3-4\,b)} \ |\phi_0|^2 \ .
\eeq
This asymptotic expression for $M_H^2$ can be compared with the behaviour of $M_W^2$ and $M_Z^2$ for small $|\phi_0|^2$, as implied in \eqref{MassGaugeBosons}. The comparision leads to the mass ratios
\beq \label{AsymptoticMassRatio}
\frac{M_H}{M_Z} \ = \   \frac{M_H}{2\, M_W}   \  \thicksim \ \sqrt{ \, \frac{18 + 16 \,d - 63\, b + 30\, b^2 - 24 \,b^3}{3\, (3-4\,b)} } \ .
\eeq
This is the asymptotic value of the ratios when the constant $a$ in the potential tends from above to the critical value $(39 - 36\,b) / (6 - 8 \,b)$. In other words, when the constant $a$ is chosen so that $V(|\phi|^2)$ attains its absolute minima for positive but very small $|\phi_0|$, as suggested by \eqref{EstimatePhi}. The asymptotic value of the ratio depends on the behaviour of the function $\lambda(|\phi|^2)$ near the origin, reflected here in the presence of the coefficients $b$ and $d$ coming from expansion \eqref{ExpansionLambda}. In the case of a constant function $\lambda(|\phi|^2) = \lambda_0$, the coefficients $b$ and $d$ vanish, so we get that $M_H / M_Z \thicksim 1.41 $ for very small $|\phi_0|$, in agreement with the numerical values on top of table \ref{tab:table1}.  In the case of a function $\lambda(|\phi|^2)$ defined by \eqref{DefinitionLambda2} with constant $q = -1/5$, the expansion coefficients are $b = 3/5$ and $d = 27/25$, so we get an asymptotic mass ratio of $M_H / M_Z \thicksim 1.31$, in agreement with the values on top of table \ref{tab:table2}. When the behaviour of $\lambda(|\phi|^2)$ near the origin is determined by coefficients $b$ and $d$ such that the numerator of \eqref{AsymptoticMH} is negative, or $b = 3/4$, the derivation of \eqref{AsymptoticMH} is not valid and the formula is not applicable.

\newpage

\section{Further investigations}

\subsection*{Higher-dimensional equations of motion}
\addcontentsline{toc}{subsection}{Higher-dimensional equations of motion}

In section 2.3 we defined the family of left-invariant metrics $g_\phi$ on $K=\SU$ and studied several of its properties. Subsequently, in section 2.6, we looked at higher-dimensional metrics $g_P$ on the product $P = M_4 \times K$ that coincide with $g_\phi$ when restricted to the fibres K. The parameter $\phi(x) \, \in \CC^2$ was allowed to depend on the fibre in question, i.e. it was allowed to depend on the coordinate $x \in M_4$. Finally, we studied the fibre-integral of the higher-dimensional density $R_P - 2 \Lambda_P$ and showed that it defines a Lagrangian on $M_4$ with terms very similar to those found in the Standard Model Lagrangian. These similarities include the presence of a Higgs-like field $\phi(x)$ with its usual covariant derivative; the four-dimensional Yang-Mills terms, as in the familiar Kaluza-Klein calculation; and the existence of a potential term that, in some cases, has absolute minima for non-zero values of $\phi$, leading to spontaneous symmetry breaking and a vacuum metric with $\mathrm{U}(1) \times \SU$ isometry group, which in turn produces the usual massless gauge bosons. 

However, we have not really justified the initial choice of metric $g_\phi$ on the internal space, other than pointing to its nice features and to the similarities of the resulting geometrical model with the bosonic part of the Standard Model. More importantly, having always worked with fibre-integrals leading to effective Lagrangians in four dimensions, we have not investigated whether the internal metrics $g_\phi$ would be stable in a fully dynamical higher-dimensional theory. The potential $V(|\phi|^2)$ may govern the dynamics of the parameter $\phi$ within the restricted family of metrics $g_\phi$, so that a minimum of the potential corresponds to a metric that is stable within the family. But nothing was said about stability in the space of all metrics on $P$. 
If all the coefficients of the internal metric were allowed to be dynamical, besides the parameter $\phi(x)$, what would prevent an initial metric $g_\phi$ to evolve over time to a metric outside that family, according to the higher-dimensional, classical equations of motion?

If the higher-dimensional equations of motion are determined by the Lagrangian $R_P - 2 \Lambda_P$ on $P$, then the classical solutions are the Einstein metrics. But a cartesian product of metrics $g_M \times g_K$ is Einstein on $M_4 \times K$ if and only if both $g_M$ and $g_K$ are Einstein, with the same constant, on the respective spaces. Thus, our vacuum metric $g_M \times g_{\phi_0}$ cannot be a solution of the full equations of motion, since the left-invariant metrics $g_\phi$ are not Einstein on $K$, except for the bi-invariant metric at $\phi = 0$. 

To justify the last assertion, recall that a metric on an $n$-dimensional compact manifold $K$ is Einstein if and only if it is a critical point of the normalized functional
\beq \label{Functional1}
{\mathcal E} (g) \ := \  (\Vol_g \, K)^{(2-n) / n}\,  \int_K \,  R_g\ \vol_g  \ .
\eeq
The left-invariant metric $g_\phi$ has constant scalar curvature, so the integral above is equal to $(\Vol_{g_\phi} \, K)^{2/n} \, R_{g_\phi}$. Putting $n=8$ and using formula \eqref{ScalarCurvature} we obtain
\bal  \label{Functional2}
{\mathcal E} (g_\phi) \ &= \  (\Vol_{g_\phi} \, K)^{1/4} \ R_{g_\phi}  \linebr
&=  \  6\, \big( \sqrt{3}\,\pi^5 \big)^{1/4} \ \frac{4 - 25\, \mph^2 + 33\, \mph^4 - 8 \,\mph^6}{(1-\mph^2)^{7/4} \, (1 - 4\mph^2)^{7/8}}  \nonumber \ .  % Formula corrected in arXiv-v2.
\end{align}
If a particular $g_{\phi_0}$ is a critical point of functional \eqref{Functional1} for general variations of the metric, then it must define a stationary point of function \eqref{Functional2} under variations of $\phi$, since these are just a special kind of variation of the metric. But a simple plot shows that the only stationary point of ${\mathcal E} (g_\phi)$ as a function of $|\phi|$ happens at $|\phi| = 0$. Therefore, the only possible Einstein metric in the family $g_\phi$ is the bi-invariant metric, which is well-known to be Einstein. Notice how the scaling factor $\lambda (|\phi|^2)$ of the metric $g_\phi$ is absent from \eqref{Functional2}, therefore the argument is valid for any choice of scaling function.

The stability of vacuum metrics under higher-dimensional dynamics is an important and challenging topic in Kaluza-Klein theories, as already mentioned in the Introduction. It has been extensively studied and discussed in the literature. See for instance the reviews in \cite{Bailin, Duff, Witten81}. Within the small realm of the present model, after recognizing that the metrics $g_\phi$ are not Einstein, once could try to address the problem in several, non-exclusive ways. The first would be to propose that the higher-dimensional dynamics may be governed not by the Lagrangian $R_P - 2 \Lambda_P$, but by a more elaborate scalar density whose associated equations of motion could have something like $g_M \times g_{\phi_0}$ as a classical solution. A second way would be to study vacuum metrics that are not pure cartesian products $g_M \times g_{\phi_0}$, for example letting $\phi_0$ have a slight dependence on the $x$ coordinate, and see if this concession leads to an Einstein metric that could be a reasonable candidate for the vacuum. A third approach, probably the most natural within the limited scope of our model, would be to slightly adjust the definition of the metric $g_\phi$ and accept additional parameters besides $\phi$ and the scaling factor $\lambda$. The hope would be to find a solution of the Einstein condition in this enlarged family of metrics, which in turn would help to fix the values of the additional parameters. We will now elaborate on this third route.

\subsection*{A more precise version of the model}
\addcontentsline{toc}{subsection}{A more precise version of the model}

Motivated by the inaccuracy of the classical relation $M_Z = 2 M_W$, obtained in section 4.2, as well as the previous discussion about the instability of product metrics $g_M \times g_{\phi_0}$ under the higher-dimensional equations of motion, we will now adjust the definition of the metric $g_\phi$ on the internal space by including additional deformation parameters that may help mitigate those problems. The additional parameters essentially correspond to the three different gauge coupling constants of the Standard Model, so it sounds reasonable to let them be adjustable.

Recall that the metric $g_\phi$ on $\SU$ was defined as the left-invariant extension of the inner-product on $\su$ determined by \eqref{DefinitionMetric1}. This formula uses the Ad-invariant product $\beta(u,v) = \lambda \Tr (u^\dag \, v)$ on $\su$ and, in fact, the deformation $g_\phi$ is defined to coincide with $\beta$ when restricted to the subspaces $\iota(\utwo)$ and $\iota(\CC^2)$. Let us now relax these definitions by renouncing to $\beta$, the most general $\Ad_{\SU}$-invariant product on the Lie algebra $\su$, and use instead the general $\Ad_{\Utwo}$-invariant product on $\su$, which we will call $\tbeta$. Decomposing vectors in $\su = \utwo \oplus \CC^2 = \mathfrak{u}(1) \oplus \sutwo \oplus \CC^2$ as
\beq \label{Decomposition2}
v \ = \ v' \, +\, v'' \ = \ v_Y\, +\, v_W \, + \, v'' \ ,
\eeq
the product $\tbeta$ on $\su$ can be written as a sum
\beq \label{DefinitionTBeta}
\tbeta(u,v) \ := \ \lambda_1 \Tr(u_Y^\dag\, v_Y) \ + \ \lambda_2 \Tr(u_W^\dag\, v_W) \ + \ \lambda_3 \Tr\big[(u'')^\dag \, v'' \big]
\eeq
for positive constants $\lambda_1$, $ \lambda_2$ and $ \lambda_3$. So the new product $\tbeta$ is a version of $\beta$ with an independent rescaling factor in each component of $\su$. Using the finer decomposition \eqref{Decomposition2}, the formula for the $\Ad_{\Utwo}$-action on $\su$ can be written as
\beq \label{AdjointAction2}
\Ad_{\iota(a)} (v) \ = \ \bmatr -\Tr(v_Y) &  - [\,(\det a)\, a\, v'' \,]^\dag \\[0.6em]   (\det a)\, a\, v''  &  v_Y + \Ad_a (v_W) \ematr  \ ,
\eeq
instead of \eqref{AdjointAction}, for all matrices $a\in \Utwo$. It is not difficult to convince oneself that $\tbeta$ is indeed the most general inner-product on $\su$ invariant under such transformations. The new deformed metric $\tg_\phi$ can then be defined in terms of $\tbeta$ by a formula entirely analogous to the definition of $g_\phi$ in terms of $\beta$, namely
\bal \label{DefinitionMetric3}
\tg_\phi (u,v) \ &:= \ \tbeta (u,v) \, +\, \tbeta \left( \,[u', v''] + [v', u''] ,\, \phi\, \right)  \linebr
                    & \ = \ \tbeta (u,v) \, + \, \tbeta \left(\, [\Ad_\theta u ,\, v] , \, \phi \, \right)  \ . \nonumber
\end{align}
This definition implies that the product $\tg_\phi$ coincides with $\tbeta$ when restricted to the subspaces $\utwo$ and $\CC^2$ of the larger $\su$, although these subspaces are not $\tg_\phi$-orthogonal to each other. One can check that the orthogonal complements $(\CC^2)^\perp$ for the new product $\tg_\phi$ coincides with that calculated for the product $g_\phi$, so is still given by \eqref{orthogonal2}. Formula \eqref{orthogonal1} for $\utwo^\perp$ is no longer valid, however, due to the different rescalings inside $\utwo$. The transformation rule of $\tg_\phi$ under the $\Ad_{\Utwo}$-action on $\su$ remains as calculated for $g_\phi$, namely
\beq
(\Ad_{\iota(a)^{-1}})^\ast \,\tg_\phi \ = \ \tg_{(\det a) a\, \phi} \  
\eeq
for any $a \in \Utwo$. The arguments of section 2 carry over to show that when we extend $\tg_\phi$ to a left-invariant metric on $K$, it has an ${\rm U}(1) \times \SU$ isometry group. The electromagnetic ${\rm U}(1)$ is generated by the same left-invariant vector field as before, namely $\gamma_\phi^\LL$, where the matrix $\gamma_\phi$ is given by \eqref{GammaVector} and is the unique element in the subspace $\iota(\utwo)$ of $\su$ that satisfies $[\gamma_\phi ,\, \iota(\phi)] = 0$, up to normalization. The norm of this matrix is now
\beq
\tg_\phi (\gamma_\phi, \, \gamma_\phi) \ = \ \tbeta (\gamma_\phi, \, \gamma_\phi) \ = \ (\lambda_1 + 3 \, \lambda_2) \, /\, 2 \ .
\eeq

\subsubsection*{Orthonormal basis and volume form}

Let $\{  u_0, \ldots, u_3, w_1, \ldots, w_4 \}$ be a $\tbeta$-orthonormal basis of $\su = \mathfrak{u}(1) \oplus \sutwo \oplus \CC^2$ such that the vectors $\{w_j\}$ span the subspace $\iota(\CC^2)$ of $\su$; the vectors $\{u_1, u_2, u_3\}$ span the subspace $\iota(\sutwo)$; and $u_0$ is the vector
\beq
u_0 \ = \  \frac{1}{\sqrt{6\, \lambda_1}} \,  \diag(-2i, i, i) \ = \ \frac{1}{\sqrt{6\, \lambda_1}} \,  \iota(i I_2) \ ,
\eeq
that spans $\iota(\mathfrak{u}(1))$. We want to use these vectors to define a $\tg_\phi$-orthonormal basis of $\su$. The subset $\{w_1, \ldots, w_4 \}$ automatically defines an orthonormal basis of $\iota(\CC^2)$, since $\tg_\phi$ coincides with $\tbeta$ on that subspace. The extension of identity \eqref{NiceIdentityMetric} to the new setting is
\beq \label{NiceIdentityMetric2}
\tg_\phi \big( u' + [u', \phi], \, v' + [v', \phi] \big) \ = \ \big(1\, - \, \lambda_3 \lambda_2^{-1} \, |\phi|^2 \big) \ \tbeta (u', v')    \ 
\eeq
for any vectors $u'$ and $v'$ in $\iota(\sutwo)$. It follows that the vectors
\beq \label{V2orthonormalbasis1}
v_j \ := \  \frac{1}{\sqrt{1\, - \, \lambda_3 \lambda_2^{-1} \, |\phi|^2}} \, \left( u_j + [u_j , \, \phi] \right) \qquad {\rm{for\ }}\  j = 1, 2, 3,  
\eeq
are $\tg_\phi$-orthonormal and are also orthogonal to the $w_j$. An explicit calculation then shows that the desired $\tg_\phi$-orthonormal basis of $\su$ can be completed with the vector
\bal  %\label{V2orthonormalbasis2}
v_0  =  \sqrt{ \frac{ \lambda_2 \lambda_3^{-1} - |\phi|^2}{\lambda_2 \lambda_3^{-1}  -  \big(1+  3\, \lambda_2 \lambda_1^{-1} \big) |\phi|^2}} \, u_0  +  \frac{\sqrt{3}\ \iota \big(\, 2i \, \phi \phi^\dag \,  - \, i |\phi|^2 I_2  \, +\, \lambda_2 \lambda_3^{-1} i \phi \, \big) }{\sqrt{ 2 \lambda_1 \big(\lambda_2 \lambda_3^{-1} - |\phi|^2\big) \big[\lambda_2 \lambda_3^{-1} - \big(1+ 3 \, \lambda_2 \lambda_1^{-1} \big) |\phi|^2 \big]}}  \nonumber .
\end{align}
This is the analog of formula \eqref{orthonormalbasis2} for the new metric $\tg_\phi$, instead of $g_\phi$. Using the orthonormal basis $\{  v_0, \ldots, v_3, w_1, \ldots, w_4 \}$ of $\su$ that has just been constructed, a derivation entirely similar to that of section 2 leads to the volume form
\bal \label{V2VolumeForm}
\vol_{\, \tg_\phi} \ &= \  \big(1\, - \, \lambda_3 \lambda_2^{-1} \, |\phi|^2 \big) \,\sqrt{ 1 - \lambda_3 \, \big(\lambda_2^{-1}+ 3 \, \lambda_1^{-1} \big) |\phi|^2 } \, \ \vol_{\tbeta} \linebr  
&=\ \sqrt{\lambda_1 \, \lambda_2^{\,3} \, \lambda_3^{\, 4}} \ \big(1\, - \, \lambda_3 \lambda_2^{-1} \, |\phi|^2 \big) \,\sqrt{ 1 - \lambda_3 \, \big(\lambda_2^{-1}+ 3 \, \lambda_1^{-1} \big) |\phi|^2 } \, \ \vol_{\beta_0} \ . \nonumber     % Formula corrected in arXiv-v2.
\end{align}
This expression reduces to \eqref{VolumeForm} in the special case where $\lambda_1 = \lambda_2= \lambda_3 =: \lambda$, of course.

\subsubsection*{Yang-Mills terms}

The substitution of the products $\beta$ and $g_\phi$ on $\su$ by the more general $\tbeta$ and $\tg_\phi$ demands very few changes in the derivation of the four-dimensional Yangs-Mills terms, as obtained by fibre-integration of the higher-dimensional scalar curvature. One point that does need to be adapted, however, is the calculation of the fibre-integral of products of right-invariant vector fields, as \eqref{ProductInvariantVectors3} is no longer valid. Now we will go through these calculations and, at the end, record the correspondence between the parameters $\lambda_1$,  $\lambda_2$ and $\lambda_3$ of the product $\tg_\phi$ and the gauge coupling constants of the model.

On the general grounds of \eqref{ProductInvariantFields} we know that, for any group element $h\in K$,
\beq \label{V2ProductInvariantVectors4}
\tg_\phi (u^\RR, v^\RR) \ |_{h} \ = \  \tbeta (\Ad_{h^{-1}} u, \Ad_{h^{-1}} v)  \, + \, \tbeta\big(\,[\Ad_{\theta  h^{-1}}u, \Ad_{h^{-1}} v ] , \, \phi \, \big) \ .
\eeq
This formula is not as simple as \eqref{ProductInvariantVectors4} because $\tbeta$, unlike $\beta$, is not $\Ad_{\SU}$-invariant. But $\tbeta$ is still invariant under the adjoint action of the element $\theta = \diag (1, -1, -1)$, so the calculations immediately below \eqref{ProductInvariantVectors4} carry over to show that
\beq \label{V2ProductInvariantVectors6}
\int_{h \in K} \, \tg_\phi (u^\RR, v^\RR) \ \vol_{\, \tg_\phi}  \ = \  \int_{h \in K} \, \tbeta (\Ad_{h^{-1}} u, \Ad_{h^{-1}} v)  \ \vol_{\, \tg_\phi} \ .
\eeq
Since the right-hand side of this equation integrates the $\Ad_h$-action over all $h \in K$, the resulting integral must be invariant under $\Ad_{\SU}$-transformations of the vectors $u$ and $v$. In other words, the resulting integral must be proportional to the Cartan-Killing product $\Tr ({\rm{ad}}_u \, {\rm{ad}}_v)$ on $\su$. To determine the constant of proportionality it is enough to calculate the integral in the case where $u$ and $v$ are both equal to the diagonal matrix $e_0 := \diag(-2i, i, i)$ in $\su$. For any element $h \in \SU$, a direct computation with matrix components yields
\beq
 \Ad_h e_0 \ = \ h \,e_0 \, h^\dag \ = \  -3\, i \bmatr  | h_{11}|^2 - 1/3    & \, h_{11} \hb_{21}    & \,   h_{11} \hb_{31}    \linebr
 \hb_{11} h_{21}    &  \,  | h_{21}|^2 - 1/3   &  \, h_{21}  \hb_{31}   \linebr
 \hb_{11} h_{31}    & \,  h_{31}  \hb_{21}    & \,  | h_{31}|^2 - 1/3  \ematr \ .
\eeq
The components of $ \Ad_h e_0$ defined by decomposition \eqref{Decomposition2} can be easily read from the right-hand side matrix. In terms of the usual isomorphism $\iota : \mathfrak{u}(1) \oplus \sutwo \oplus \CC^2 \to \su$ we have that
\bal
(\Ad_h e_0)_Y \ &= \ \frac{3}{2} \, \left(  h_{11}|^2 - 1/3  \right)  e_0   \linebr
(\Ad_h e_0)_W \ &= \ -3\, \iota\left( \,i\, \bmatr     | h_{21}|^2 + \big(| h_{11}|^2 -1 \big)/2       &        h_{21}  \hb_{31}     \linebr   h_{31}  \hb_{21}      &     | h_{31}|^2 + \big(| h_{11}|^2 -1 \big)/2   \ematr   \,      \right)  \nonumber \linebr
(\Ad_h e_0)'' \ &= \ -3\, \iota\left( \, i\, [\, \hb_{11} h_{21} \quad \  \hb_{11} h_{31}\, ]^T \, \right)  \ . \nonumber 
\end{align}
The definition of the inner-product $\tbeta$, as given in \eqref{DefinitionTBeta}, can then be directly applied to calculate that
\begin{multline}
\tbeta \big(\Ad_h e_0, \, \Ad_h e_0 \big) \ = \ \frac{3\, \lambda_1}{2}\, \big(\,  9 | h_{11}|^4 - 6 | h_{11}|^2  + 1 \,\big) \ + \ 18\, \lambda_3 \, \big(\, | h_{11} h_{21}|^2 + | h_{11} h_{31}|^2 \, \big) \\
+ \ 9 \,\lambda_2 \left(   | h_{21}|^4    + | h_{31}|^4 - \frac{1}{2}\, | h_{11}|^4 + 2  \,| h_{21} h_{31}|^2  + | h_{11}|^2 -  \frac{1}{2} \right) \ . \nonumber
\end{multline}
But integrals over $\SU$ of complex polynomials in the variables $h_{11}$, $h_{21}$ and $h_{31}$ are computed in appendix A.1 of \cite{Baptista}. Repeated usage of those results yields that
\bal
\int_{h \in K}\, \tbeta \big(\Ad_h e_0, \, \Ad_h e_0 \big)  \ \vol_{\, \tg_\phi}  \ &= \ \frac{6}{8}\, \big( \lambda_1  + 3 \,\lambda_2 + 4\, \lambda_3 \big) \ \Vol(K, \tg_\phi)  \nonumber \linebr
 &= \ \frac{1}{8}\, \Tr(e_0^\dag\, e_0) \,  \big( \lambda_1  + 3 \,\lambda_2 + 4\, \lambda_3 \big) \ \Vol(K, \tg_\phi)  \nonumber \ .
\end{align}
Since integrating with the variables $h$ or $h^{-1}$ is the same for a bi-invariant volume form such as $ \vol_{\, \tg_\phi} $, it follows from identity \eqref{V2ProductInvariantVectors6} and the comments thereafter that
\beq \label{V2ProductInvariantVectors5}
\int_{h \in K} \, \tg_\phi (u^\RR, v^\RR) \ \vol_{\, \tg_\phi}  \ = \  \frac{1}{8}\, \big( \lambda_1  + 3 \,\lambda_2 + 4\, \lambda_3 \big) \, \Tr(u^\dag\, v) \,  \Vol(K, \tg_\phi) \ ,
\eeq
for general matrices $u$, $v$ in $\su$. This is the analog of \eqref{ProductInvariantVectors3} for the stretched metric $ \tg_\phi$ and reduces to that formula when $\lambda_1 = \lambda_2= \lambda_3 =: \lambda$. 

Having adapted formula \eqref{ProductInvariantVectors3} to the new metric $\tg_\phi$, the rest of the derivation of the four-dimensional Yang-Mills terms induced by the higher-dimensional curvature $R_P$ is entirely analogous to the work done in section 2. The generalization of the main integral \eqref{IntegralNormF} is just
\begin{multline} \label{V2IntegralNormF}
\int_K \, |\FF|^2 \ \vol_{\, \tg_\phi} \ = \ \frac{1}{4} \, g_M^{\mu \nu} \, g_M^{\sigma \rho}  \ \Big\{ \   \tg_\phi(e_j, e_k)\,  (F^j_{A_\LLL})_{\mu \sigma} \, (F^k_{A_\LLL})_{\nu \rho}  \\ + \  \tilde{\lambda}\, \Tr(e_j^\dag \, e_k)\, (F^j_{A_\RRR})_{\mu \sigma} \, (F^k_{A_\RRR})_{\nu \rho} \  \Big\} \ \Vol (K, \tg_\phi) \ ,
\end{multline}
where we have simplified the notation by defining the positive constant
\beq \label{TLambda}
\tilde{\lambda} \ : = \ \frac{1}{8}\, \big( \lambda_1  + 3 \,\lambda_2 + 4\, \lambda_3 \big) \ .
\eeq
Just as in section 2, in the case where the one-forms $A_\LL$ have values in the electroweak subalgebra $\utwo$ of $\su$, then the coefficient $\tg_\phi(e_j, e_k)$ in front of the curvature components $F_{A_\LLL}$ are equal to $\tbeta(e_j, e_k)$, since the metric $\tg_\phi$ coincides with $\tbeta$ on that subspace. So for the restricted gauge algebra $\utwo \oplus\su$, the expression for the norm of $\FF$ is
\begin{multline} \label{V2IntegralNormF2}
\int_K \, |\FF|^2 \ \vol_{\, \tg_\phi} \ = \ \frac{1}{4} \, g_M^{\mu \nu} \, g_M^{\sigma \rho}  \ \bigg\{ \,   \sum_{j, k = 1}^4 \tbeta(e_j, e_k)\,  (F^j_{A_\LLL})_{\mu \sigma} \, (F^k_{A_\LLL})_{\nu \rho}  \\ + \ \tilde{\lambda} \, \sum_{j, k = 1}^8 \Tr(e_j^\dag \, e_k)\,  (F^j_{A_\RRR})_{\mu \sigma} \, (F^k_{A_\RRR})_{\nu \rho} \,  \bigg\} \ \Vol (K, \tg_\phi) \ .
\end{multline}
This is the analog of formula \eqref{IntegralNormF2} for the new metric $\tg_\phi$. The coefficients in front of the electroweak curvature $F_{A_\LLL}$, which has values in $\utwo$, are proportional to the stretched products $\tbeta(e_j, e_k)$. Inspecting the definition of $\tbeta$ in \eqref{DefinitionTBeta}, we recognize that the parameters $\lambda_1$ and $\lambda_2$ play the expected role in the Yang-Mills Lagrangian: they are inversely proportional to the squares of the coupling constants $g'$ and $g$ of electroweak theory \cite{Hamilton, Weinberg}. The strong coupling constant, on its turn, is related to the combination $\tilde{\lambda}$ given by \eqref{TLambda}. The precise relations between the gauge coupling constants and the parameters $\lambda_j$ are calculated in section 2 of \cite{Baptista}. The result is
\bal
\frac{g'}{2} \ &= \ \sqrt{\frac{ 3}{ \, \lambda_1}}   &   e\  &= \  \frac{2\, \sqrt{3}}{\sqrt{ \lambda_1 +  3\, \lambda_2}}    \linebr   
\frac{g}{2} \ &= \    \frac{1}{\sqrt{\lambda_2}}      &  \frac{g_s}{2} \ &= \ \frac{2\,\sqrt{2}}{\sqrt{ \lambda_1 +  3\, \lambda_2 + 4\, \lambda_3 }} \ . \nonumber
\end{align}

\subsubsection*{Scalar curvature of $\tg_\phi$}

Section 2.6 was dedicated to the calculation of the scalar curvature of the left-invariant metric $g_\phi$. It used the general formula \eqref{GeneralScalarCurvature} applied to the $g_\phi$-orthonormal basis constructed in section 2.5. Since the calculation is long, most of the explicit work was omitted in that section and only the main results were recorded.

The scalar curvature of the new metric $\tg_\phi$ can be calculated in an entirely similar fashion, using \eqref{GeneralScalarCurvature} and the $\tg_\phi$-orthonormal basis of $\su$ constructed before \eqref{V2VolumeForm}. The explicit calculation, however, is even longer than that of section 2.6, so it will not be carried out here. The final formula for $R_{\tg_\phi}$ must generalize \eqref{ScalarCurvature} and, at the same time, reduce to the scalar curvature of $\tbeta$ in the case of vanishing $\phi$. The latter scalar curvature is much quicker to compute, because the usual $\tbeta$-orthonormal basis of $\su$ is simpler to manipulate when applied to the general formula \eqref{GeneralScalarCurvature}. Using such a basis, we get that
\bal
R_{\tbeta} \ &= \ 3 \left( \frac{1}{\lambda_2} +   \frac{4}{\lambda_3} -  \frac{\lambda_1 + \lambda_2}{2\, \lambda_3^2}  \right) \ .         % Formula corrected in arXiv-v2.
\end{align} 
Not having a simple and explicit formula for the scalar curvature of $\tg_\phi$ is particularly unfortunate in light of the discussion of section 5.1. Such a formula could be plugged into the normalized functional \eqref{Functional1}, together with the volume \eqref{V2VolumeForm}, and be used to test whether the parameters $|\phi|^2$ and $\lambda_1$, $\lambda_2$, $\lambda_3$ can be chosen to define a critical point of that functional, as this would correspond to a metric in the family $\tg_\phi$ with a chance of satisfying the Einstein condition. In fact, finding a stable Einstein metric on $K$ with isometry group ${\rm U}(1) \times \SU$ would probably be the most desirable development among all the additional investigations suggested here. It is known that the bi-invariant metric on $\SU$ is only a saddle point of the normalized Einstein-Hilbert functional, not a maximum \cite{Jensen}. We also know from \eqref{ScalarCurvature} that the scalar curvature explodes to minus infinity near the boundary of parameter space defined by a finite value of $|\phi|$. So the existence of a genuine maximum of the normalized Einstein-Hilbert functional at a left-invariant metric with small $|\phi|$ does not sound entirely impossible.

\subsubsection*{The fibre's second fundamental form}

In the discussion of section 3, the covariant derivative $\dd^A \phi$ of the Higgs-like parameter appeared in the calculation of the second fundamental form of the fibres, denoted there as $S$. It was the norm $|S|^2$ that gave rise to the term $|\dd^A \phi|^2$ in the four-dimensional Lagrangian density. Subsequently, in section 4, the classical masses of the Higgs-like and gauge bosons were calculated from the equations of motion determined by that same Lagrangian. 

Thus, at this point the natural task is to replicate the calculations of $|S|^2$ and the simpler $|N|^2$ using the new fibre metric $\tg_\phi$, instead of the old $g_\phi$. Unfortunately, once again the explicit calculation of $|S|^2$ is straightforward but lengthy, more so now than in section 3, and hence will not be carried out here. Once performed, these calculations will yield a Lagrangian density analogous \eqref{DensityM} with explicit expressions for coefficient functions $\tilde{C}_\phi$ and $\tilde{D}_\phi$ in terms of the parameters of the metric $\tg_\phi$, i.e. in terms of $|\phi|^2$ and the positive constants $\lambda_1$, $\lambda_2$ and $\lambda_3$. With these expressions at hand, the customary arguments described in sections 4.1 and 4.2 can be employed to calculate the classical masses of the Higgs-like and the $Z$ and $W$ bosons, as determined by the new Lagrangian density. The mass calculation also requires the explicit coefficients of the Yang-Mills terms associated to $\tg_\phi$, but these have already been computed in \eqref{V2IntegralNormF2}.

Although we do not offer here the generalized expressions for the bosons masses, there is one instance where the calculations are shorter and can be readily performed. This is the calculation of the mass ratio of the $Z$ and $W$ bosons. In fact, improving the classical ratio $M_Z = 2M_W$ obtained for the fibre metric $g_\phi$ was one of the motivations to introduce and study the new metrics $\tg_\phi$. Going through the calculations done back in section 4.2, we recognize that the linearized equations of motion for the components of the one-form $A_\LL$ is generalized from \eqref{EqMotionGaugeBosons} to the new expression
\beq
 \tbeta(e_k, e_j) \ g_M^{\mu \nu} \, g_M^{\sigma \rho} \ \nabla_\nu (F_{A_\LLL}^k)_{\mu \sigma} \ - \  \tilde{C}_{\phi_0} \ g_M^{\mu \rho} \, (A_\LL^k)_\mu \,    \Tr \big(  [e_j, \,\phi_0]^\dag \, [e_k,\, \phi_0]  \big)  \ = \ 0 \ ,
\eeq
where $ \tilde{C}_{\phi_0}$ is a function of $|\phi_0|^2$ and the constants $\lambda_j$ that we do not calculate here, as explained.  We have also used identity \eqref{IdentityQuadraticForm} to write the quadratic form in a more $\su$-like appearance. Therefore, picking a basis $\{ e_k \}$ of the subspace $\utwo \subset \su$ that simultaneously diagonalizes the product $\tbeta$ and the quadratic form $\Tr \big(  [e_j, \,\phi_0]^\dag \, [e_k,\, \phi_0]  \big)$, the equations of motion imply that the mass of the gauge bosons is given by
\beq \label{MassGaugeFields2}
 \left[{\rm{Mass}}\  (A_\LL^k)_\mu \right]^2 \ = \  \frac{  \tilde{C}_{\phi_0} \,  \Tr \big(  [e_k, \,\phi_0]^\dag \, [e_k,\, \phi_0]  \big) }{\tbeta\big( e_k, \, e_k\big) }   \ ,
\eeq
where no sum over the index $k$ is intended. One such basis is explicitly constructed in appendix A.1. It comprises the four $\tbeta$-orthogonal vectors $\{\gamma_\phi, \,\tilde{z}_\phi,\, w_\phi^1,\, w_\phi^2 \}$. If the components of the one-form $A_\LL$ on that basis are denoted by
\[
A_\LL \ = A^\gamma \, \gamma_\phi \ + \  Z \, \tilde{z}_\phi \ + \ W^1 \, w^1_\phi \ + \ W^2 \, w^2_\phi \ ,
\]
then the classical mass associated to each component field follows directly from \eqref{MassGaugeFields2}. Although we do not have an explicit expression for $\tilde{C}_{\phi_0}$, this factor cancels out in the ratio $M_Z / M_W$. Thus, using algebraic identities \eqref{A.5} and \eqref{A.8} to calculate the remaining factors of \eqref{MassGaugeFields2} in the case of the $w_\phi^a$-components, for $a=1,2$, and using identities \eqref{A.12} and \eqref{A.12} to calculate the same factors in the case of the $\tilde{z}_\phi$-components, we finally obtain that mass ratio of the $Z$ and $W$ bosons is simply
\bal
\frac{M_Z}{M_W} \ = \ \sqrt{1 \, +  \, 3\, \lambda_1^{-1} \, \lambda_2} \ . 
\end{align}
So the introduction of the positive parameters $\lambda_j$ in the definition of $\tbeta$ and $\tg_\phi$ allows for adjusting the mass ratio, as happens in the Standard Model. The parameters $\lambda_1$ and $\lambda_2$ of $\tg_\phi$ are of course essentially equivalent to the usual electroweak gauge coupling constants.

\subsection*{Full $\SU \times \SU$ gauge fields}  % Title changed in arXiv-v2.
\addcontentsline{toc}{subsection}{Full $\SU \times \SU$ gauge fields}

\subsubsection*{Additional bosons and their masses}
%\addcontentsline{toc}{subsubsection}{Additional bosons and their masses}

One point where the calculations in this study have not gone far enough is in investigating the consequences of having gauge fields $A_\LL$ and $A_\RR$ with values in the natural Lie algebra $\su \oplus \su$, instead of the Standard Model algebra $\utwo \oplus \su$. Recall that the higher-dimensional metric $g_P$ was defined in \eqref{DefinitionBundleMetric} using an horizontal distribution $\HH$. This distribution was made more explicit in formula \eqref{DefinitionHorizontalDistribution}, which defines the basic horizontal vector fields $X^\HH$ on $P$ in terms of one-forms $A_\LL$ and $A_\RR$ on the four-dimensional $M_4$. In principle, those one-forms can have values in the full space of left or right-invariant vector fields on $K$, each identifiable with the algebra $\su$. However, in order to reproduce the usual features of the Standard Model, in many of the calculations  we considered the special case where $A_\RR$ has values in $\su$ but $A_\LL$ has values in the subspace $\utwo$ of $\su$. This was done, for example, when calculating the expression for the fibres' second fundamental form, whose norm $|S|^2$ produced a term $| \dd^{A_\LL} \phi|^2$ similar to the norm of the covariant derivative of the traditional Standard Model's Higgs field.

The main step that used the restriction to $\utwo$ was taken after \eqref{Aux3.1}. Had we kept one-forms $A_\LL$ with values in the full $\su$, then formula \eqref{ResultT} would be substituted by the slightly more involved expression
\begin{align}
 \label{ExplicitTensorT3}
2\, g_P ( \, S_{u^\LLL} v^\LL, \,  X) \ &= \  - \, (\Lie_X \,  g_\phi)(u, v)  \ - \  A_\LL^k (X) \  (\Lie_{e_k^\LLL}\, g_\phi)(u.v)       \\[.5em]
  &= \ - \, \beta \big(\, [u', v''] + [v', u''], \, \dd \phi (X) \,\big) \ -\  g_\phi \big( \, v,  [\, A_\LL (X),  \, u \, ] \, \big)   \nonumber \linebr
   &\ \qquad \qquad \qquad \qquad \qquad \quad - \ g_\phi \big( \, u,  [\, A_\LL (X),  \, v \, ] \, \big)  \ - \ ( \Lie_X \log \lambda) \  g_\phi ( u,  v) \, ,    \nonumber
\end{align}
valid for any $u, v$ in $\su$ and any tangent vector $X$ in $TM \subset TP$. This formula does not display the covariant derivative of the traditional Higgs field $\phi$, as happens with \eqref{ResultT} combined with \eqref{CovariantDerivativePhi3}, but it still determines the tensor $S$. Using the definition of the product $g_\phi$ and the orthonormal basis of section 2, one can use the formula above to calculate the norm $|S|^2$ by methods similar to those employed in section 3. The calculation seems to be straightforward but considerably longer than that of section 3, now that $A_\LL$ has values in $\su$ rather than $\utwo$. In particular, we will not be able to offer here a formula for $|S|^2$ as explicit as \eqref{NormT}. This is unfortunate, because it prevents the direct calculation of the masses of all the gauge bosons associated to an $\su$-valued one-form $A_\LL$.

For now, we register a geometrically natural, though hardly explicit, formula for the norm of the fibres' second fundamental form. Denote by $\left\langle \cdot \, , \, \cdot \right\rangle$ the inner-product on the space of symmetric 2-tensors $\mathrm{Sym}^2 [\su^\ast]$ induced by the product $g_\phi$ on $\su$. It can be defined explicitly as
\[
\left\langle h_1, \, h_2 \right\rangle_{g_\phi} \ := \ \sum_{j,\, k} \, h_1(e_j, e_k) \ h_2(e_j, e_k) \ , 
\]
where $\{ e_k \}$ is any $g_\phi$-orthonormal basis of $\su$. Then formula \eqref{ExplicitTensorT3} implies the general decomposition
\begin{multline}
\label{NormT2}
|S|^2 \ = \   \frac{1}{4} \ g_M^{\mu \nu}  \left\langle \Lie_{X_\mu}\, g_\phi,  \ \Lie_{X_\nu}\, g_\phi \right\rangle \ + \  \frac{1}{2} \ g_M^{\mu \nu} \ A_\LL^k (X_\mu)  \left\langle \Lie_{e_k^\LLL}\, g_\phi,  \ \Lie_{X_\nu}\, g_\phi \right\rangle \linebr + \   \frac{1}{4} \ g_M^{\mu \nu} \ A_\LL^k (X_\mu)\  A_\LL^j (X_\nu)   \left\langle \Lie_{e_k^\LLL}\, g_\phi,  \ \Lie_{e_j^\LLL}\, g_\phi \right\rangle \ .
\end{multline}
This expression shows how the fibres' second fundamental form, after fibre-integration, gives rise to the quadratic terms in the gauge fields $A_\LL^k$ that are essential to mass generation, through spontaneous symmetry breaking, in the four-dimensional Lagrangian. Quite naturally, the coefficients of these terms are determined by the Lie derivatives of the fibres' metric along different directions. So the components of $A_\LL$ along Killing vector fields satisfying $\Lie_{v^\LLL} g_\phi = 0$ disappear entirely from $|S|^2$ and correspond to massless bosons. The classical mass of a gauge boson is a measure of how much the internal metric changes along the flow generated by the corresponding invariant vector field. Formula \eqref{NormT2} remains valid when the fibres of the higher-dimensional spacetime $P$ are equipped with arbitrary left-invariant metrics $g_K$, not necessarily in the family $g_\phi$. 

Notice from \eqref{ExplicitTensorT3} how the natural objects
\beq \label{CovariantDerivativeFibreMetric}
(\dd^A g_K) (X) \ := \  \Lie_X \,  g_K  \ + \  A_\LL^k (X) \  (\Lie_{e_k^\LLL}\, g_K)
\eeq
are essentially equivalent to the second fundamental form of the fibres. They can be regarded as the ``covariant derivative'' of the left-invariant fibre metric $g_K$ along a vector field $X$ in $M_4$. The fibres of $P$ are totally geodesic if and only if their metrics $g_K$ are ``covariantly constant'' along $M_4$, in the sense that \eqref{CovariantDerivativeFibreMetric} vanishes for all vectors $X$. The gauge fields $A_\RR$ do not appear in \eqref{CovariantDerivativeFibreMetric} because the Lie derivatives $\Lie_{v^\RRR}\, g_K$ are identically zero for left-invariant fibre metrics.

Observe also that, for arbitrary left-invariant metrics on $K$, the fibres' mean curvature vector $N$ continues to be independent from the one-forms $A_\LL$, even for gauge fields with values in the larger algebra $\su$. This is manifest in formula \eqref{ExplicitVectorN1}, for instance, which was deduced using the unimodularity of $K$. Thus, the terms in the Lagrangian proportional to $|N|^2$ and $\check{\delta} N$ still do not involve gauge fields.

Let us now come back to the discussion of the full $\su$-gauge bosons. In section 4.2 we calculated the masses of the components of $A_\LL$ with values in the subspace $\utwo$ of $\su$. These components correspond to the four electroweak gauge bosons. A one-form $A_\LL$ with values in the full $\su$ would imply the existence of four additional bosons. All of these would be massive in the present model, since $\gamma_\phi \in \utwo$ generates the only left-invariant Killing field of $g_\phi$, up to normalization. The classical mass of the additional bosons should be computable using an orthonormal basis applied to \eqref{ExplicitTensorT3} and \eqref{NormT2}, as was done in section 3 for the $Z$ and $W$ bosons, although the calculation will be longer in this case. It would be very interesting to carry it out explicitly; check how the usual arguments about the unitary gauge can fit in; and investigate the conditions necessary for the additional four bosons to be significantly heavier than their electroweak counterparts. 

If no significant obstacles are found in the calculation of the masses of the four additional bosons but, at the end, they turn out not to be heavier than the $Z$ and $W$ bosons, this would of course be bad news for the present model, as no additional gauge bosons have been experimentally observed at low energies. One way out would be the usual route of adjusting the model by introducing a mechanism to spontaneously break the left $\SU$ down to $\Utwo$, and therefore make the new bosons heavier. This could be achieved using a Higgs-like field $\Phi: M_4 \to \su$ in the adjoint representation, which can also be regarded as a simple left-invariant vector field on $P$, and adding the norm of its covariant derivative and a new potential $U(\Phi)$ to the higher-dimensional Lagrangian density. For example, take the $\Ad_{\SU}$-invariant potential
\beq
\label{GlobalPotential}
U(\Phi) \ := \ \frac{\alpha}{4}  \left[ \ \Tr (\Phi^\dag \Phi )\ - \ 6\, \tau \ \right]^2 \ 
\eeq
with positive constants $\alpha$ and $\tau$. It is clear that this potential has absolute minima when the matrix $\Phi$ is in the conjugation class of $\Phi_0 = \sqrt{\tau} \, \diag(-2i, i, i )$ inside $\su$. The ``vacuum vector'' $\Phi_0$ is preserved by the usual subgroup $\Utwo$ of $\SU$, so the potential $U(\Phi)$ would provide the necessary mechanism to make the additional four bosons heavier without affecting the masses of the $Z$ and $W$ bosons calculated before. However, after spending considerable effort trying to obtain the all the bosonic components of the Standard Model Lagrangian from natural objects such as the higher-dimensional scalar curvature, and therefore suggesting a more geometrical origin for the usual Higgs covariant derivative and potential, the introduction in the model of new ad hoc fields and potentials, such as those in \eqref{GlobalPotential}, would not be the most favoured option.

\subsubsection*{Additional fermionic interactions}
%\addcontentsline{toc}{subsubsection}{Additional fermionic interactions}

The model for fermions described in \cite{Baptista} associates them to spinorial functions on the spacetime $P$ having a prescribed behaviour along the internal space $K$. This behaviour determines the Lie derivatives of the functions along vertical vector fields, which in turn determine the fermionic gauge representations obtained in the four-dimensional Lagrangian, after integration of the Dirac kinetic terms along the fibres. Using the explicit vertical behaviour suggested in sections 2.2 and 2.3 of \cite{Baptista}, it is possible to calculate how the four-dimensional fermions would couple to gauge fields $A_\LL$ with values in the full algebra $\su$. In fact, the necessary work is already done in the aforementioned section 2.3. It can be summarized by the formulae
\beq \label{CovariantDerivative}
\nabla^A \,\Psi \ := \ \dd \Psi \ + \  \sum_{j} \, A_\LL^j \ \big[\, \rho_{e_j}^\LL (\psi_+)   \ \ \    \rho_{\conj{e_j}}^\LL (\psi_-) \, \big]   \ + \  A_\RR^j \ \big[\, \rho_{e_j}^\RR (\psi_+)   \ \ \    \rho_{\conj{e_j}}^\RR (\psi_-) \, \big]      \ ,
\eeq 
with the coupling to the $A_\LL$ gauge fields determined by
\bal \label{RhoLAction}
\rho^\LL_v (\psi_\pm) \ = \  \rho^\LL_v  \bmatr  a & c^T \linebr b & D  \ematr \ = \ \bmatr  0 & -2\, v_{11} \, c^T \linebr \big(  2\, v_{11}\, I_3 \, + \, v \big)\, b & v\, D  \ematr  
\end{align}
for all matrices $v$ in $\su$. Here $a$ is a single Weyl spinor; $b$ and $c$ are 3-vectors of Weyl spinors; $D$ is a $3\times3$ matrix of Weyl spinors. They can be identified with the first generation of fermions according to the rule
\beq \label{FermionicIdentification1}
\bmatr  a & c^T \linebr b & D  \ematr \ = \ 
\bmatr \nu_\RR  & \, u_\RR^r & \, u_\RR^g & \, u_\RR^b   \\[0.4em]
 e_\RR^{-} & \, d_\RR^r & \, d_\RR^g & \, d_\RR^b  \\[0.4em]
 \nu_\LL  & \, u_\LL^r & \, u_\LL^g & \, u_\LL^b   \\[0.4em]
 e_\LL^{-} & \, d_\LL^r & \, d_\LL^g & \, d_\LL^b  \quad 
\ematr \ .
\eeq
For matrices $v$ with values in the subalgebra $\iota(\utwo)$ of $\su$, the transformation \eqref{RhoLAction} gives the usual fermionic couplings to the electroweak gauge group, with the correct hypercharges and weak isospin. If $A_\LL$ is taken to have values in the full $\su$, the same formula \eqref{RhoLAction} suggests what the additional fermionic couplings should be like. The components of $A_\LL$ with values in the subspace $\iota(\CC^2)$ of $\su$ are associated to matrices $v$ of the form
\beq \label{NoLabel}
v \ = \ \bmatr  & - y^\dag  \\  y &  \ematr \quad \in \ \su \, ,
\eeq
with $y \in \CC^2$. Since the entry $v_{11}$ is zero, the new components of $A_\LL$ do not couple to the vector $c$ of \eqref{RhoLAction}, that is, to the right-handed up quark. However, the matrix $v$ does act on the vector $b$ and on the columns of the matrix $D$ by mixing their top entries with the middle and bottom ones. In other words, the new components of $A_\LL$ would mix the right-handed electron with the left-handed electron and neutrino. They would also mix the right-handed down quark with the left-handed up and down quarks. The mixing would be analogous for anti-particles. Thus, the higher-dimensional model described in \cite{Baptista} suggests that the interactions generated by the additional components of $A_\LL$ would conserve the baryon number but not parity.

It is also appropriate to recall from \cite{Baptista} that once the gauge algebra is extended from the Standard Model's $\utwo \oplus \su$ to the larger $\su \oplus \su$, the action on spinors $\rho^\LL + \rho^\RR$, described in section 2.3 of that study, no longer defines a Lie algebra homomorphism from the gauge algebra to the algebra $\mathfrak{su} (\Delta_{12} )$ of transformations in spinor space. It would be interesting to better understand the implications of this fact.

\newpage

\vspace{.3cm}

\appendix

\section{Appendices}

\subsection{A $\phi$-rotated basis of $\su$ }
%\addcontentsline{toc}{subsection}{A $\phi$-rotated basis of $\su$}

Given a non-zero vector $\phi = [\phi_1 \  \phi_2]^T$ in $\CC^2$, consider the orthogonal vector defined by $\tphi := \big[\, \bar{\phi}_2 \  -\bar{\phi}_1  \big]^T$. It satisfies
\bal  \label{A.1}
\phi^\dag \tphi \ = \ \tphi^\dag \phi \ &= \ 0  \linebr
\phi \phi^\dag \ + \ \tphi \tphi^\dag \ &= \ \mph^2 \, I_2   \ , \nonumber
\end{align}
where $\dag$ denotes the hermitian conjugate and $I_2$ is the identity matrix. Define the $\phi$-oriented Pauli matrices by
\bal  \label{A.2}
\sigma^1_\phi \ &:= \ \mph^{-2} \, (\phi \tphi^\dag \, + \, \tphi \phi^\dag)    \linebr
\sigma^2_\phi \ &:= \ i \mph^{-2} \, (\phi \tphi^\dag \, - \, \tphi \phi^\dag)    \nonumber \linebr
\sigma^3_\phi \ &:= \ \mph^{-2} \, (\tphi \tphi^\dag \, - \, \phi \phi^\dag)     \ . \nonumber 
\end{align}
These are traceless hermitian matrices that satisfy the usual algebraic relations
\beq  \label{A.3}
\sigma^a_\phi \, \sigma^b_\phi \ = \ \delta^{ab} \, I_2 \ + \ i \varepsilon^{abc}\, \sigma^c_\phi
\eeq
and coincide with the Pauli matrices when $\phi = [0 \ \, 1]^T$. They can be regarded as a rotated version of the latter matrices.

The matrices $\sigma^j_\phi$ can be used to write down a basis of $\su$ that simultaneously diagonalizes the Ad-invariant inner-product and the quadratic form 
\beq \label{QuadraticForm}
(u, v) \ \mapsto \ \Tr \left(  \big[u,\, \iota(\phi) \big]^\dag \, \big[v, \, \iota(\phi)\big]   \right) 
\eeq
on the Lie algebra. Such a basis is useful in the calculation of the mass of the Z and W gauge bosons worked out in section 3.
Fix the vector $\phi \in \CC^2$ and recall the usual vector space isomorphism $\iota: \utwo \oplus \CC^2 \to \su$. We can define four different $3\times3$-matrices in the subalgebra $\iota(\utwo)$ of $\su$ through the formulae
\bal  \label{A.4}
w^1_\phi \ &:= \ \iota(\, i \sigma^1_\phi \,)  \ \   & z_\phi   \ &:= \  \frac{1}{2}\, \iota(\, i I_2 \, -  \, i \sigma^3_\phi \,)  \linebr
w^2_\phi \ &:= \ \iota(\, i \sigma^2_\phi \,)             \ \    & \gamma_\phi   \ &:= \  \frac{1}{2}\, \iota\Big(\, \frac{i}{\sqrt{3}} \, I_2\, +\, i \sqrt{3} \, \sigma^3_\phi \, \Big)         \nonumber \ .
\end{align}
One can readily check that these matrices are orthogonal to each other with respect to the Ad-invariant inner-product on $\su$, so they span the subspace $\iota(\utwo)$. Their norm in $\su$ is simply
\beq \label{A.5}
\Tr(\gamma_\phi^\dag \, \gamma_\phi) \ = \ \Tr(z_\phi^\dag \, z_\phi) \ = \ \Tr \! \big( \, ( w^a_\phi)^\dag \, w^a_\phi \,  \big) \ = \ 2 \ .
\eeq
The commutators in $\su$ of these four matrices are
\bal  \label{A.6}
[z_\phi , \, \gamma_\phi ]\ &= \ 0  \ \    \linebr
\big[w^1_\phi, \, w^2_\phi \big] \ &= \   z_\phi  - \sqrt{3} \, \gamma_\phi         \ \        \nonumber  \linebr     
\big[w^1_\phi, \, \gamma_\phi \big] \ &= \   \sqrt{3} \, w^2_\phi \ = \   \sqrt{3} \, \big[z_\phi, \, w^1_\phi \big]        \ \    \nonumber    \linebr   
\big[w^2_\phi, \, \gamma_\phi \big] \ &= \  - \sqrt{3} \, w^1_\phi \ = \   \sqrt{3} \, \big[z_\phi, \, w^2_\phi \big]        \ .         \nonumber 
\end{align}
The commutators of these matrices with the element $\iota(\phi)$ in $\su$ can be checked to be
\bal  \label{A.7}
\big[\gamma_\phi , \, \iota(\phi) \big] \ &= \ 0  \ \   & \big[z_\phi , \, \iota(\phi) \big]  \ &= \  2\, \iota(\,i \phi \,)   \linebr
\big[w^1_\phi , \, \iota(\phi) \big] \ &= \ \iota(\, i \tphi \,)             \ \    & \big[w^2_\phi , \, \iota(\phi) \big]   \ &= \ \iota(\, \tphi \,)            \nonumber \ .
\end{align} 
The latter commutators are vectors in the subspace $\iota(\CC^2)$ of $\su$. They are orthogonal to each other and to $\iota(\phi)$ with respect to the Ad-invariant inner-product on $\su$. So the three non-zero commutators together with $\iota(\phi)$ span the whole subspace $\iota(\CC^2)$. The norm in $\su$ of $\big[\gamma_\phi , \, \iota(\phi) \big]$ is zero, whereas the other commutators have norm
\bal  \label{A.8}
\Tr \Big( \big[z_\phi , \, \iota(\phi) \big]^\dag \, \big[z_\phi , \, \iota(\phi) \big]      \Big) \ &= \ 8 \, \mph^2   \linebr
\Tr \Big( \big[w^a_\phi , \, \iota(\phi) \big]^\dag \, \big[w^a_\phi , \, \iota(\phi) \big]      \Big) \ &= \ 2 \, \mph^2 \nonumber \ .
\end{align}  
It is clear from \eqref{A.7} that $\{\gamma_\phi, \,z_\phi,\, w_\phi^1,\, w_\phi^2 \}$ is a basis of $\iota(\utwo)$ that diagonalizes the quadratic form \eqref{QuadraticForm} on that subspace of $\su$. Moreover, we have the relations
\bal
\big[\,\iota(\phi),\, \iota(\phi) \, \big] \ &= \ 0     &  \big[\,\iota(i\phi),\, \iota(\phi) \, \big] \ &= \ \,|\phi|^2\, \iota( i\sigma_\phi^3) \, - \, \iota(i I_2)   \nonumber \linebr
\big[\,\iota(\tphi),\, \iota(\phi) \, \big] \ &= \ -i \,|\phi|^2\, \sigma_\phi^2    &  \big[\,\iota(i\tphi),\, \iota(\phi) \, \big] \ &= \ -i \,|\phi|^2\, \sigma_\phi^1  \ .
\end{align}
All these commutators are orthogonal to each other with respect to the Ad-invariant product on $\su$. So we recognize that the vectors $\iota(\phi)$, $\iota(i\phi)$, $\iota(\tphi)$ and $\iota(i\tphi)$ form a basis of $\iota(\CC^2)$ that simultaneously diagonalizes the Ad--invariant product and the quadratic form \eqref{QuadraticForm} on that subspace of $\su$. Since the subspaces $\iota(\utwo)$ and $\iota(\CC^2)$ are orthogonal to each other with respect both to the Ad-invariant product and the quadratic form, we conclude that $\{\gamma_\phi, \,z_\phi,\, w_\phi^1,\, w_\phi^2,\, \iota(\phi), \iota(i\phi), \, \iota(\tphi),\, \iota(i\tphi) \}$ is a basis of $\su$ with the desired properties.

Section 5.2 describes an inner-product $\tbeta$ on $\su$ that is not fully $\Ad_{\SU}$-invariant, only $\Ad_{\Utwo}$-invariant. We will now describe a basis of $\su$ that simultaneously diagonalizes the product $\tbeta$ and the quadratic form \eqref{QuadraticForm}. In fact, this basis can be taken to coincide with the preceding one, except that the vector $z_\phi$ should now be substituted by the deformation
\beq
\tilde{z}_\phi   \ := \  \frac{1}{2}\, \iota \big(\, i I_2 \, -  \, \lambda_1 \lambda_2^{-1}\, i \sigma^3_\phi \, \big) \ , 
\eeq
where the positive constants $\lambda_1$ and $\lambda_2$ are those in the definition \eqref{DefinitionTBeta} of the product $\tbeta$. The new vector $\tilde{z}_\phi$ is orthogonal to $\gamma_\phi$, $w_\phi^1$ and $w_\phi^2$ with respect to the product $\tbeta$. Its norm is
\beq \label{A.12}
\tbeta(\tilde{z}_\phi, \, \tilde{z}_\phi) \ = \ \frac{\lambda_1}{ 2}\, \left(3 + \frac{\lambda_1}{\lambda_2} \right) \ .
\eeq
The commutators in \eqref{A.6} involving $z_\phi$ are now changed to
\bal
[\tilde{z}_\phi , \, \gamma_\phi ]\ &= \ 0    &   [\tilde{z}_\phi , \, w_\phi^1 ]\ &= \ \lambda_1 \, \lambda_2^{-1} \, w_\phi^2   \ \    \linebr
[\tilde{z}_\phi , \, w_\phi^2 ]\ &= \ - \lambda_1 \, \lambda_2^{-1} \, w_\phi^1   &  [w_\phi^1 , \, w_\phi^2 ]\ &= \ \frac{4}{3 + \lambda_1 \, \lambda_2^{-1}} \, \left(\tilde{z}_\phi  - \sqrt{3} \, \gamma_\phi \right) \ . \nonumber
\end{align}
Since the commutator with $\iota(\phi)$ is now
\[
[\, \tilde{z}_\phi , \, \iota(\phi) \,] \ = \ \frac{1}{2} \,  \left(3 + \lambda_1 \, \lambda_2^{-1} \right) \, \iota( i \phi) \ , 
\]
the quadratic form \eqref{QuadraticForm} applied to $\tilde{z}_\phi$ has the new value
\beq \label{A.14}
\Tr \Big( \big[ \tilde{z}_\phi , \, \iota(\phi) \big]^\dag \, \big[ \tilde{z}_\phi , \, \iota(\phi) \big]      \Big) \ = \  \frac{1}{2} \,  \left(3 + \lambda_1 \, \lambda_2^{-1} \right)^2  \, \mph^2   \ .
\eeq
The new formulae involving $\tilde{z}_\phi$ reduce to the old ones when $\lambda_1 = \lambda_2$, of course.

\subsection{A proof about the Killing fields of $g_\phi$}
%\addcontentsline{toc}{subsection}{A proof about the Killing fields of $g_\phi$}

In the context of section 2.4, the aim of this discussion is to show that if a left-invariant vector field $u^\LL$ is Killing for the metric $g_\phi$ on $\SU$, then the vector $u$ is necessarily in the subalgebra $\utwo$ of $\su$. By formula \eqref{LieDerivativeMetric2}, the condition that $u^\LL$ is Killing for $g_\phi$ is equivalent to the condition that
\beq \label{C.1}
 \beta \big( \,[v', v''],\,  [u', \phi]\, \big) \ + \ \beta \big( \,[[v', u''], v'] \, + \, [[u'', v''], v''],\,  \phi \, \big) \  = \ 0  \nonumber
\eeq
for all vectors $v$ in $\su$. In particular, choosing $v$ in the subspace $\iota(\CC^2)$ of $\su$, which implies $v' = 0$, we must have that
\beq \label{C.2}
 \beta \big( \, [[u'', v], v],\,  \phi \, \big) \  = \  \beta \big( \, u'',\, [v, [ v, \phi]] \, \big) \  = \ 0  
\eeq
for all vectors $v$ in $\iota(\CC^2)$. From now on, we shall think of $u''$, $v$ and $\phi$ as complex vectors in $\CC^2$ and explicitly write the map $\iota: \CC^2 \to \su$ whenever it is necessary to regard them as matrices in $\su$. Then a short calculation using matrices of the form \eqref{phi} leads to the general identity in $\iota(\CC^2)$
\bal \label{C.3}
 \Big[\, \iota(v), \, \big[ \iota(v) ,\, \iota(\phi)  \big] \,  \Big] \ &= \ \iota\Big[\,  \big(  2\, \phi^\dag v - v^\dag \phi  \big) \, v \, - \, |v |^2 \, \phi  \,\Big]     \linebr
&= \ \iota\Big[\,    \langle v, \phi  \rangle  \, v  \, + \, 3\,  \langle v, i \phi  \rangle  \,i v  \, - \,    \langle v, v \rangle \, \phi   \,\Big]  \nonumber \ ,
\end{align}
where the brackets on the left-hand side are commutators of matrices in $\su$ and $\langle \cdot , \cdot \rangle$ is the canonical real product on $\CC^2$. The Ad-invariant product $\beta$ is proportional to the product $\beta_0$ written down in \eqref{DecompositionInnerProduct}. Thus, following that formula, we recognize that $\beta$ can be identified with $\langle \cdot , \cdot \rangle$, up to normalization, when restricted to vectors in the subspace $\iota(\CC^2)$. This means that condition \eqref{C.2} applied to the vector coming from \eqref{C.3} is equivalent to the equation
\beq \label{C.4}
\langle v, \phi  \rangle  \,  \langle u'',  v \rangle  \, + \, 3\,  \langle v, i \phi  \rangle  \,   \langle u'', i v\rangle  \, - \,    \langle v, v \rangle \, \langle u'', \phi\rangle   \ = \ 0 \ 
\eeq
for all vectors $v \in \CC^2$. Choosing a non-zero $v$ orthogonal both to $\phi$ and $i \phi$ in $\CC^2$, the first two terms of the equation vanish and the condition reduces to $\langle u'', \phi\rangle = 0$. Thus, any Killing field $u^\LL$ must have $u''$ orthogonal to $\phi$ in $\CC^2$. Assume that this is true and now choose $v = \alpha_1 \phi + \alpha_2 i \phi$ with real constants $\alpha_a$. Substituting this vector $v$ into equation \eqref{C.4} yields the condition
\beq \label{C.5}
4\,\alpha_1\, \alpha_2\,  \langle \phi, \phi \rangle \, \langle u'', i \phi \rangle   \ = \ 0 \ .
\eeq
This is satisfied for all scalars $\alpha_a$ only if $u''$ is orthogonal also to $i \phi$, besides being orthogonal to $\phi$. Assume that this is true and choose $v = \alpha_1 \phi + \alpha_2  \tphi$, where the new vector $\tphi \in \CC^2$, defined before \eqref{A.1}, is orthogonal both to $\phi$ and $i\phi$. Substituting this vector $v$ into equation \eqref{C.4} yields the condition
\beq \label{C.6}
\alpha_1\, \alpha_2\,  \langle \phi, \phi \rangle \, \langle u'', \tphi \rangle   \ = \ 0 \ 
\eeq
for all $\alpha_a$. So a Killing field $u^\LL$ will have $u''$ orthogonal to the span of $\{  \phi, i\phi, \tphi  \}$. Finally, assuming that $u''$ satisfies this, choose $v = \alpha_1 \phi + \alpha_2 i \tphi$ and substitute it into equation \eqref{C.4}. Since $i \tphi$ is also orthogonal to $\phi$ and $i\phi$, this yields the last condition
\beq \label{C.7}
\alpha_1\, \alpha_2\,  \langle \phi, \phi \rangle \, \langle u'', i \tphi \rangle   \ = \ 0 \ ,
\eeq
which shows that $u''$ must also be orthogonal $i \tphi$. Since the vectors $\{  \phi, i\phi, \tphi , i\tphi \}$ span the whole $\CC^2$, we conclude that $u''$ must in fact be zero, and hence $u$ must belong to the subspace $\utwo$ of $\su$. The main discussion in section 2.4 then goes on to show that $u$ must be proportional to the matrix $\gamma_\phi$ defined in \eqref{A.4}.

\subsection{Weyl rescaling of $g_P$}
%\addcontentsline{toc}{subsection}{Weyl rescaling of $g_P$}

Let $\pi: (P, g_P) \to (M, g_M)$ be a Riemannian submersion with fibre $K$. Denote by $n$, $m$ and $k$ the real dimensions of these manifolds, so that $n = m+k$. Let $\Omega_M: \, M \to \mathbb{R}^+$ be any positive function on the base and let $\Omega := \pi^\ast\, \Omega_M$ be the corresponding lift to $P$ as a function constant along the fibres. The Weyl rescalings of the metrics $g_P$ and $g_M$ are defined by
\[ 
\tilde{g}_P\  := \  \Omega^2\, g_P  \qquad \qquad  \qquad  \tilde{g}_M\ := \ \Omega_M^2\, g_M \ .
\]
Then the projection $\pi: (P, \tilde{g}_P) \to (M, \tilde{g}_M)$ is still a Riemannian submersion. The volume forms on $P$, $K$ and $M$ transform according to
\beq  \label{B.1}
\vol_{\tg_\PPP}  \ = \  \Omega^n \, \vol_g    \qquad \qquad  \vol_{\tg_\KKK}  \ = \  \Omega^k  \, \vol_{g_\KKK}  \qquad \qquad  \vol_{\tg_\MMM}  \ = \  \Omega_M^m \, \vol_{g_\MMM} \ .
\eeq
A well-known formula for the transformation of the scalar curvature under a rescaling of the metric says that \cite{Wald}
\beq \label{B.2}
R_{\tg_\PPP} \ = \ \Omega^{-2} \, \big[ \,  R_{g_\PPP} \, - \, 2(n-1)\, \Delta_{g_\PPP} ( \log \Omega)  \, - \, (n-1)(n-2)\,   |\dd  \log \Omega |^2_{g_\PPP}  \, \big] \ ,
\eeq
where $\Delta_{g_\PPP}$ denotes the scalar Laplacian on $P$ defined by the metric $g_P$. Moreover, from the general formula \eqref{ExplicitVectorN1} it can be  deduced that the mean curvature vector of the fibres transforms as
\beq \label{B.3}
\tilde{N} \ = \ \Omega^{-2} \, \big[ \,  N \, - k\, \grad_{g_\PPP} (\log \Omega)  \, \big] \ ,
\eeq
which is also as well-known formula. It implies that the norm of $N$ transforms as
\bal \label{B.4}
| \tilde{N} |^2_{\tg_\PPP} \ &= \  \Omega^2\, | \tilde{N} |^2_{g_\PPP} \ = \ \Omega^{-2}\, \big| \,  N \, - k\, \grad_{g_\PPP} (\log \Omega)  \, \big|^2_{g_\PPP} \linebr
&= \ \Omega^{-2}  \left\{  \, |N |^2_{g_\PPP}  \, + \, k^2 \,  |\dd \log \Omega |^2_{g_\PPP}  \, - \, 2\, k \,  (\dd \log \Omega)(N) \, \right\} \ . \nonumber
\end{align}
To compute the transformation rule of $\check{\delta} N$, start by recalling expression \eqref{DivergenceN2} with all the pull-backs $\pi^\ast$ explicitly written
\beq \label{B.5}
\check{\delta}_g N  \ = \   -  \pi^\ast \, \left[ \, \divergence_{g_\MMM} (\pi_\ast N)  \, \right]  \ .
\eeq
The right-hand side depends on the metric both through $N$ and through the divergence operator. For a fixed vector field $X$ in $M_4$, it follows from the general relation $\Lie_X \vol_g = (\divergence_g X) \vol_g $ and the rescaling rule for volume forms that the divergence of $X$ transforms under Weyl rescalings as
\beq \label{B.6}
\divergence_{\tg_\MMM} X \ = \ \divergence_{g_\MMM} X \ + \ m\, (\dd \log \Omega_M)(X) \ .
\eeq
At the same time, it follows directly from \eqref{B.3} that the push-forward $\pi_\ast N$ transforms as
\beq \label{B.7}
\pi_\ast\, \tilde{N} \ = \ \Omega_M^{-2} \, \big[ \,  \pi_\ast N \, - k\, \grad_{g_\MMM} (\log \Omega_M)  \, \big] \ .
\eeq
Combining \eqref{B.6} and \eqref{B.7}, a short calculation then shows that
\begin{multline} \label{B.8}
\divergence_{\tg_\MMM} (\pi_\ast \tilde{N})  \ = \  \Omega_M^{-2} \, \Big[ \, \divergence_{g_\MMM} (\pi_\ast N) \, -\, k\, \Delta_{g_\MMM} (\log \Omega_M) \linebr  + \, (m-2) (\dd \log \Omega_M) (\pi_\ast N) \, + \, k(2-m) \big|\dd \log \Omega_M \big|^2_{g_\MMM} \, \Big] \, .
\end{multline}
This is a function on the base $M$ and, according to \eqref{B.5}, we only have to pull it back to $P$ to obtain the desired formula for $\check{\delta}_{\tg_\MMM}  \tilde{N}$. Since $\pi$ is a Riemannian submersion and $\Omega = \pi^\ast\, \Omega_M$, it is clear that the last two terms pull-back very simply:
\bal \label{B.9}
\pi^\ast\, \Big[ \, (\dd \log \Omega_M )(\pi_\ast N)  \,  \Big]  \ &= \ (\dd \log \Omega)(N)  \nonumber  \linebr
\pi^\ast\,  |\dd \log \Omega_M |^2_{g_\MMM}  \ &= \ |\dd \log \Omega |^2_{g_\PPP} \ .
\end{align}
In addition, a formula obtained in \cite{Besson} says that the Laplacians in a Riemannian submersion are related by
\bal \label{B.10}
\Delta_{g_\PPP} (\log \Omega) \ &= \  \pi^\ast \big[ \,  \Delta_{g_\MMM} (\log \Omega_M) \, \big] \ + \  \Delta_{g_\KKK} (\log \Omega) \ - \ (\dd \log \Omega)(N) \nonumber \linebr
&= \  \pi^\ast \big[ \,  \Delta_{g_\MMM} (\log \Omega_M) \, \big] \ - \  (\dd \log \Omega)(N) \ ,
\end{align}
where the last equality uses that $\Omega$ is constant on the fibres $K$. Combining these formulae with definition \eqref{B.5}, we finally conclude that
\beq \label{B.11}
\check{\delta}_{\tg_\PPP}  \tilde{N} \ = \  \Omega^{-2} \, \Big[ \, \check{\delta}_{g_\PPP}  N \, + \, k\, \Delta_{g_\PPP} (\log \Omega) \linebr  - \, (m-3) (\dd \log \Omega) (N) \, + \, k \, (m-2) \big|\dd \log \Omega \big|^2_{g_\PPP} \, \Big] \, .
\eeq
Taking the preceding formulae for $R_{\tg_\PPP}$, $| \tilde{N} |^2_{\tg_\PPP}$ and $\check{\delta}_{\tg_\PPP}  \tilde{N}$, which all contain the same basic components, one can look for real constants $\alpha_1$ and $\alpha_2$ such that
\beq
R_{\tg_\PPP} \ + \ \alpha_1\,  | \tilde{N} |^2_{\tg_\PPP} \ + \ \alpha_2 \, \check{\delta}_{\tg_\PPP}  \tilde{N} \ = \ \Omega^{-2} \, \Big[ \,   R_{g_\PPP} \ + \ \alpha_1\,  | N |^2_{g_\PPP} \ + \ \alpha_2 \, \check{\delta}_{g_\PPP}  N  \, \Big]
\eeq
for every rescaling function $\Omega$. This defines a system of linear equations for $\alpha_1$ and $\alpha_2$ that has a solution for
\beq
\alpha_1 \ = \ \frac{n-1}{k} \, \Big(\, 2 \, - \, \frac{n-2}{k} \, \Big)   \qquad \qquad \qquad   \alpha_2 \ = \ 2\  \frac{n-1}{k} \  .
\eeq
Therefore the function on $P$
\beq
W_{g_\PPP} \ := \  R_{g_\PPP} \ +\ \frac{n-1}{k} \, \Big( 2  -  \frac{n-2}{k} \, \Big) \  \big| N \big|^2_{g_\PPP} \, + \, 2\, \frac{n-1}{k} \ \check{\delta}_{g_\PPP}  N 
\eeq
transforms simply as $W_{\tg_\PPP} \, = \, \Omega^{-2} \, W_{g_\PPP}$ under a rescaling of the submersion metric $g_P$.

\newpage

%%%%%%%%%%%%%%%%%%%%%%%%%%%%%%%%%%%%%%%%%%%%%%%%%%%%%%%%%%%%%%

\vspace{1cm}


\begin{thebibliography}{ZZZZ}
                           
\bibitem[AY]{Abe} K. Abe and I. Yokota: Volumes of compact symmetric spaces, \textit{Tokyo J. Math.} \textbf{20} (1997), 87--105.

\bibitem[BL]{Bailin} D. Bailin and A. Love: Kaluza-Klein theories, \textit{Rep. Prog. Phys.} \textbf{50} (1987), 1087--1170.

\bibitem[Ba]{Baptista} J. Baptista: Higher-dimensional routes to the Standard Model fermions, {\tt arXiv:2105.02901 [hep-th]}.

\bibitem[Bes]{Besse} A. Besse: \textit{Einstein manifolds}, Classics in Mathematics, Springer-Verlag, 1987.

\bibitem[Be]{Besson} G. Besson: A Kato type inequality for Riemannian submersions with totally geodesic fibers, \textit{Ann. Glob. Anal. Geom.} \textbf{4} (1986), 273--289.

\bibitem[Ble]{Bleecker} D. Bleecker: \textit{Gauge theory and variational principles}, Addison-Wesley, 1981. 

\bibitem[BD]{BD} T. Brocker and T. Dieck: \textit{Representations of compact Lie groups}, Graduate texts in Mathematics, Springer-Verlag, 1985.

\bibitem[CJ]{Coq} R. Coquereaux and J. Jadczyk: Geometry of multidimensional universes, \textit{Commun. Math. Phys.} \textbf{90} (1983), 79--100.

\bibitem[DNP]{Duff} M. Duff, B. Nilsson and C. Pope: Kaluza-Klein supergravity, \textit{Physics reports} \textbf{130} (1986), 1--142.

\bibitem[EBH]{EBH} F. Englert and R. Brout:  \textit{Phys. Rev. Lett.} \textbf{13} (1964), 321. \\
                                 G. Guralnik, C. Hagen and T. Kibble: \textit{Phys. Rev. Lett.} \textbf{13} (1964), 585. \\
                                 P. Higgs: \textit{Phys. Lett.} \textbf{12} (1964), 132. 

\bibitem[Ham]{Hamilton} M. Hamilton: \textit{Mathematical gauge theory: with applications to the Standard Model of particle physics}, Universitext, Springer International Publishing, 2017.

\bibitem[Her]{Hermann} R. Hermann: A sufficient condition that a mapping of Riemannian manifolds be a fibre bundle, \textit{Proc. Amer. Math. Soc.} \textbf{11} (1960), 236--242.

\bibitem[Ho]{Hogan} P. Hogan: Kaluza-Klein theory derived from a Riemannian submersion, \textit{J. Math. Phys.} \textbf{25} (1984), 2031--2035.

\bibitem[Jen]{Jensen} G. Jensen: The scalar curvature of left-invariant Riemannian metrics, \textit{Indiana U. Math. J.} \textbf{20} (1971), 1125--1144.

\bibitem[K]{Original} T. Kaluza: \textit{Sitzungsber. Preuss. Akad. Wiss. Berlin Math. Phys.} \textbf{K1} (1921), 966. \\   % Happy 100th birthday!
                           O. Klein: \textit{Z. Phys.} \textbf{37} (1926), 895. \\
                           B. DeWitt: \textit{Lectures at 1963 Les Houches School}, Gordon and Breach, 1964. \\
                           R. Kerner:  \textit{Ann. Inst. H. Poincar\'e}  \textbf{9} (1968), 143. \\
                           A. Trautman: \textit{Rep. Math. Phys.} \textbf{1} (1970), 29. \\
                           Y. Cho: \textit{J. Math. Phys.} \textbf{16} (1975), 2029. \\
                           Y. Cho and P. Freund: \textit{Phys. Rev.} \textbf{D12} (1975) 1711. \\
                           J. Scherk and J. Schwarz: \textit{Phys. Lett.} \textbf{57B} (1975), 463. \\
                           E. Cremmer and J. Scherk: \textit{Nucl. Phys.} \textbf{B108} (1976), 409.

\bibitem[Mil]{Milnor} J. Milnor: Curvature of left invariant metrics on Lie groups, \textit{Adv. Math.} \textbf{21} (1976), 293--329.

\bibitem[O'Ne]{ONeill} B. O'Neill: The fundamental equations of a submersion, \textit{Michigan Math. J.} \textbf{13} (1966), 459--469.

\bibitem[OW]{WessonOverduin} J. Overduin and P. Wesson: Kaluza-Klein gravity, \textit{Physics reports} \textbf{283} (1997), 303--380.

\bibitem[MS]{MS} N. Manton and P. Sutcliffe: \textit{Topological Solitons}, Cambridge Univ. Press, 2004. 

\bibitem[Wald]{Wald} R. Wald: \textit{General relativity}, Chicago Univ. Press, 1984. 

\bibitem[Wei]{Weinberg67} S. Weinberg:  A model of leptons, \textit{Phys. Rev. Letters} \textbf{19} (1967), 1264--1266. 

\bibitem[Wei2]{Weinberg} S. Weinberg: \textit{The quantum theory of fields, vol. 2}, Cambridge Univ. Press, 1996. 

\bibitem[Wi1]{Witten81} E. Witten: Search for a realistic Kaluza-Klein theory, \textit{Nucl. Phys.} \textbf{B 186} (1981), 412--428.

\bibitem[Wi2]{Witten83} E. Witten: Fermion quantum numbers in Kaluza-Klein theory, in \textit{Shelter Island II, Proceeding of the 1983 Shelter Island conference}, MIT Press, 1985, 227--277.

\end{thebibliography}
\end{document}